\documentclass[]{aa} 
\usepackage{amsmath}
\usepackage{graphicx}
\usepackage{caption}
\usepackage{subcaption}
\usepackage{txfonts}
\usepackage{natbib}
\usepackage{xcolor}
\usepackage{multirow}
\usepackage{soul}

\graphicspath{{../figures/}}

\bibpunct{(}{)}{;}{a}{}{,}

\def\teff{\ifmmode T_{\rm eff} \else $T_{\mathrm{eff}}$\fi}

\def\ltsima{$\buildrel<\over\sim$}
\def\lsim{\lower.5ex\hbox{\ltsima}}

\newcommand{\ha}{\ifmmode {\rm H}\alpha \else H$\alpha$\fi}
\newcommand{\hb}{\ifmmode {\rm H}\beta \else H$\beta$\fi}
\newcommand{\lya}{\ifmmode {\rm Ly}\alpha \else Ly$\alpha$\fi}

\newcommand{\ebv}{\ifmmode E_{\rm B-V} \else $E_{\rm B-V}$ \fi}
\def\micron{$\mu$m}

\def\msun{\ifmmode \rm{M}_{\odot} \else M$_{\odot}$\fi}
\newcommand{\msunyr}{\ifmmode \rm{M}_{\odot} {\rm yr}^{-1} \else M$_{\odot}$ yr$^{-1}$\fi}
\def\zsun{\ifmmode Z_{\odot} \else Z$_{\odot}$\fi}
\def\lsun{\ifmmode \rm{L}_{\odot} \else L$_{\odot}$\fi}

\def\mup{\ifmmode M_{\rm up} \else M$_{\rm up}$\fi}
\def\mlow{\ifmmode M_{\rm low} \else M$_{\rm low}$\fi}

\def\aap{A\&A}

\def\apjl{ApJL}
\def\apjs{ApJS}

\newcommand{\oh}{\ifmmode 12 + \log({\rm O/H}) \else$12 + \log({\rm
O/H})$\fi}
\def\hyperz{{\em Hyperz}}
\def\flyf{\ifmmode f_{\rm Lyf} \else $f_{\rm Lyf}$\fi}
\def\pz{\ifmmode P(z) \else $P(z)$\fi}
\def\ki2{\ifmmode \chi^2_\nu \else $\chi^2_\nu$\fi}
\def\zphot{\ifmmode z_{\rm phot} \else $z_{\rm phot}$\fi}

\newcommand{\xphot}{\ifmmode x_\gamma \else $v_\gamma$\fi}
\newcommand{\xobs}{\ifmmode x_{\rm obs} \else $x_{\rm obs}$\fi}
\newcommand{\xcmf}{\ifmmode x_{\rm CMF} \else $x_{\rm CMF}$\fi}
\newcommand{\vexp}{\ifmmode V_{\rm exp} \else $V_{\rm exp}$\fi}
\newcommand{\vmax}{\ifmmode V_{\rm max} \else $V_{\rm max}$\fi}
\newcommand{\nh}{\ifmmode N_{\rm HI} \else $N_{\rm HI}$\fi}
\newcommand{\dv}{\ifmmode \Delta v({\rm em-abs}) \else $\Delta v({\rm em}-{\rm abs})$\fi}

\def\fesc{\ifmmode f_{\rm esc} \else $f_{\rm esc}$\fi}

\def\frellya{\ifmmode f^{\rm rel}_{\rm{Ly}\alpha} \else $f^{\rm rel}_{\rm{Ly}\alpha}$\fi}

\newcommand{\mstar}{\ifmmode M_\star \else $M_\star$\fi}
\newcommand{\muv}{\ifmmode M_{1500} \else $M_{1500}$\fi}
\newcommand{\luv}{\ifmmode L_{\rm UV} \else $L_{\rm UV}$\fi}
\newcommand{\lfir}{\ifmmode L_{\rm IR} \else $L_{\rm IR}$\fi}
\newcommand{\lir}{\ifmmode L_{\rm IR} \else $L_{\rm IR}$\fi}
\newcommand{\lbol}{\ifmmode L_{\rm bol} \else $L_{\rm bol}$\fi}
\newcommand{\liruv}{\ifmmode L_{\rm IR+UV} \else $L_{\rm IR+UV}$\fi}
\newcommand{\liroveruv}{\ifmmode L_{\rm IR}/L_{\rm UV} \else $L_{\rm IR}/L_{\rm UV}$\fi}

\newcommand{\sfrbc}{\ifmmode {\rm SFR}_{\rm SED} \else SFR$_{\rm SED}$\fi}	
\newcommand{\sfrir}{\ifmmode {\rm SFR}_{\rm IR} \else SFR$_{\rm IR}$\fi}
\newcommand{\sfruv}{\ifmmode {\rm SFR}_{\rm UV} \else SFR$_{\rm UV}$\fi}
\newcommand{\tdust}{\ifmmode T_{\rm dust} \else $T_{\rm dust}$\fi}
\newcommand{\tpeak}{\ifmmode T_{\rm peak} \else $T_{\rm peak}$\fi}
\newcommand{\av}{\ifmmode A_{\rm V} \else $A_{\rm V}$\fi}
\newcommand{\afuv}{\ifmmode  A_{\rm FUV} \else $A_{\rm FUV}$\fi}
\newcommand{\auv}{\ifmmode  A_{\rm UV} \else $A_{\rm UV}$\fi}
\newcommand{\sfruvir}{\ifmmode \rm SFR_{\rm UV+IR} \else $ \rm SFR_{\rm UV+IR} $\fi}	
\newcommand{\zspec}{\ifmmode \rm z_{spec} \else  $\rm z_{spec}$\fi}
\newcommand{\taf}{\ifmmode \tau \else $\tau$\fi}
\newcommand{\uvj}{\emph{UVJ}}

\begin{document}
  \title{Insights on star-formation histories and physical properties of $1.2 \leq z \lesssim 4 $ \textit{Herschel}-detected galaxies} 
  \subtitle{}
  \author{P. Sklias\inst{1,6}, D. Schaerer\inst{1,2}, D. Elbaz\inst{3}, M. Pannella\inst{4}, C. Schreiber\inst{5}, A. Cava\inst{1}  }
  \institute{
Observatoire de Gen\`eve, Universit\'e de Gen\`eve, 51 Ch. des Maillettes, 1290 Versoix, Switzerland
\and
CNRS, IRAP, 14 Avenue E. Belin, 31400 Toulouse, France
        \and
Laboratoire AIM-Paris-Saclay, CEA/DSM/Irfu - CNRS - Universit\'e Paris Diderot, CEA-Saclay, pt courrier 131, F-91191 Gif-sur-Yvette, France
\and
Faculty of Physics, Ludwig-Maximilians Universit\"at, Scheinerstr.\ 1, 81679 Munich, Germany 
\and
Leiden Observatory, Leiden University, NL-2300 RA Leiden, The Netherlands
\and
Departamento de Astrof\'isica, Facultad de CC. F\'isicas, Universidad Complutense de Madrid, E-28040 Madrid, Spain}

\authorrunning{P. Sklias et al.}
\titlerunning{Insights on star-formation histories of $1.2 \leq z \lesssim 4 $ galaxies with GOODS-\textit{Herschel}}

\date{Received 18 February 2016; accepted 27 April 2017}

\abstract{}
{We aim to test the impact of using variable star-forming histories (SFHs) and the use of the IR luminosity as a constrain on the physical 
parameters of high redshift dusty star-forming galaxies. 
We explore in particular the properties (SFHs, ages, timescales) of galaxies depending on their belonging to the ``main sequence'' of star-forming galaxies (MS).
}
{ We performed spectral energy distribution (SED) fitting of the UV-to-NIR and FIR emissions of a moderately large sample of GOODS-\emph{Herschel} galaxies, for which rich
multi-wavelength, optical to IR observations are available. We tested different SFHs and the impact of imposing  energy conservation in the 
SED fitting process, to help with issues like the age-extinction degeneracy and produce SEDs consistent with observations. 
}
{Our simple models produce well constrained SEDs for the broad majority of the sample (84\%), with the notable exception of the very high \lir\ end, for which 
we have indications that the energy conservation hypothesis  cannot hold true for a single component population approach. We observe trends in the preferences in SFHs among 
our sources depending on stellar mass \mstar\ and $z$. Trends also emerge in the characteristic timescales of the SED models depending on the location on the 
SFR -- \mstar\ diagram. We show that whilst using the same available observational data, we can produce galaxies less star-forming than classically inferred,
if we allow rapidly declining SFHs, while properly reproducing their observables. These sources, representing 7\% of the sample, can be post-starbursts undergoing quenching, 
and their SFRs are potentially overestimated if inferred from their \lir. Based on the trends observed in the rising SFH fits
we explore a simple evolution model for stellar mass build-up over the considered time period.
}
{ Our approach successfully breaks the age-extinction degeneracy, and enables to evaluate properly the SFRs of the sources in the SED fitting process. 
Fitting without the IR constrain leads to a strong preference for declining SFHs, while its inclusion increases the preference of rising SFHs, more so at high $z$,
in tentative agreement with the cosmic star-formation history (CSFH), although this result suffers from poor statistics.
Keeping in mind that the sample is biased toward high luminosities and intense star formation,
the evolution shaped by our model appears as both bursty (in its early stages) and steady-lasting (later on). 
The SFH of the sample considered as a whole follows the CSFH with a surprisingly small scatter, 
and is compatible with other studies supporting that the more massive galaxies have built 
most of their mass earlier than lower mass galaxies.
}


  \maketitle

\section{Introduction}
 
Recent years have seen significant advance in the understanding of the star-formation history of the Universe. Large extragalactic surveys 
covering various intervals of the electromagnetic spectrum have helped defining the broad lines
of the cosmic star-formation history (CSFH), and its role in galaxy evolution. It is now established that star formation in the Universe has followed
a particular evolution, rising steeply 
since the Big Bang, reaching a peak between redshifts $z=$ 1.5 and $z=$ 3, 
and steadily declining since
\citep[e.g.,][for a review]{1998ApJ...498..106M,perez2005,2008MNRAS.388.1487L,madau2014}.
Observations at different wavelengths are necessary to probe the total star-forming activity of galaxies, especially at high-$z$, where it was proven
that star formation is increasingly dust-obscured \citep{CE01,2005ApJ...632..169L}.
In the pre-\textit{Herschel} era, studies utilizing restframe UV to mid infrared (MIR) photometry have highlighted the existence of a tight
correlation between stellar mass (\mstar) and star-formation rate (SFR) \citep[e.g.,][]{2007ApJ...660L..43N,2007ApJ...670..156D,elbaz2007},
commonly addressed today as the ``main sequence'' (MS) of star-forming galaxies (SFGs).
More recently, thanks to \textit{Herschel}  
it has been made possible to observe and characterize the bulk of star-forming activity  
in the far infrared (FIR) up to relatively high redshifts \citep[e.g.,][]{2010A&A...518L..25R,2010MNRAS.409L...1B,2012ApJ...744..154R},
and to better constrain the MS and galaxy properties, and their evolution with $z$ \citep[e.g.,][]{2011ApJ...739L..40R,pannella2015,schreib2014,2015arXiv151006072T}.

Traditionally, the SFR in a galaxy is estimated through the conversion of fluxes measured in certain wavelength intervals or nebular line emission
measurements \citep[e.g.,][]{erb2006,2010ApJ...712.1070R,2010ApJ...721..193P}, with the most popular calibrations found in \cite{K98}. Today, thanks to the broad
coverage of the electromagnetic spectrum made possible by observatories like HST, \emph{Spitzer, Herschel} and many others, the joint consideration
of the dust-obscured UV-inferred and the IR-inferred SFRs allows to properly assess the total star formation that occurred over the past hundred million years
or so, prior to the observation.

A common method to characterize the stellar population of a galaxy and constrain its physical parameters (such as \mstar, age, attenuation, etc.)
is to perform spectral energy distribution (SED) fitting of its photometry. Various approaches exist for that, and usually some assumptions
are made on the star-formation histories (SFHs) and extinction laws that can impact more or less the estimation of the parameters \citep[some studies
highlighting this fact using various samples are:][among others]{2012A&A...541A..85M,2012ApJ...754...25R,2012ApJ...745...86W,debarros2014,sklias2014}.

When considering variable SFHs, past studies often opted preferably for SFHs with exponentially declining SFRs. Such SFHs are a ``natural'' choice
when working in the low-$z$ Universe ($z\lesssim1$) as they follow the cosmic trend, but for  higher redshift galaxies, many authors argue based on both
theoretical and observational considerations that rising SFRs are best suited to characterize them 
\citep{papovich2004,2009ApJ...698L.116P,renzini2009,maraston2010,finlatoretal2011}.
When variable SFHs are used in SED fitting, the inferred ongoing SFR (the instantaneous value provided by the SED fit) can often be very different from the 
observation-inferred SFR \citep[e.g.,][]{2011ApJ...738..106W,2012ApJ...754...25R,2013A&A...549A...4S}. 
Indeed, degeneracies between age and extinction, or age and the e-folding timescales can emerge and cast
uncertainty on the rest of the parameters and the SFH. A purely observation-based approach is not 
hindered by this as the SFRs, mass, and attenuation can be estimated to a first order by the observations themselves, 
but not much insight on the SFH can be obtained from that. 
Some studies tackle this issue by imposing a maximal age or slowly evolving timescales to obtain plausible quantities \citep[in terms of 
age for example,][]{maraston2010,2012A&A...545A.141B} or a better agreement between the observation-inferred and the SED-inferred SFRs \citep{2011ApJ...738..106W}.
 
In the present work we wish to explore the question of variable SFHs in a slightly different approach. 
By working on a relatively large sample of galaxies individually detected with \textit{Herschel} at $z >1$ we have access
to the two main observables from which the SFR is traditionally estimated, namely the UV and IR luminosities. 
With little limitation on the choice of SFHs, we aim to produce stellar populations that are coherent with these observables,
by imposing energy conservation in the fitting procedure. In this way we are not limited in aiming to reproduce the observation-inferred 
SFRs, although this is expected to happen for the fraction (or majority) of the sample for which the assumptions in calibrations 
such as that of \cite{K98} hold true. By allowing even rapidly varying SFHs we explore to which extent we find solutions that are different
than the Kennicutt inferred values, while remaining plausible concerning the reproducibility of the observables.
We focus on the effect the prior knowledge of the IR luminosity (\lir) can have on the traditional SED fitting procedures and on the
characterization of the galaxies' stellar population properties, and explore how the IR constraint helps in breaking degeneracies
like the ones mentioned in the previous paragraph.

A central question we wish to address in this work is the following: do galaxies tend to follow --
statistically speaking -- the trend of the CSFH on an individual level? 
Can we, with the help of the IR-constraint and the rich multi-wavelength photometry, distinguish
preferences in the type of SFH best fitting our sources, depending on their observed redshift? 
Ultimately, how does the IR-constraint affect the SFHs best characterizing the stellar populations of
our sources, and what can we learn about timescales
, in absence of spectroscopic information
(i.e., nebular emission measurements that allow to probe timescales well below the order of 100 Myr)? 
 
\begin{figure}[htb]
\hspace{-0.18 in}
\includegraphics[width=0.55\textwidth]{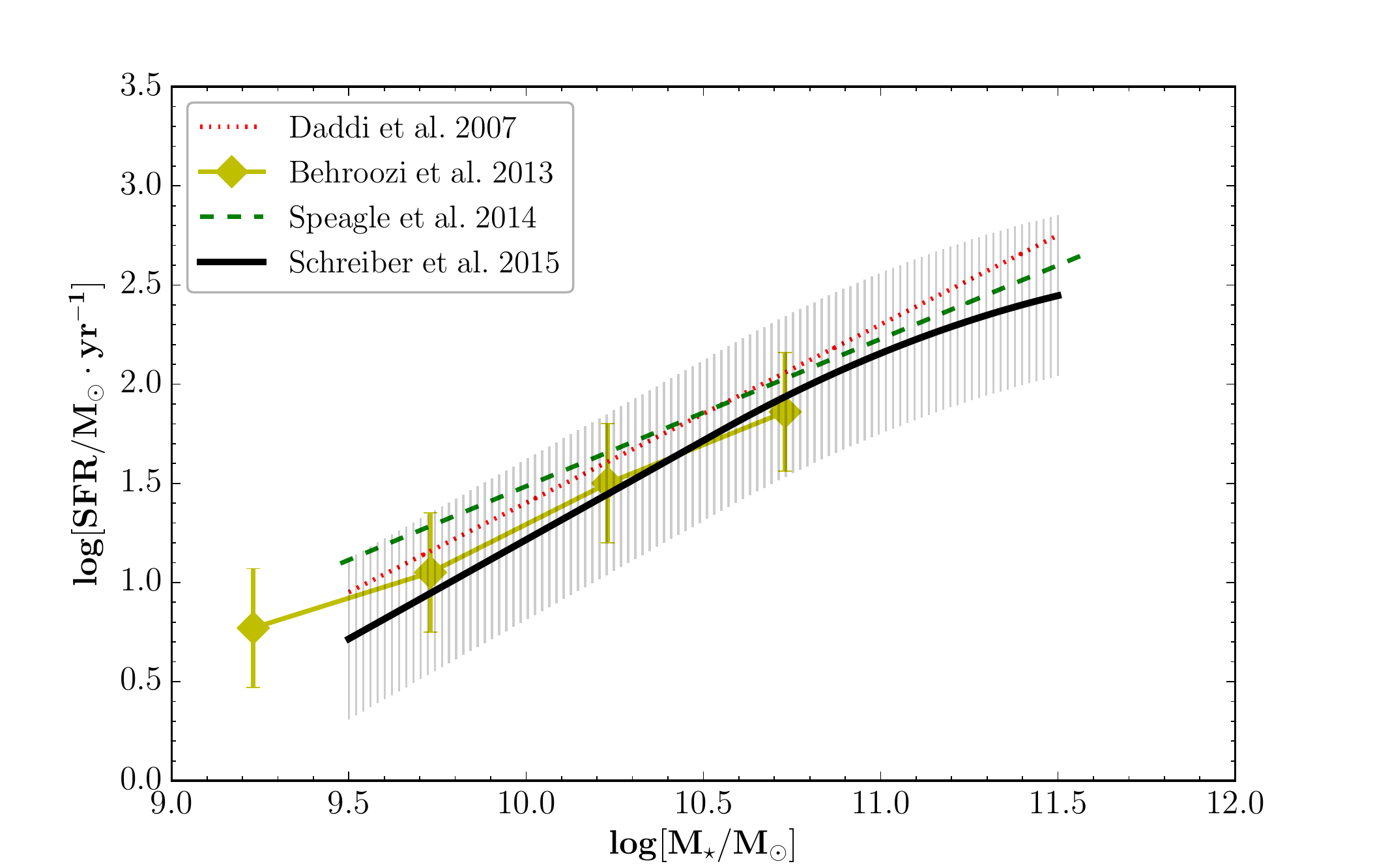}
\caption{ Various calibrations of the SFR -- Mass relation at $z\sim 2$, corrected for IMF differences. 
The hatched area indicates the band 4 times above and below the relation \citep[presently that of ][]{schreib2014}, 
in which we consider sources as MS galaxies. 
}
\label{f_ms_summary}
\end{figure}

We do not attempt to produce a reanalysis of the MS or its evolution; we use the latest MS parametrization 
of \cite{schreib2014} as reference, to describe our models in the \mstar -- SFR plane and as a function of redshift. 
The adopted MS is shown in Fig. \ref{f_ms_summary} together with other recent parameterizations
from the literature \citep{2007ApJ...670..156D, 2013ApJ...770...57B, speagle14}.
Based on the definition by \cite{2011ApJ...739L..40R}, galaxies found within 4 times the SFR of their corresponding MS are 
considered MS galaxies and those above it, starbursts. In our energy conserving models, we define a third category
that we find of special interest, the sources for which their SED-inferred SFR is found to be 4 or more times smaller than the
classically UV+IR inferred value and that are mainly below the MS. 
For simplicity, we label them as ``quenching''.
Later on, we discuss more in detail the choice of this definition, and the context from which it emerges.

Our paper is organized as follows: 	
in Sect. \ref{s_sample} we describe the sample used for this work, and
in Sect. \ref{s_method} we present the SED fitting method, the explored SFHs, and definitions. 
In Sect.  \ref{s_results}, we go through the results by first discussing how well the \lir\ is reproduced 
before and after the IR information is used as a constraint and how 
the physical parameters are affected by this. Then we discuss the timescales and the general picture of 
star formation as shaped by our method, and the trends between the different SFHs and other parameters such as
\mstar\ and $z$. Implications of the results are discussed in Sect. \ref{s_implic},
notably on how the galaxies would evolve if their SFH followed the trends we observe in the solutions with rising 
SFRs, how the sample in its whole is perceived to have evolved since the very early Universe 
with the SFHs we obtain, and lastly a discussion on the quenching sources and the possibility of
their spectro-photometric discrimination from the main population of actively star-forming galaxies.  
Our summary and conclusions follows in Sect.  \ref{s_conclu}.
Finally, there are two short appendices, one that discusses the robustness of the obtained solutions, and one presenting examples of 
SED fits, to illustrate how the explored approach impacts the fitting and interpretation of our sample and to highlight some of the discussed
aspects.

For our mass estimates we use a Salpeter IMF, from 0.1 to 100 \msun.
We adopt a $\Lambda$-CDM cosmological model with $H_{0}$=70 km s$^{-1}$ Mpc$^{-1}$,
$\Omega_{m}$=0.3, and $\Omega_{\Lambda}$=0.7.


\section{Observations - Selection}	\label{s_sample}

We use the public GOODS-\textit{Herschel} \citep{elbaz2011} source catalogs for both GOODS-N and GOODS-S fields.
For GOODS-N the catalog consists of $K_s$-band detected sources with \textit{Herschel} counterparts \citep[the details
for its creation and the sources of the various bands used can be found in][]{pannella2015}.
The optical data are from the Subaru's Suprime-Cam bands BVRIzY, complemented by KPNO's \textit{U}-band, 
and the NIR data are from CFHT's Wircam, Subaru's MOIRCS and \textit{Spitzer}'s IRAC (all available bands).
For GOODS-S we use the catalog of the official CANDELS release \citep{2013ApJS..207...24G}, created from $H$-band detected
sources. The utilized bands in our work are the \textit{U}-band from CTIO and VLT/VIMOS, 
HST/ACS and WFC3's F435W, F606W, F775W, F814W, F850LP, F105W, F125W, F160W 
broad bands and the F098M medium band, the $K_s$ bands from ISAAC and Hawk-I, and again all IRAC bands.
 
\begin{figure}[htb]
\centering
\includegraphics[width=0.45\textwidth]{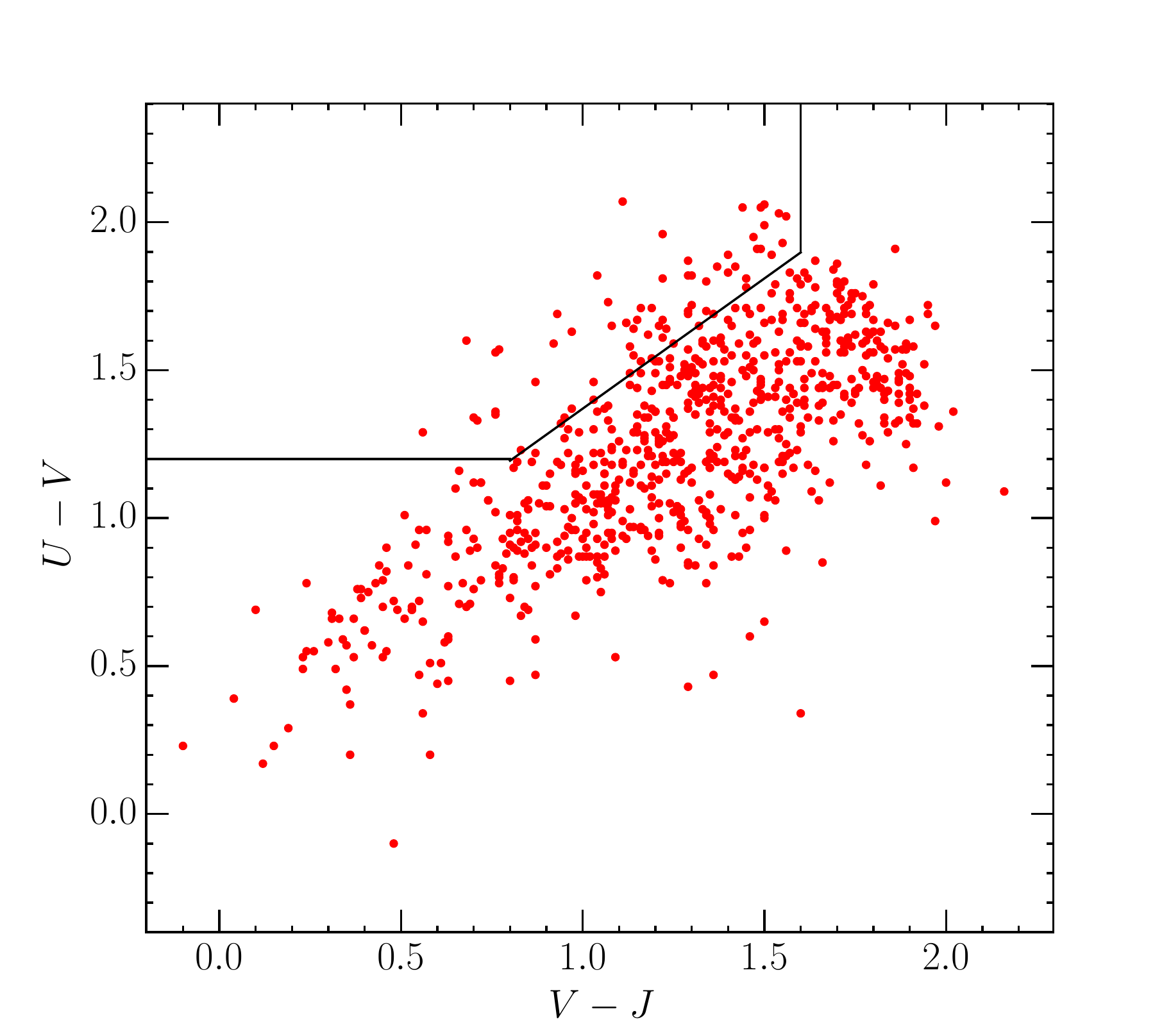}
\caption{\uvj\ diagram of the \textit{Herschel}-detected sources with $z \geq 1.2$.
The restframe magnitudes are obtained by photometric interpolation and originate from the work of \cite{schreib2014}.
The line separating the actively star-forming sources (bottom-right) from quiescent (top left) is the \cite{williams09}
adopted definition for $z \in [1,2]$.}
\label{uvj-fig}
\end{figure}
 
\begin{figure}[htb]
\hspace{-0.2 in}
\includegraphics[width=0.55\textwidth]{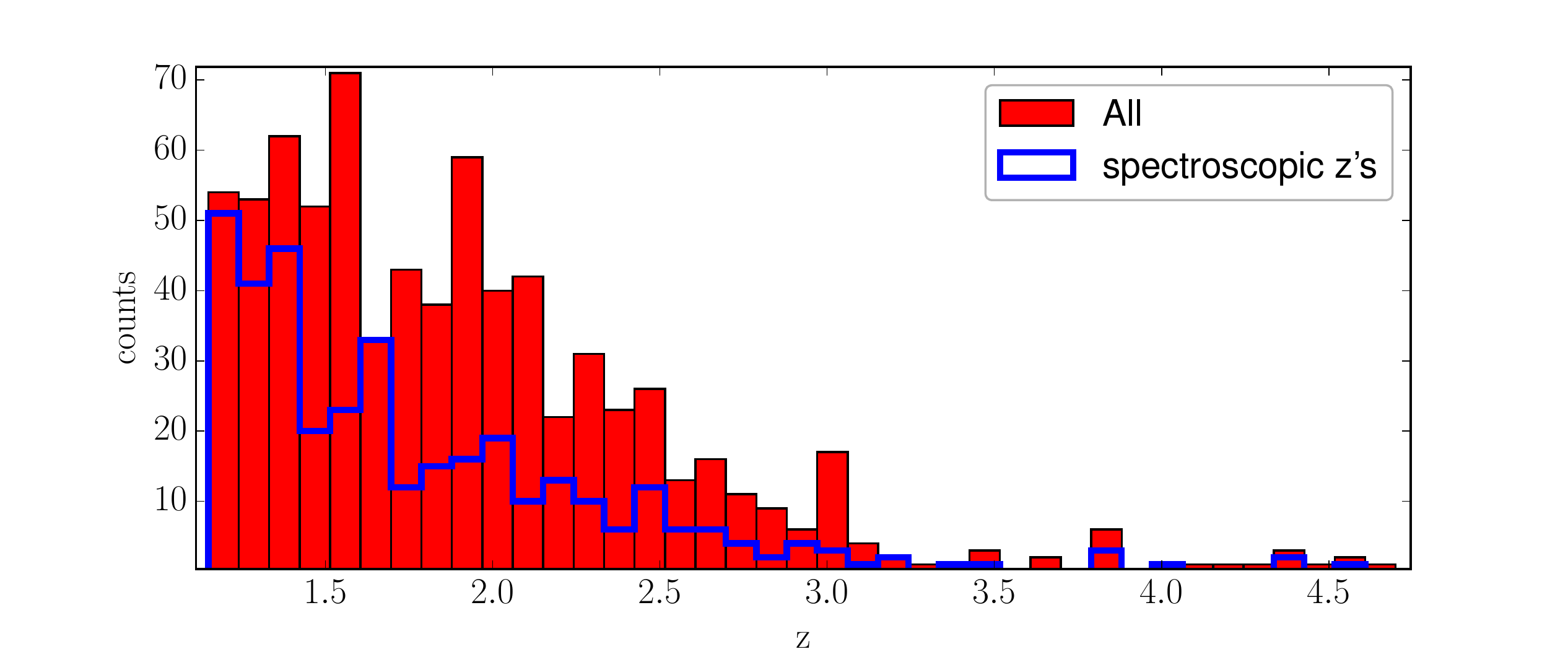}
\caption{Redshift distribution of the initial sample, with in blue the \zspec's. Sources are selected
above  $z\sim 1.2$ and reach out to $z \sim 4.7$. The photometric redshifts used are obtained from the present work.}
\label{z-fig}
\end{figure}

We select all sources with a spectroscopic redshift \zspec\ $\geq 1.2 $ 
as well as objects with well-defined photometric redshifts \zphot\ $\geq 1.2 $ 
(cf. Sect. \ref{s_method}). The \zspec\ measurements are from \cite{barger2008} and Stern et al., \emph{in prep.} for GOODS-N,
and from \cite{2005A&A...437..883M}, \cite{2008A&A...482...21C} and \cite{2008A&A...478...83V} for GOODS-S.
The choice of $z = 1.2$ as the minimum redshift allows for the best sampling of the UV emission, that is sufficiently redshifted to 
the available observing bands. This way accurate estimations of the UV luminosity (\luv) can be made for all sources.
 We do not wish to reduce our selection to actively star-forming galaxies only, as we are also interested in modeling 
IR-bright galaxies that might undergo more moderate star formation. To that end we do not \textit{de facto} exclude  sources that
are classified as quiescent by the \uvj\ color-color criterion of \cite{williams09}\footnote{The removal of such sources is important in studies using the stacking
of sources \citep[as in][among others]{pannella2015,schreib2014,2015arXiv151006072T} to reach fainter luminosities and more mass-complete samples. 
If the quiescent sources are not removed then, the stacked flux is ``diluted'' incorrectly. 
In the present study, this is not an issue as we focus only on individually detected sources.}. 
Based on this criterion 8\% of the 
selected sources qualify as quiescent, but marginally, not forming a separate group on the $U-V$ vs $V-J$ plane 
(see Fig. \ref{uvj-fig}). An important fraction of them   
is also located at $z\geq3$ which is very unlikely for non-AGN \emph{Herschel}-detected sources. 
Misclassifications are expected because of the uncertainties in the \uvj\ colors \citep{schreib2014}, which evidently increase
with $z$. However, at $z \leq 1.5$ a small fraction of  weakly star-forming galaxies is expected to be detected with \emph{Herschel}. 
We thus start with the complete sample of \emph{Herschel}-detected sources,
which includes 753 sources, 365 sources with  \zspec.
The redshift distribution is shown in Fig. \ref{z-fig}.

Our sample is \textit{Herschel}-selected, meaning we require at least one detection in either PACS or SPIRE bands. 
All galaxies are detected in the \textit{Spitzer}  MIPS 24\micron\ band by construction, since the \textit{Herschel}  photometry was 
extracted using the position of 24\micron\ detections as prior information.
As \textit{Herschel}'s SPIRE bands have very coarse resolution and hence the photometry of some sources can be significantly blended,
we do not use this instrument's photometric data for sources that are flagged as ``non-clean'' \citep[according to the criterion introduced by][]{elbaz2011}.

In total, up to seven bands in the MIR-FIR are available to 
derive the luminosities: \textit{Spitzer}'s  16\micron\ and  24 \micron,  
and \textit{Herschel}'s  100, 160,  250, 350, and 500 \micron\ bands.

Naturally, a sample constructed in this way is luminosity-limited, which introduces a bias toward high \lir\ at
high $z$. The sample is not meant to be mass-complete, especially at higher $z$, so it cannot be considered as being representative 
of all galaxies or star formation across various redshifts.
Being that we require the galaxies to be individually detected by \emph{Herschel}, this yields a sample with a larger fraction of starbursts 
than in a mass-selected sample \cite[e.g., ][]{2011ApJ...739L..40R}.
\cite{schreib2014} present a more detailed discussion on the matter of the bias of individually detected galaxies against the general population.
However, its study allows for some insightful comments in the more general behavior of 
star-forming galaxies as will be discussed later on.

For the SED fits described below, we have imposed a minimal error of 0.1 mag to all data. Although for some observations (e.g., in
the HST bands) the precision is better than this, it is more appropriate  to account for the uncertainty on the calibration and other aperture corrections
when combining the photometry from many different instruments. 

\section{Method} \label{s_method}

\subsection{SED modeling}	\label{sed_mod}

We used a modified version of the \hyperz\ photometric redshift code of \cite{bolzonellaetal2000}, 
described in \cite{schaerer&debarros2009,schaerer&debarros2010,sklias2014}. Primarily designed to derive
redshifts from broad-band SED fits of UV to NIR photometry, our version is adapted in order to use data up
to the submillimeter range with corresponding templates of UV to submm defined libraries. It also has the 
ability to include line emission when performing stellar population SED fitting.

Overall we followed the approach of \cite{sklias2014}, performing separate fits of the stellar and dust emission.
Stellar population SED fitting is done using the \cite{BC2003} library (BC03, hereafter), using variable star-formation histories,
including constant SFR (CSFR hereafter), exponentially declining and rising (sometimes called \textit{direct-}$\tau$ and \textit{inverted}-$\tau$
models in the literature, respectively, for example, \cite{maraston2010}), and  delayed ( $\propto t/\tau \cdot e^{-t/\tau} $), with various e-folding
timescales\footnote{The available timescales for each SFH are: [0.03, 0.05, 0.07, 0.1, 0.3, 0.5, 0.7, 1, 3] Gyr 
for the declining; [0.03, 0.05, 0.07, 0.1, 0.3, 0.5, 0.7, 1, 2, 3, 4, 5] Gyr for the delayed; and 
[0.03, 0.05, 0.07, 0.1, 0.3, 0.5, 0.7, 1, 2, 3] for the rising.} $\tau$, starting at 30 Myr. 
For all our stellar SED fits we fix a minimal age of 50 Myr for the populations, to avoid extremely young solutions 
deemed unphysical in comparison with the dynamical timescale of galaxies \citep{2012ApJ...754...25R,2012ApJ...745...86W}. 
This and the maximal age allowed at a given $z$ by the age of the Universe are the only priors applied, and the steps in \taf\ and $t$ have
flat probability distributions. 
We note that these choices do not limit in any way the reproduction of the photometry of IR-bright galaxies; extremely young ages
can be of use in UV-selected samples \citep[e.g.,][]{debarros2014}, and the non-impact of \taf\ values below 30 Myr is discussed in Sect. \ref{s_timesc}. 
From this we obtain stellar masses, ages (defined the age as the time passed since the beginning of star formation), 
\luv, the UV continuum slope $\beta$, and instantaneous SFR values.
The UV luminosities  that we use in our analysis are taken as
 $ \lambda\cdot F_{\lambda} $, averaged over 1400-2200 \AA, with $ \lambda_{\rm eff} = 1800$ \AA, 
and the slope $\beta$ over the interval of 1300-1800 \AA. 
The measure of $\beta$ from SED fits  is model-depended to a certain extent, as it can be
  affected by the SFH, the attenuation law, and also on the wavelength interval considered
  For this reason they should not be considered as substitutes of spectroscopically-measured slopes.
  In Sect. \ref{s_free} we compare our estimates with the slopes obtained thanks to the method proposed by \cite{2012ApJ...756..164F}.

%

Throughout the text we use the abbreviations DECL, RIS, DEL, and CSFR, for the declining, rising, delayed, and constant SFR SFHs mentioned, respectively.
The stellar populations produced as described above are referred in the text as single component populations, 
in contrast with multiple-component populations, or SFHs superimposed
with bursts used in other works. They are not to be confused though with what is often defined as simple
 stellar population (SSP) in the literature, which
refers to a population issued from an instantaneous burst at a given time, without any star formation occurring afterwards.
Dust attenuation is applied following the Calzetti law \citep{2000ApJ...533..682C} on the whole sample and for the main body of this work. 
The Small Magellanic Cloud (SMC) and Milky Way laws \citep[][respectively]{prevot84,seaton79} have also been  explored for a small subsample 
for which the former yielded some incompatible results (Sect. \ref{s_overpred}).
  For all our fits  the metallicity is kept to \zsun\ to avoid the degeneracies that rise from leaving it a free parameter. IR-bright galaxies with important
amounts of dust are not expected to have ongoing star-formation of metal poor stars. However, \cite{pannella2015} show in their work
that a metallicity evolution does exist with $z$ and the \liroveruv\ ratio for such galaxies. Using the fundamental metallicity relation of \cite{mannucci2010}
we have checked that 90\% of our sample is confined in the range of $12+\log({\rm O/H}) = [8.4, 8.9]$, that is, ranges from slightly sub-solar to 
super-solar.

The redshifts were fixed to \zspec\ for the spectroscopically confirmed sources. 
For the others, we performed an additional step where we first fit the SED to derive \zphot. The redshift range for this is from 0 to 5 
(no \emph{Herschel} detections are expected beyond that).
%
The relative accuracy we reach for the sources with \zspec\ (defined as $\Delta z = (\zphot - \zspec ) / (1+z)$) is less than 1\%, with 8\%\ of sources
having $\Delta z \geq 0.2$.  Sources without a \zspec\ for which the \zphot's obtained from fits with different SFHs varied substantially (shallow or multi-peaked
probability distributions, with typical 68\%\ widths  reaching 0.1 in $z$) were excluded from the study (about ten
objects, mostly faint and with poor photometry).
In the following, the redshift of each galaxy was fixed either to its \zspec\ when available, or to the \zphot\ of the best fit otherwise.
This is important as redshift must not vary between the stellar emission fits and the IR fits, for the \lir\ estimates to be accurate and coherent.

A central point of the present work is that we performed two kinds of fits on the stellar emission of the sample (covering the range from the \textit{U}-band
to IRAC 8 \micron), and explored how the derived parameters are affected by the prior knowledge of the observed \lir:
\begin{itemize}
 \item[\emph{a)}] SED fits that use extinction as a free parameter (\ki2\ minimization), with \av\ ranging from 0 to 4, in steps of 0.1. 
This is the most common way SED fitting is performed on UV-NIR photometry in the 
 literature \citep[e.g.,][]{schaerer&debarros2009,maraston2010,2011ApJ...738..106W,schreib2014} 
 \item[\emph{b)}] SED fits where the extinction is fixed for each source through the observed \liroveruv\ ratio to ensure energy balance between
the stellar model and the IR luminosity \citep[as presented in][]{sklias2014}. 
The \liroveruv\ ratio is known to be an effective tracer
of UV attenuation \citep[e.g.][]{burgarellaetal2005,2010MNRAS.409L...1B,2012A&A...545A.141B,2013MNRAS.429.1113H}. 
By making use of the relation between \liroveruv\ 
and $A_{\rm UV} $  presented in \cite{2013A&A...549A...4S}, and by assuming a given extinction law, we constrain Av within 0.1 mag, in order to
produce solutions that account for the observed \lir. In \cite{sklias2014} we have shown that it can be very useful in breaking 
-- at least partially -- the degeneracies that can occur when modeling sources with variable SFHs.

\end{itemize}
Our version of \emph{Hyperz} calculates
the luminosity that is absorbed by dust in the interval [$912\AA - 3\mu$m]. Assuming basic energy conservation, this luminosity
is expected to be emitted in the IR  \citep[f.][]{2013A&A...549A...4S}. It is what is referred hereafter as the {\em predicted} \lir.

Based on the available photometry from 16 to 500 \micron, we derived robust estimates of the observed \lir\
used for our energy conserving fits, defined as the integrated flux over the restframe interval $ [8-1000]$ \micron.
To do so we used the following template libraries:
\begin{itemize}
\item  \cite{CE01}: a set of synthetic templates of varying IR luminosity;
\item  \cite{2009ApJ...692..556R}: a set of templates containing observed SEDs of local purely star-forming LIRGs and 
ULIRGs, and some models obtained by combining the aforementioned; 
\item  \cite{2008A&A...484..631V}: a set of templates produced from observed LIRGS and ULIRGS, starburst dominated;
\item  \cite{2013A&A...551A.100B}: a set of UV-to-submm \textit{Herschel}-motivated templates, created from various types of galaxies.
\end{itemize}
Multiple libraries were used to increase fit quality and for each source the best fit to the IR photometry provides the observed \lir.
The values obtained in this way were compared to the ones obtained by \cite{pannella2015} and are in good agreement
with a mean offset of 0.03 dex and a standard deviation of $\sim$0.14. After performing some fits on Monte-Carlo variations of the shallower GOODS-N
catalog, we note that 98\% of the sources have their \lir\ constrained with an uncertainty of less than $\pm0.2$ dex (while 86\% at less
than $\pm0.1$ dex), at the 68\% confidence level.

From the initially selected sample we retained a subsample for which we obtained reasonably good fits,
cutting at $\chi^2 = 10$ for the unconstrained extinction solutions applied on the CSFR models\footnote{The CSFR models yield on average
the largest \ki2's as they have one less degree of freedom with respect to the others.}. That leaves us with 704 sources,
about 93\%\ of the sample, which are well fitted using the standard approach.
Another advantage of this cut is that it eliminated sources that are very strongly affected by AGN. 
Indeed, for most of the redshift range considered in the present work, in the presence of strong AGN contamination 
the IRAC bands included in the stellar emission fits cannot be fitted well by a pure stellar component, as they can be increasing
as a power law longwards of the 1.6 \micron\ bump. This leads to large \ki2\ values 
that will therefore be flagged out by our \ki2\ selection introduced above.
This cleaned sample of 704 sources is used in Sections \ref{s_free} and \ref{s_caveat} to explore
how the initially well-fitted sources fare in reproducing the \lir\ and how the constrain on \av\
affects them.
Later on, when we focus on the energy conserving fits we apply a second cut on top of the \ki2\ selection.
Our energy conservation approach does not allow to accurately reproduce  the \lir\ for all the sources in the sample 
(this is discussed in Sections \ref{s_caveat} and \ref{s_overpred}). For our analysis we keep only the sources that reproduce
\lir\ within $\pm0.2$ dex, as motivated by the uncertainties in the estimation of the observed IR luminosity.
This leaves us with a final sample of 633 (84\% of the initial) sources 
which we will refer to as the \emph{energy conserving sample} hereafter. It is the focus of this work from Sect. \ref{s_ms} and on.

\subsection{Other definitions -- tools}

We use the redshift-dependent main sequence relation of \cite{schreib2014} as our reference, notably
for defining a distance from the MS for each source (we take the geometrical
distance from the curve in the log(\mstar) -- log(SFR) plane):

\begin{equation}
\begin{split}
  \log(\mathrm{SFR}[\msunyr]) = \log(\mathit{\mstar}[\msun]) - 9.5 + 1.5\log(1+\mathit{z})   \\  
 - 0.3[\mathrm{max}(0, \log(\mathit{\mstar}[\msun]) - 9.36 - 2.5\log(1+\mathit{z}))]^2    	\\
 \end{split}		\label{eq_ms}
\end{equation}

The distance from the MS can be interpreted as a proxy for the present-over-past SFR ratio, 
which is related to $t/\tau$ in exponentially declining SFHs. 
In the present work instead of showing SFR -- \mstar\ diagrams typically used in the 
literature \citep[e.g.,][]{2007ApJ...670..156D,2011ApJ...739L..40R,schreib2014} we opt to present them under the form 
of  MS-normalized SFR versus \mstar, meaning that the SFR will be divided by SFR$_{\rm MS}$. This is advantageous for several reasons: it
eliminates the scatter due to the width of the redshift spread of the sample, allows for the MS to be highlighted for all redshifts, and to
reveal immediately the position of any given source in relation with the MS, which is very useful to our discussion. It will be still referred to as the
SFR -- mass diagram, for short.
%
We also use the definition of ``starburstiness'' as defined in \cite{elbaz2011}:

\begin{equation}
  R_{\rm SB} = {\rm sSFR / sSFR_{\rm MS} }		\label{eq_rsb}
\end{equation}

where $\rm sSFR = SFR /$\mstar\ is the specific star-formation rate. It is a way to measure the 
distance at which a galaxy is from the MS.

The SFRs obtained from the observed UV and IR luminosities, \sfruv\ and \sfrir, are based on the calibration of 
\cite{K98}:
\begin{equation}
  \sfruv(\msunyr) = \frac{\luv (\lsun)} {3.08\times 10^{9}} \, ,\:   \sfrir(\msunyr) = \frac{\lir (\lsun)} {5.8\times 10^{9}}
\end{equation}

From the above, we take $ \sfruvir = \sfruv + \sfrir $ as the observation-derived total SFR, which we confront to 
the SED-derived \sfrbc\ of our models.


\section{Results}	\label{s_results}

First, we examined how well fits to the stellar SED can predict the IR luminosities and if/how the different
SFHs affect the prediction (Sect. \ref{s_free}). Then we focused on the energy conserving models, their limitations,
the impact on physical parameters and the general landscape of star formation for our sample (Sect. \ref{s_energy_cons}), 
ages and timescales (Sect. \ref{s_timesc}), and the star-formation histories (Sect. \ref{s_sfh}), where we explored in particular
if the SFHs best suiting our sample tend to follow the cosmic star-formation history.
 
\subsection{On the reproducibility of the observed \lir}	\label{s_free}

\begin{figure}[htb]
\centering
\includegraphics[width=0.5\textwidth]{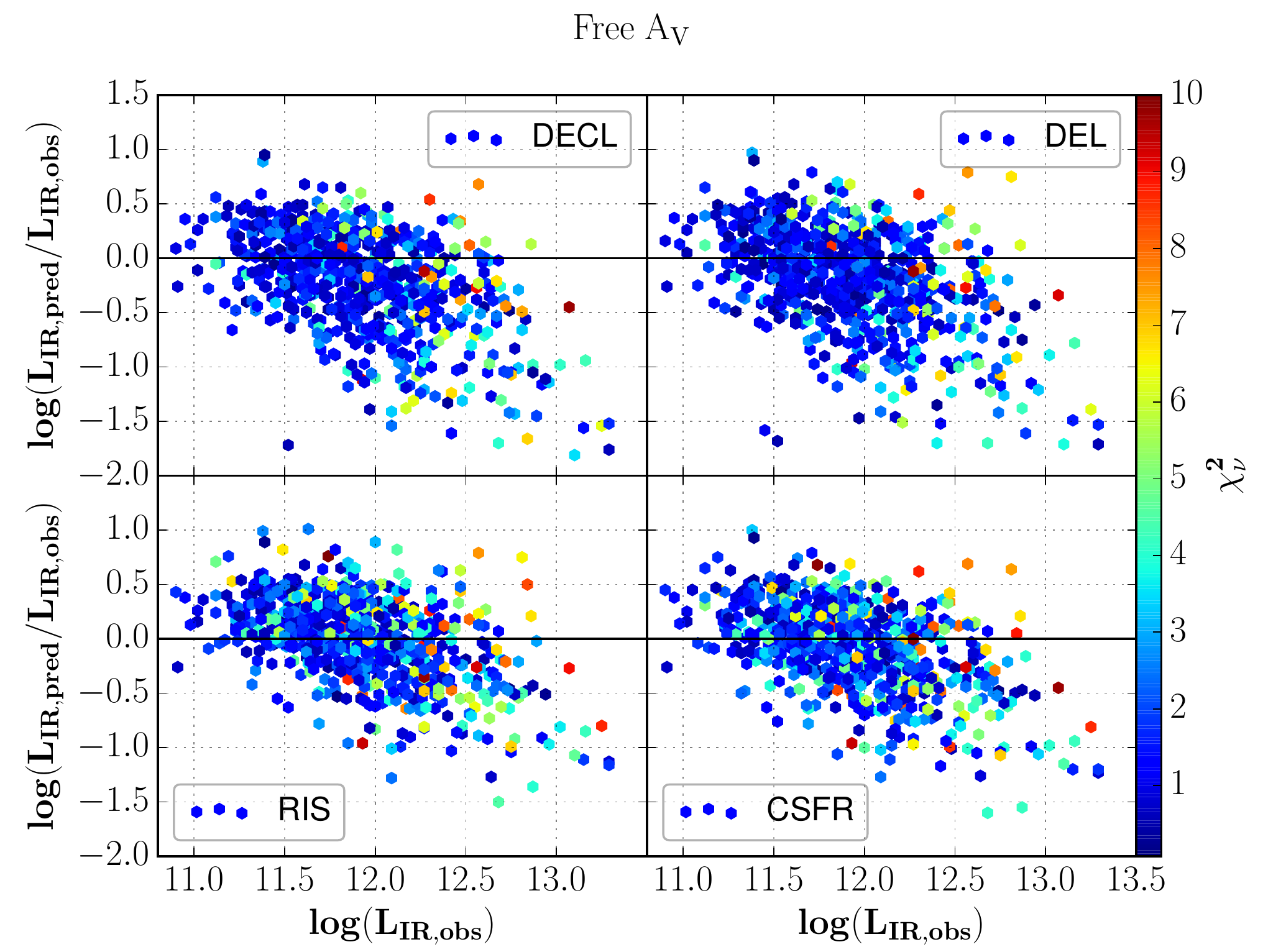}
\caption{ Ratio  of predicted/observed \lir\ vs observed \lir\ for the four SFH models explored, with the extinction fitted as a free parameter.
	  Color indicates \ki2, dark blue is $\ki2 \lesssim 0.5$ and dark red is $\ki2 \gtrsim 10$.
	  The tendency to underestimate the observed \lir\ is more pronounced when the SFR is allowed to decline (declining and delayed models).
	  This behavior correlates with \lir. The highest luminosity sources are also less well fitted in terms of \ki2.  
}
\label{obspredchi2-fig}
\end{figure}

In Fig. \ref{obspredchi2-fig} we compare the observed and predicted \lir\ for each of the SFHs we explore.
Although the bulk of the sample is well fitted in terms of \ki2, reproducing the observed \lir's does not work for the whole sample. 
Depending on the SFH considered, some 34\% to 42\% of the sample matches the observed
\lir's within 0.2 dex. The slightly higher percentage is obtained with the rising and constant SFHs.

Models that allow for the SFR to decline (the declining and the delayed) can lead to a strong underestimation of the \lir.
This behavior is found also for the constant and rising SFR models, albeit to a lesser extent. 
In all cases, the underestimation is strongly correlated with the observed
\lir, with the most IR-bright sources having their luminosities strongly underestimated. The offset is reasonably small up to luminosities of $10^{12} \lsun$,
allowing us to correctly recover the \lir\ in this range, on average. 
This  underprediction also correlates
with redshift, but as the sample is luminosity-limited at high-$z$, and highest luminosity sources are found from 
$z \sim2$ and on, this trend is less striking. 
From the modeling perspective, this underprediction of the \lir\ is linked to the fact that the preferred solutions apply
less extinction than what would be needed to match it. We also note (Fig. \ref{obspredchi2-fig}, color indicates \ki2) that the high
luminosity sources tend to be less well fitted than the rest.

\begin{figure*}[htp!]
\hspace{-0.25 in}
    \includegraphics[width=0.57\textwidth]{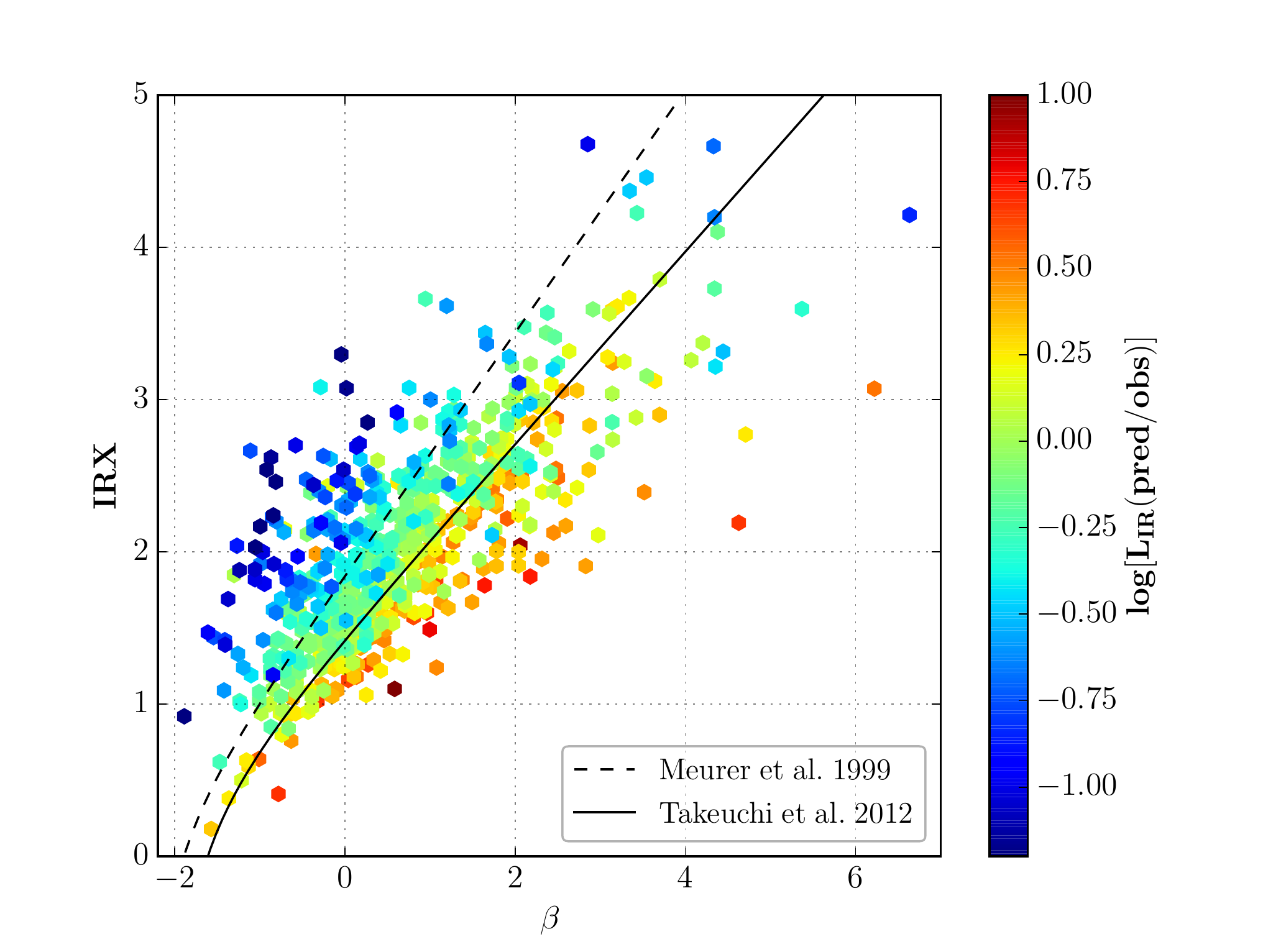} 
\hspace{-0.4 in}    
    \includegraphics[width=0.57\textwidth]{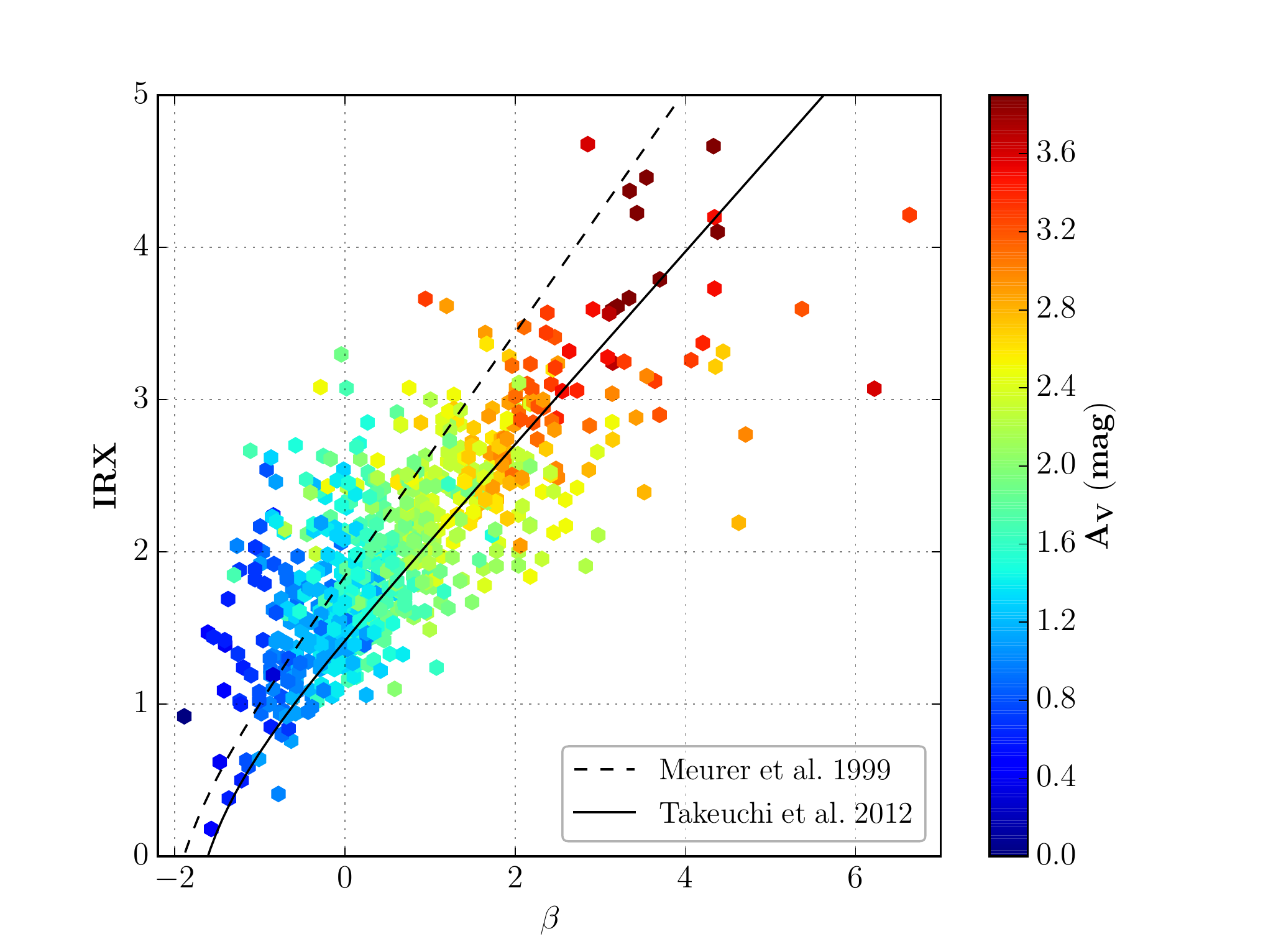}
\caption{{\bf Left:}  IRX-$\beta$ diagram of our sample (\ki2 cut applied on the 
free \av\ models), with the colorbar indicating the ratio between predicted and observed \lir's. 
Also shown are the \citet{1999ApJ...521...64M} relation (dashed) and its more
recent revision by \citet{2012ApJ...755..144T} (solid).
Surprisingly, although the sample is IR-selected, it shows a wide
range of UV slopes, with an important number of sources appearing ``too blue'' for their \liroveruv\ ratios. 
We can see that the sources 
for which the \lir\ is most underpredicted (blue in the colorbar), are found predominantly on the
left of the diagram, meaning they have the bluest slopes. 
{\bf Right:}
Same diagram, this time colored according to the extinction \av\ for the same model. Sources on the left of the Meurer relation 
tend to have less extinction than those at same IRX on the relation.
}
\label{IRX-beta-fig}
\end{figure*}

%

Observationally speaking, this underestimation of \lir\ in the fits is linked to the fact that the concerned sources tend
to have bluer slopes than what one expects from their luminosities.
Fig. \ref{IRX-beta-fig} shows the so-called IRX-$\beta$ diagram, where IRX = log(\liroveruv), is  plotted against the UV slope $\beta$.
IRX is computed from the observed luminosities, and the colorbar shows the ratio of the predicted versus observed \lir\ (left) and the extinction
\av\ (right) for each source (\av\ here is a free parameter, to explore how the IR predictions are distributed on this plane).
For convenience we used the predicted values from the CSFR models, other models give quite similar distributions\footnote{
The CSFR models, being that they have one degree of freedom less than the rest,
offer a more visual illustration of the way the highlighted parameters (predicted \lir\ and \av) behave. The same tendencies are present
in the other models, but can be less pronounced or more noisy.
}.
Indeed, we can see that most 
of the sources with the highest underestimation are found on the left-most part of the IRX-$\beta$ plane (blue-colored points), whether
for a moderate or a high IRX ratio. Although all sources in this IR-selected sample have quite high IRX ratios, the ones with 
the strongest \lir\ are rarely the ones with the strongest IRX, but are intrinsically systems larger in size. To put it in similar
words as presented by \cite{2014arXiv1409.2093O}, for typical UV-slopes (i.e., not the extremely red ones beyond $\beta\sim2$), 
we see more IR emission  at a given $\beta$ than what is expected from galaxies in the local Universe.
 In Fig. \ref{IRX-beta-fig}b we see that \av\ (obtained from the UV-to-NIR fits) correlates with IRX at first order, 
but we can see that at the left of the \cite[][M99 hereafter]{1999ApJ...521...64M} relation, sources with same extinction as on the relation can be found 
at higher IRX, up to almost 1 dex. This indicates that an increased IRX does not always  correspond to an overall stronger
dust attenuation.
In fact, the also plotted 
\citet{2012ApJ...755..144T} relation (revised version of the M99, corrected for aperture effects), acts
more like a lower envelope for ULIRGs.
Studies have shown that IR-selected or otherwise very IR-luminous galaxies often lie above the Meurer relation, both in the local
Universe \citep{goldader02,2010ApJ...715..572H}, and at high-$z$ \citep{2010ApJ...712.1070R,2012ApJ...759...28P,2014ApJ...796...95C}. \cite{2014ApJ...796...95C}
in particular have studied a large \emph{Herschel}-selected sample in the COSMOS field \citep{scov2007} and notice indeed
that the majority of their sources lie above the Meurer relation. 
 Their sample of star-forming galaxies is more high-\lir\ biased than
ours (as COSMOS is shallower than GOODS), and occupies about the same area on the IRX-$\beta$ diagram as our brightest sources.

This, combined to the fact that they are fitted with less extinction than same IRX sources on the  Meurer
relation is strongly suggestive that their UV spectrum is dominated by less obscured stars, that cannot account for the IR
emission. \citet{2014ApJ...796...95C} attribute this to young blue stars originating from recent star formation
and patchy dust geometry. 
We have examined the image data of $\sim 30$ sources that underpredict the IR by the largest factor.  A few 
are in crowded environments in the image plane, that might cause some blending in the \emph{Herschel} fluxes 
which may not be properly accounted for, but this cannot be the explanation for all of them. 
 They share no common traits in terms of geometry, among the few that are resolved enough, some seem to be 
perturbed-asymmetrical objects, while others are disk-like. Of course, AGN contribution cannot be excluded either, although
we are confident in having removed the strong AGN-dominated sources.
A possibility also exists, that some may result from fortuitous alignments of unresolved background IR
sources, magnified by lensing, something that can occur with $\sim 10^{13} \lsun$ detections  \citep[e.g.][]{2013Natur.495..344V}.


We note that sources for which energy conservation occurs ``naturally'' (i.e., that reproduce the observed \lir\ 
before constraining the extinction, green points in the diagram) are mostly distributed about the Meurer and Takeuchi 
relations.
On the other side of the Takeuchi relation, 
we find most of the sources that tend to overestimate strongly the IR emission,
that overall are a small number (red points in the plot). 
After  verification we note that they are largely
very well constrained, very red sources, at redshifts below 2. 
It should be noted that in part the overprediction is due the CSFR models, and that many of the sources on the 
right of the Takeuchi relation do not overpredict (or not to the same extent) the \lir\ when fitted with the declining
SFHs, and as will be specifically discussed in Sect. \ref{s_timesc} and \ref{s_quen}, these sources are better fitted 
with declining models.
The seemingly extreme values of $\beta$ ($\beta\geq 4$) on the right of the diagram are mostly sources with no detections/upper limits on their restframe
UV slopes, so their $\beta$ are ill-constrained. 
This is also the case for sources with very high IRX (above $\sim4$), that have very few detections in the UV-optical bands. 
%

At this point, we wish to make a small discussion on our estimations of $\beta$, so that meaningful 
comparisons with other studies more specific to the UV emission and slopes can be made.
Following \cite{2012ApJ...756..164F}, we have performed power law fits on the UV continuum of our obtained SEDs, with the help of
the UV windows defined by \cite{calzetti94} to avoid the absorption features present in the 1250 -- 2600 \AA\ interval.
We note a mean difference between the slopes measured by \hyperz\ and the aforementioned method of -0.66, with a small
standard deviation of 0.38, which is acceptable given how estimates can vary depending on the method
used \citep{2012ApJ...756..164F}. This difference shows dependence on $\beta$, with the bluer slopes showing less difference and scatter 
between the two estimates than the redder ones. The resulting effect on the IRX-$\beta$ plane is a 
slight systematic shift bluewards, which does not change
how the highlighted parameters (predicted \lir\ and \av) are distributed, and is always compatible with the discussions on the $\beta$ slopes of
high-\lir\ sources discussed in the present Section.

\subsection{How does the \lir\ constraint affect the physical parameters?}	\label{s_energy_cons}
We now examine how physical parameters derived from SED fits change when energy conservation is taken into account.
Sect. \ref{s_caveat} starts with a discussion on the feasibility and limitations of the method we explored. Next, we discuss
changes in the SFR -- mass plane, the SFR indicators, and SFHs.
Subsequently, we will separate galaxies into normally star-forming galaxies and starburst (SB) galaxies and refer to them as 
MS galaxies and starbursts, respectively. The distinction is made based on the distance from the MS that corresponds to the redshift
of each galaxy. Galaxies found with an SFR within 4 times (0.6 dex) of their corresponding main sequence \citep{2011ApJ...739L..40R} are considered
MS galaxies, and those above it, starbursts. Galaxies found below the main sequence are also discussed, as they present
a particular interest.

\subsubsection{Consistency check of simple energy-conserving models} \label{s_caveat}

\begin{figure}[hbt!]
  \includegraphics[width=0.5\textwidth]{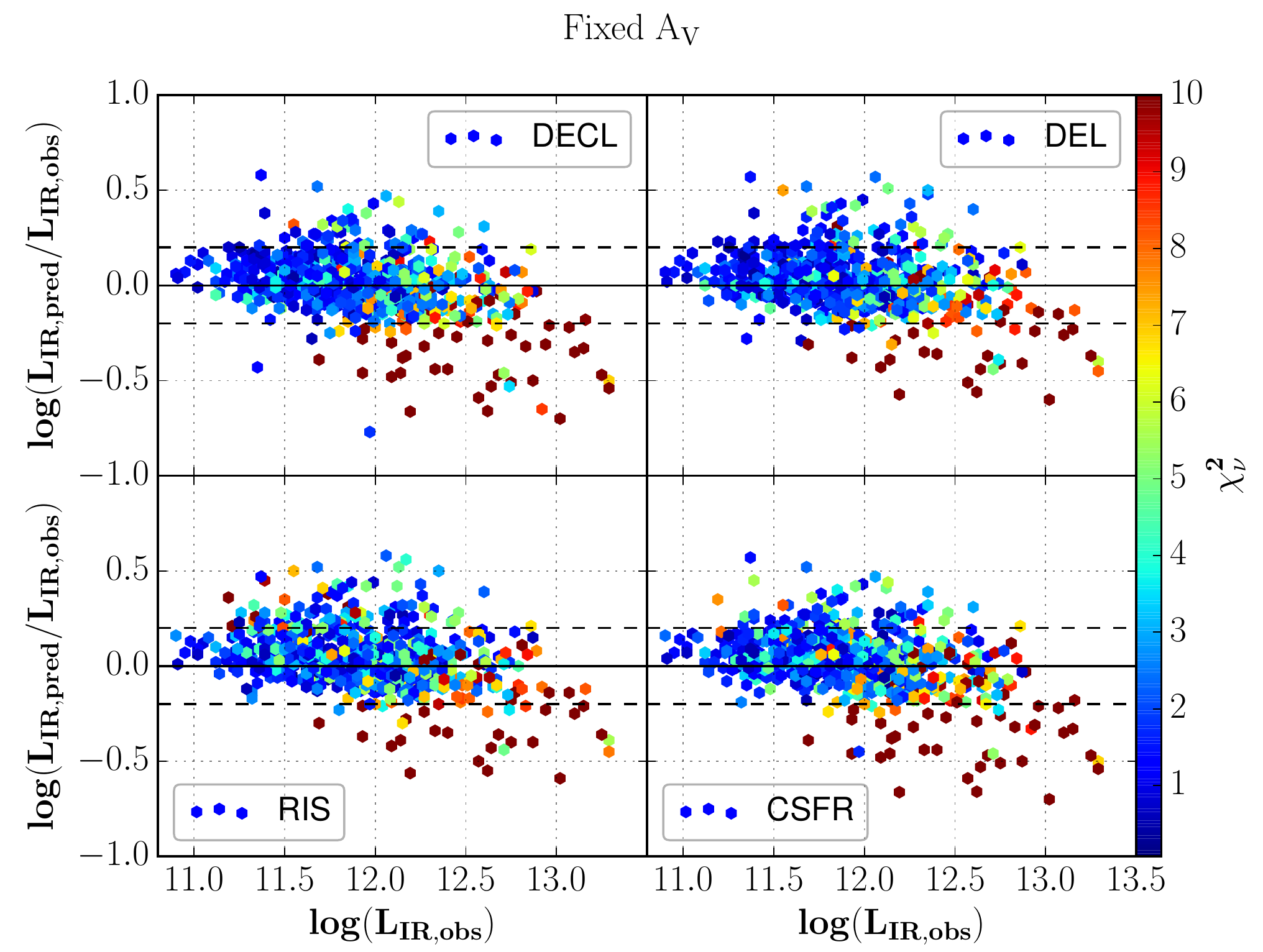}  
  \caption{The ratio of predicted over observed \lir\ plotted against observed \lir\ for the fits where the extinction
  is constrained by \liroveruv.
  The dashed lines represent the 0.2 dex threshold we set for the reproduction of the \lir.
  Color indicates \ki2, as in Fig. \ref{obspredchi2-fig}.
  We can see that this tendency of underpredicting \lir\ persists in the ULIRG and HyLIRG regime, and they also suffer a deterioration
  of their fit quality, indicating that the extinctions derived from \liroveruv\ are ill-adapted for these sources.
  }
  \label{obspred_fix-fig}
\end{figure}

To start, it is important to make a remark on the ``feasibility'' of our energy conserving fits. 
Fixing the extinction to the observed IRX-inferred value does not guarantee a perfect match for the reproduced \lir.  
Hence, our aim is to reproduce it within a reasonable scatter for the bulk of the sample.
Our method successfully reproduces  between 83\% and 86\% (depending on the SFH) of the sample's \lir's within a margin of $\pm 0.2$ dex.
Fig. \ref{obspred_fix-fig} shows the comparison between the predicted and observed \lir's, in the same way as Fig. \ref{obspredchi2-fig}, but for the 
energy conserving fits.
In contrast, when leaving the extinction unconstrained, this fraction is ~30-40\%, and 
we have seen that there is a tendency to underpredict \lir\ in Sect. \ref{s_free}, which increases with the observed IR luminosity. 
In the energy-conserving fits this tendency is reduced, but persists at high \lir, with a  fraction of successfully reproduced \lir's 
down to  $\sim 60\%$ for log(\lir) $\geq 12.5$.
All SFHs yield similar results on the whole sample: the same sources stand out at the high \lir\ end for each of them, and are equally
poorly fitted (red colored in Fig.  \ref{obspred_fix-fig}, the same subsample as Fig. \ref{obspredchi2-fig} is shown).
This shows that our approach is not appropriate for modeling these sources, which may require 
more complex hypotheses than the ones made here. Such could be varying extinction laws, or combinations of stellar populations with different
attenuations, where for example, the UV and IR emitting regions wouldn't necessarily overlap, and hence the energy conservation hypothesis
would not hold anymore. 
For example, laws grayer than Calzetti's could better account for the 
observed conflict between the UV slopes and the \lir, as suggested by \citet{2016ApJ...827...20S} for some of their $z\sim2$ IR galaxies.
 The fact that our \liroveruv-inferred extinction leads to a persisting underestimation means that it is insufficient 
to account for such IR emission.
In the same time, the large \ki2's for these fits, are mostly due 
to the UV-visible necessitating less extinction than the fixed value. 
This incompatibility is a coherent consequence of using a single component  population in our fits. 
With both the UV and IR being very bright, the observed IRX is too small to help predict all of the \lir, and simultaneously imposes more extinction
than what the emerging UV emission goes through.  
This assessment  strengthens the hypothesis that 
different stars or stellar populations are responsible for the respective UV and IR emission. 
We present a selection of SED fits for sources that have this behavior in Fig. \ref{f_seds_blu} of the appendix \ref{s_sed_figs}, 
to help visualize the preceding remarks.

Specifically for the presently discussed sources with bluer-than-expected colors and strongly underpredicted \lir's 
we have also conducted SED fits without the age prior to explore solutions with extremely young ages, as their intrinsically
bigger UV budget can in theory allow for higher \lir\ when processed by dust. 
We observe that this provides somewhat improved fits for a small number of sources (15\%\ of the discussed subsample), both in \ki2\ and matching the \lir, 
with ages of 10 -- 30 Myr, but does not affect the rest of the subsample even when ages below 50 Myr are preferred.
Tests with lower metallicity ($Z$) have also been conducted 
to exploit the full capabilities of the BC03 library, 
since such low-$Z$ stars emit more UV.
Notable changes in the predicted \lir\ (up to 0.5 dex for some) occurred only with very low $Z = 0.02 \zsun$, which conflicts with the related
aspects discussed in the introduction, in particular the expected metallicity from the FMR relation, and the fact that our sources are dust-rich. 
These very low-Z solutions however could be one more argument about multiple stellar components, since recent evidence shows the possibility
of having different metallicities among clumps of high-$z$ SFGs \citep[e.g.,][]{2015ApJ...808..139S}.

As a consequence of the remarks above, our approach should not be extended to sources with log(\lir) $ \geq 12.5$ without checking that the observables 
are reproduced. For the points made in the following sections of this study,
we restrict our sample to the 633 sources (84\% of total), that are not concerned by this issue. 
We note in passing that the majority   (77\%) of the sources discarded due to the aforementioned are starbursts.
Indeed, although the success rate of achieving energy conserving fits with our method is of 96.5\% for MS sources, it
decreases to 79\% for the starbursts.    

\subsubsection{The normalized SFR -- mass diagram}	\label{s_ms}

\begin{figure*}[htb]
  \hspace{-0.65 in}
        \includegraphics[width=1.15\textwidth]{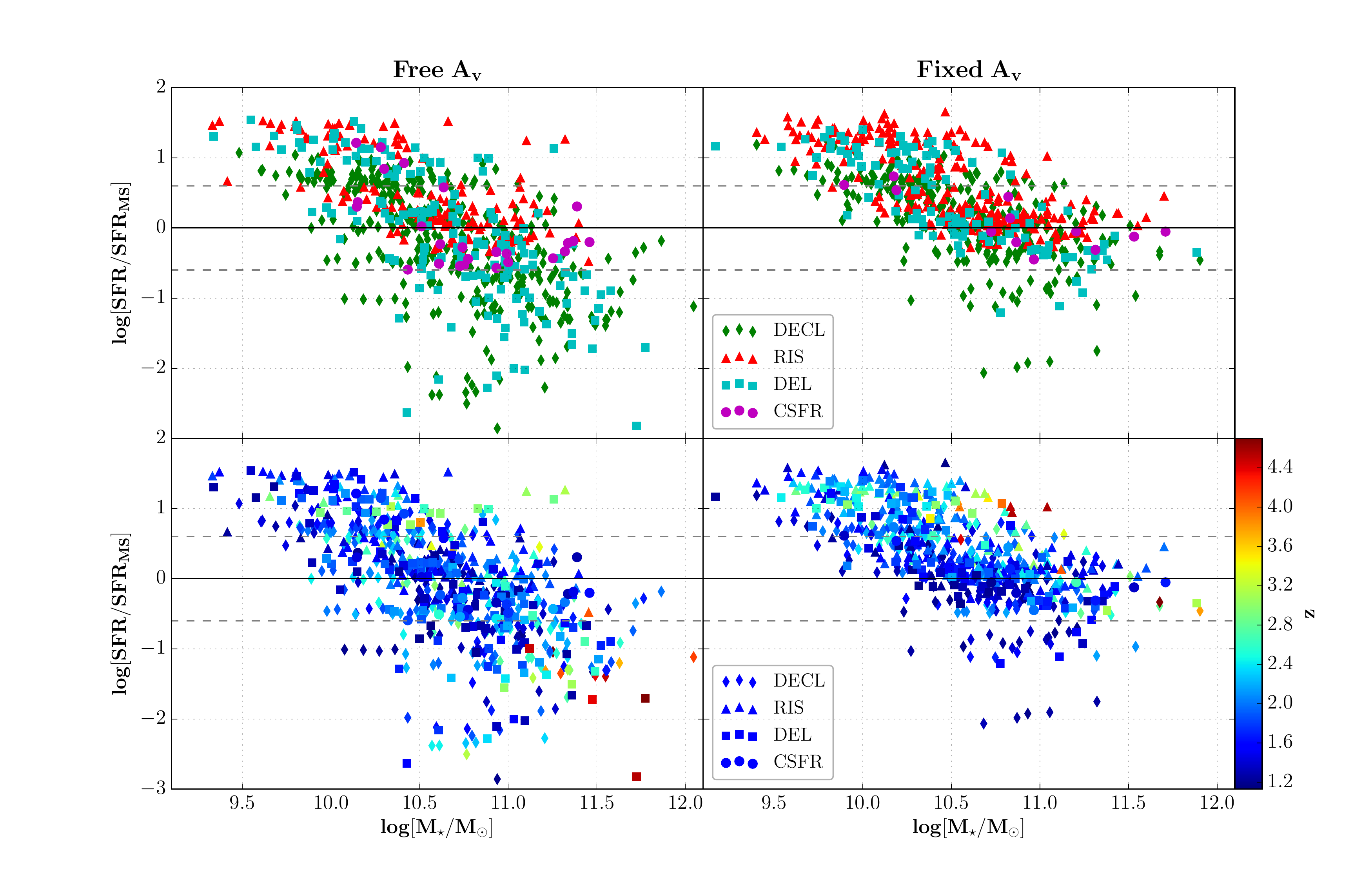} 
    \caption{ Normalized SFR -- \mstar\ diagrams before constraining the extinction from the IRX ratio (left), and after (right).
    The top panels are colored by SFH, while the bottom ones are colored by redshift.
    For each source, only the preferred SFH is plotted 
    (diamonds for declining, triangles for rising, squares for delayed, and circles for CSFR models).
    When the extinction is a free parameter we see that many of the sources are best fitted with models of reduced
    star formation, and hence can be quite below the MS (marked here by the solid black line, while the dashed lines show the distance
    4 times from it).   
    }
  \label{MS-fig}
\end{figure*}

From now and on, we focus on the energy conserving sample, meaning the subsample 
that successfully meets our \ki2 criterion and reproduces the observed \lir\ within $\pm 0.2$ dex. 
In order to better discuss and illustrate how the \lir\ constrain affects the modelization of the sample, 
the figures shown will often show two panels, one before applying the constrain (free \av) and one after (fixed \av).

Fig. \ref{MS-fig} shows the  normalized SFR -- stellar mass diagram with SED-inferred quantities, 
before (left) and after (right) constraining the extinction as 
prescribed in Sect. \ref{sed_mod}. The colorbar indicates the redshift of the sources. 
The SFR of each source is normalized by the main sequence at its corresponding $z$ and mass. 
In this way, the solid line indicates the position of a $z$-independent MS, and the dashed lines the distance 4 times (0.6 dex) from it, below
which a source is considered being part of the MS.
We recall here that given the nature of our sample (\emph{Herschel}-selected) we are naturally biased toward higher mass galaxies 
and/or starbursts at $z > 1$, leaving out the moderately star-forming low-mass galaxies that made it possible to constrain the MS
at $z \sim 1-2 $ down to $\mstar \leq 10^{10} \msun$ \citep{schreib2014}, as apparent in Fig. \ref{MS-fig}.

Allowing for variable SFHs in the SED fitting procedure of the stellar photometry produces a larger scatter around the main sequence
\citep{2013A&A...549A...4S}, as the SFR -- mass relation will depend on  more parameters than assumed in standard calibrations \citep{K98}. 
In particular, the possibility of a declining SFR can cause fitted galaxies to occupy space below the main sequence, 
as can be seen in the left panel of Fig. \ref{MS-fig}.
As a result the instantaneous SED-inferred SFRs (\sfrbc, hereafter) are found to 
underestimate the observation-inferred ones from the UV and IR (as also shown in Fig. \ref{SFR_comp-fig} (left), Sect. \ref{s_indics}). 
Such solutions describe older (often $>1$Gyr), non-star-forming populations, close to becoming ``red and dead'' galaxies.
This is the consequence of the age-extinction degeneracy that can occur when fitting red galaxies. 
 In the described cases, 
we find most of the strongly underpredicted \lir's. These models have an insufficient quantity of young massive stars and 
dust, hence the absorbed UV luminosity is modest.
Since in reality, by construction of our sample the sources are very IR-luminous, the non-star-forming - ageing populations are 
to be discarded as non physical.

This is where our energy-conserving fits come into
play, where we impose a minimal extinction necessary to reproduce the observed \lir\ as described in Sect. \ref{sed_mod}.
As it can be deduced from the above, the action of fixing the extinction based on the IRX ratio gives on average 
higher \av\ values than the ones the best fits choose when it is a free parameter.
This has a strong impact on the \sfrbc's, with most of those that previously showed very low rates being increased, as shown in Fig. \ref{MS-fig} (right).
Other effects are a slight reduction of the masses and ages (discussed further in Sect. \ref{s_timesc}).
It can be seen in the right panels of Fig. \ref{MS-fig} that the space below the MS is less populated (with the 
exception of some low-$z$ sources that we will discuss later). In the high-$z$ end, more sources
occupy the SB regime, in agreement with the IR-derived SFRs. 
Although the MS still remains somewhat scattered, 
a majority of about 70\% of the sample is distributed close to it, especially at lower $z$ where the sample is more representative
of a the general population. 

%
\begin{figure}[htb]
  \centering
  \includegraphics[width=0.5\textwidth]{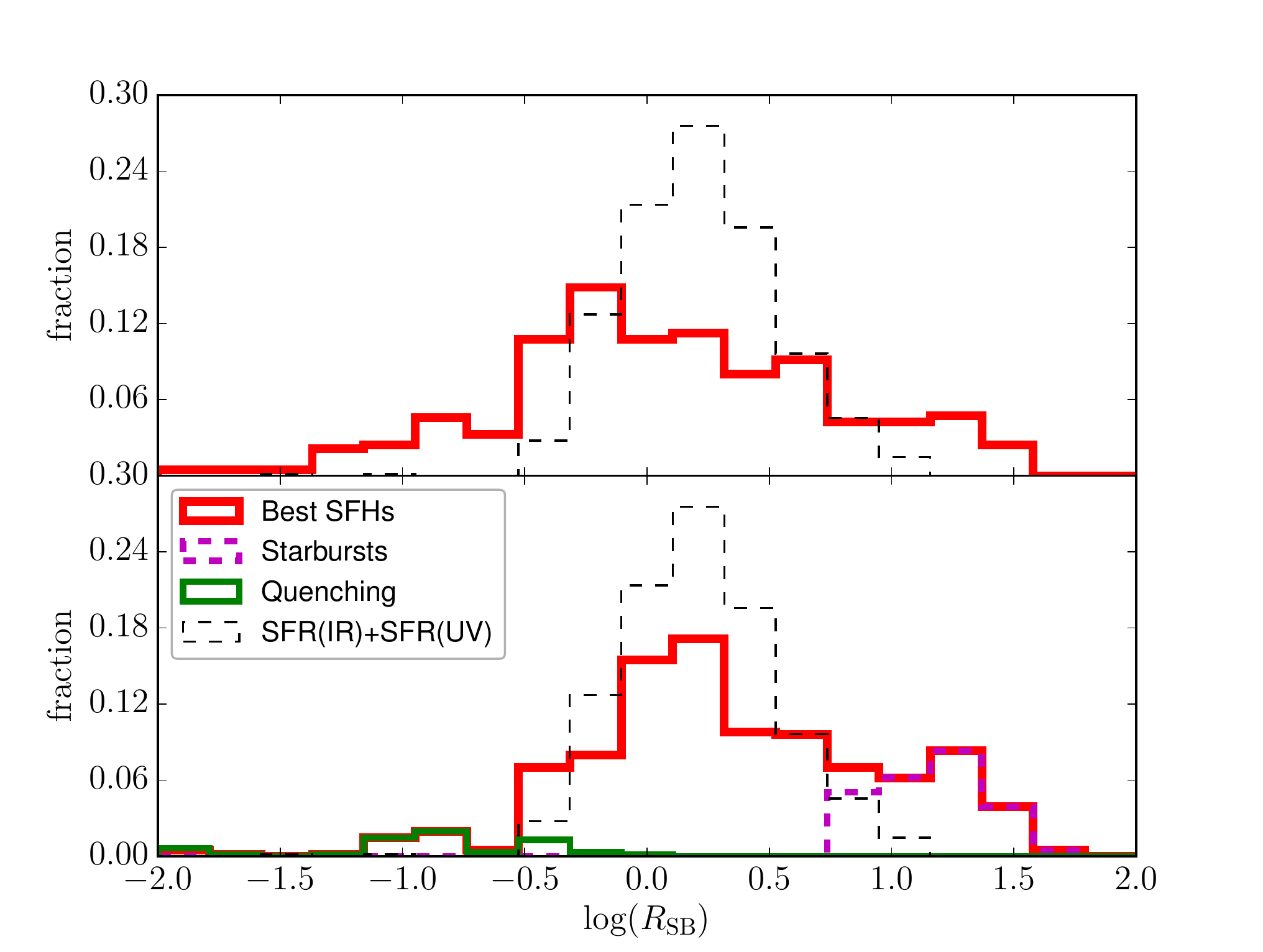}
  \caption{
  Starburstiness ($R_{\rm SB}$) distribution of our sample, representative of the scatter around the main sequence, 
  prior to constraining the extinction (top), and after constraining
  it to ensure energy balance (bottom). The thin dashed line denotes the ratios obtained with the SFRs inferred from \lir\ and \luv.
  The scatter is reduced when the extinction is constrained, especially below the MS (situated at zero in the plot). The
   sources remaining below it (green line) are discussed in Sect. \ref{s_indics} and on. The dashed magenta line shows
  the sources qualifying as starbursts, based on their distance from the MS.
  }
  \label{rsb-fig}
\end{figure}

Figure \ref{rsb-fig} shows the distribution of $R_{\rm SB}$ (cf. Eq. \ref{eq_rsb}), called starbustiness. It illustrates the scatter around the 
main sequence. The data plotted in red are obtained with the derived quantities (\mstar\ and SFR) from our best fit SEDs 
to the sample. We can see that the scatter in the unconstrained extinction fits (top) is significant, as discussed earlier
in this section. It is more reduced when the extinction is constrained, mostly below the MS.
The thin dashed line denotes the ratios based on $\sfrir+\sfruv$ and the 
masses from \cite{pannella2015} that served (together with stacked sources in the low luminosity end) 
to produce the main sequence in \cite{schreib2014}, hence is the ``canonical'' minimal scatter.



\subsubsection{Masses}	\label{s_mass}

\begin{figure}[htb]
\centering
  \includegraphics[width=0.5\textwidth]{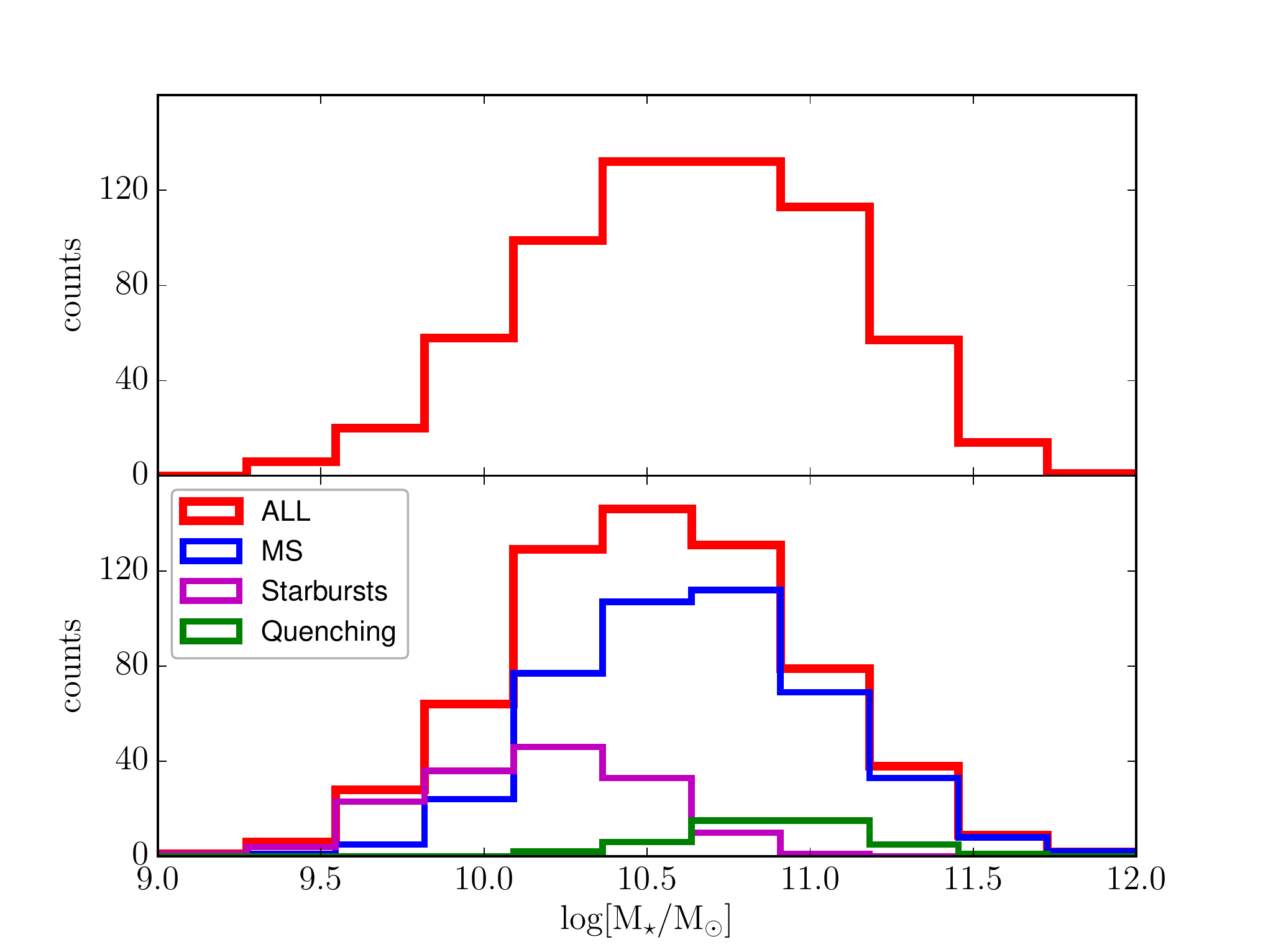}
  \caption{\mstar\ distribution of the sample, before (after) constraining \av\ on the top (bottom) panel. 
  The best fitted SFHs are kept for each source.  
  The overall distribution is affected very little by the constraining of the extinction in the fits,
  the shift toward lower masses is of the order of $\sim 0.1$ dex. 
  The additional MS, starburst and quenching (see Sect. \ref{s_indics} for this last label) sub-groups, 
  discriminated based on their position on the \mstar\ - SFR or \sfrbc\ - \sfruvir\ diagrams, 
  are discussed in Sect. \ref{s_implic}.
  }
  \label{mass_histo-fig}
\end{figure}

Although the SFR -- mass diagram changes noticeably between before and after constraining \av, the overall 
distribution of the stellar masses is impacted very little, as shown in Fig. \ref{mass_histo-fig}.  
The means are $\log (\mstar/\msun) =  10.62$ for the unconstrained fits (top), and $10.53$ for the energy conserving fits (bottom), 
which puts the overall decrease at $\lesssim 0.1$ dex. 
The standard deviation undergoes a negligible decrease. 
The median masses of the distributions are sensibly equal to the means with a precision of 0.01 dex.
The effect of fixing the \av\ is somewhat mass dependent though, with the massive-end galaxies tending to have lighter masses in the
energy conserving fits, and the low-mass end becoming slightly heavier. 

What also does not affect the masses are the different SFHs, as for most of the sample the restframe 1 to 2 \micron\ emission is well 
constrained by numerous bands.
Regardless of whether considering rising or declining models, the masses obtained are 
the same well within usual uncertainties, with a mean difference of 0.02 dex between the two, and a standard deviation of 0.14 (improved from 0.24 when 
imposing energy conservation).

 We have compared the masses that we obtained with the unconstrained extinction fits (similar to what is usually done in the literature)
with other estimates as a test of the accuracy of our solutions. 
Specifically, we compared the masses of the GOODS-North subsample with the masses derived in \cite{pannella2015}, as we share the exact same photometry extracted
for that work, and find a small systematic offset, our median mass is $\sim0.1$ dex lower.
This difference  is strongly reduced if we compare fits without line
emission. For GOODS-South we compared to the reference masses of CANDELS \citep{santini2015}, for which we have similar conclusions: among our models the masses that are
best matching the reference values are obtained when excluding nebular emission and imposing a minimal age of 100 Myr. Otherwise, our mass distribution
is within the scatter found when comparing to the various methods/codes used to produce the reference values. 
The median mass of our unconstrained models  is $< 0.1 $ dex lower from that of the CANDELS masses in GOODS-South.

Comparing the masses of the energy-conserving models with the reference values respective to the two fields, 
only brings a median difference of  $\sim 0.2 $ dex in GOODS-North and $\sim 0.15$ dex
in GOODS-South. The difference between the unconstrained models and the energy-conserving ones does not cumulate with the difference between the former
and the reference masses, partly because of the slight mass dependence described in the beginning of the section.
In conclusion, the masses of our models are on average a little smaller, but within the uncertainties common to the estimation of this quantity.

\subsubsection{SFR indicators -- evidence of quenching?}	\label{s_indics}

\begin{figure*}[htb]
 \vspace{-0.5 in}
 \hspace{-0.6 in}
 \includegraphics[width=1.15\textwidth]{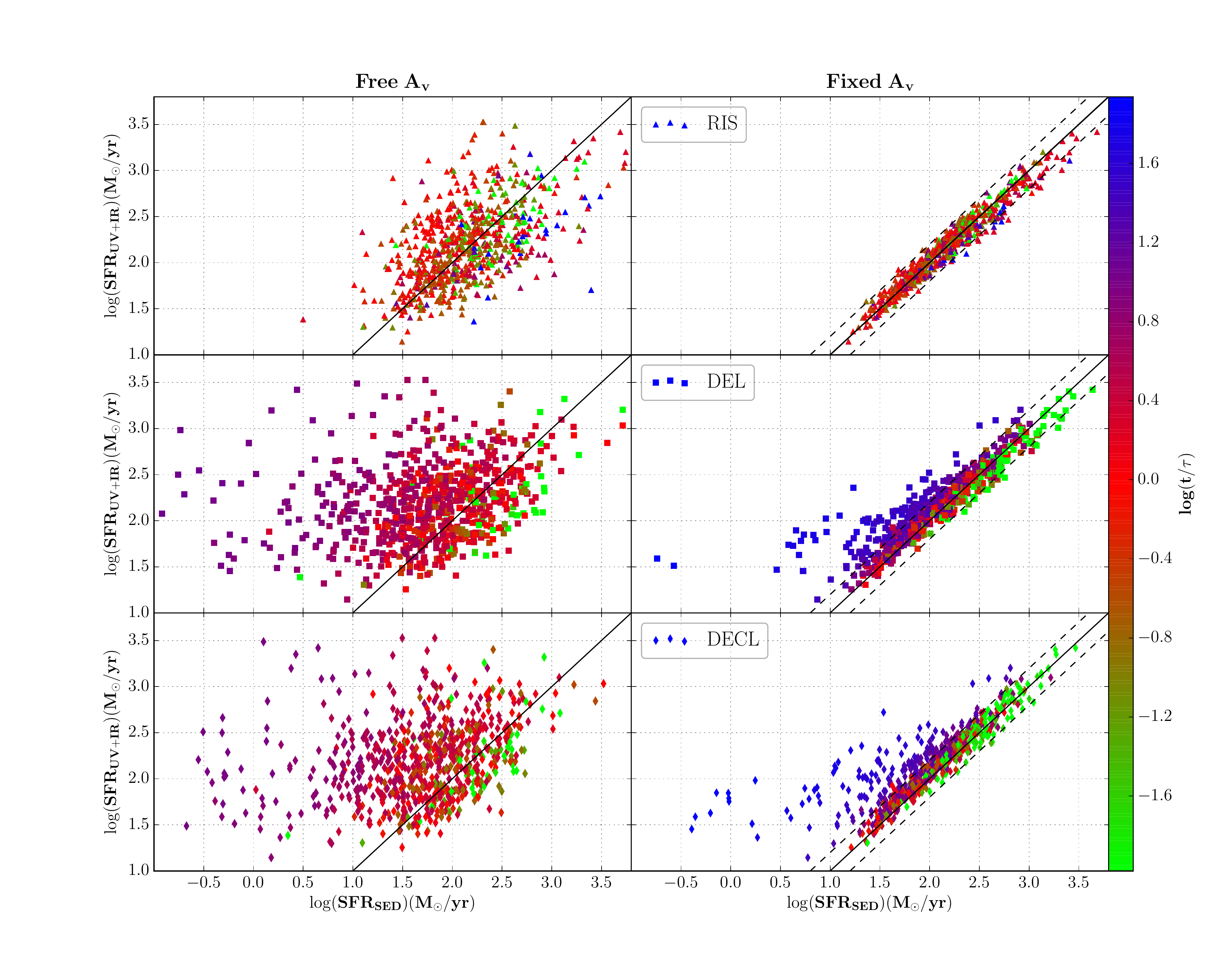}
 \vspace{-0.5 in}
 \caption{Comparison of the SED derived instantaneous SFRs with the observation derived total SFRs.
	  On the left are the fits with unconstrained extinction and on the right the ones with energy conservation.
	 Each SFH with variable SFR is plotted separately, and the colorbar indicates $t/\tau$, 
	 in logarithmic scale. Top is rising, middle is delayed	and bottom is declining SFHs. 
	 The solid line is the one-to-one relation and the dashed lines are positioned at $\pm$ 0.2 dex of it, illustrating
	the threshold of tolerance for matching the \lir.	
	We can see that for most of our sample, 
	when extinction is a free parameter (left) the photometries are 
	interpreted as weakly star-forming or having decreased SFRs in comparison with what the UV and IR-derived SFRs show
	(except the rising models that are equally distributed around the one-on-one relation).
	This is not the case in the energy conserving fits (right).
	All the rising models tightly reproduce the observed SFRs (as do the CSFR models).
	In the case of the declining and delayed models, there is an extra feature: a fraction of the 
	energy conserving fits have $\sfrbc < \sfruvir$, and a large $t/\tau$ ratio. 
      }			 \label{SFR_comp-fig}
\end{figure*}

Figure \ref{SFR_comp-fig} shows a comparison between the instantaneous SFR values obtained from the SED (\sfrbc), 
with the observationally-derived from the Kennicutt calibrations, total SFR (noted \sfruvir, short for ${\rm SFR}_{\rm IR} + {\rm SFR}_{\rm UV}$).
%
Each row shows one of the SFHs explored (from top to bottom: rising, delayed, and declining), the constant SFR solutions are not shown, but they
behave similarly to the rising ones. The colorbar indicates the age over e-folding timescale ratio, $t/\tau$.
In the left panel, we see how for the majority of the sample the \sfrbc's underestimate the \sfruvir's, something noted also in
\cite{2011ApJ...738..106W}, on PACS-detected galaxies from the GOODS-S field.
As detailed in the previous subsection, this is the result of the age-extinction degeneracy, and the fact that our fits do not 
reproduce automatically the observed \lir's. 
This is further illustrated by the gradient in $t/\tau $ observed in the subplots of the declining and delayed SFHs. 
The cases where the best fit has $t/\tau \leq 1 $ have \sfrbc's that are closer to \sfruvir (as they are close to having constant star-formation rates), 
whereas the larger the  $t/\tau $ ratio becomes the more the \sfrbc's are smaller than \sfruvir.

As can be seen in the right panels of Fig. \ref{SFR_comp-fig} where the energy-conserving models that reproduce the \lir\ within 0.2 dex are shown,
the majority of sources here is fitted by a model for which the instantaneous \sfrbc\
is comparable to \sfruvir, and hence are near or at a state of equilibrium as assumed by Kennicutt's calibrations. 
This interpretation must be more realistic as it can account for the IR emission.
In the case of the CSFR models, the matching of the SFRs is directly dependent to the reproduction of the \lir, which is what is
expected.
There are some sources best fitted with rising models with $\sfrbc > \sfruvir$ that also overpredict \lir\ (not shown in the diagram).
This tells us that the best SFH in terms of \ki2 does not guarantee an accurate reproduction
of the \lir.

A very interesting fact to notice also in this plot is the presence of sources on the left of the one-to-one relation, 
meaning with $\sfrbc < \sfruvir$, while still matching the \lir's. In the final sample (where only the best fitted SFH per source is kept)
they are 22.4\% of the sample, to have an \sfrbc\   
more than 0.2 dex below \sfruvir.
As expected they're all declining models, or delayed on the declining phase, predominantly below $z \sim 2-2.5$. 
Furthermore, an estimated 7\% (of total, 43 sources) with $\sfrbc < \sfruvir$ are found to have an \sfrbc\ 0.6 dex or more below \sfruvir. 
Based on their observed \lir, the majority of these sources is situated on the MS,
but if we consider the \sfrbc\ from our solutions, they are relocated below the MS.
A small minority (7 sources) is IR-bright enough to be in the SB regime, so our solutions lower them toward the MS, but since their
SFRs are declining, they're only transiting toward below the MS.
For this reason we will hereafter refer to these galaxies as quenching. 
Exploring deeper these cases shows how this difference between \sfrbc\ and \sfruvir\  is possible. 

%
%

The main question is: how is it possible that a SED model that accurately predicts the observed IR and UV luminosity, can
do so with an \sfrbc\ that underestimates \sfruvir\ by up to 1 dex or more, for some cases? 

The answer lies within the timescales of star formation involved. The \cite{K98} calibrations rely on the assumption 
of continuous star formation over a duration of $\gtrsim$ 100 Myr.
If this assumption breaks down, significant deviations can in principle be expected \citep[cf.][]{2012ApJ...754...25R,2013A&A...549A...4S}.
Based on simulations, \cite{2014MNRAS.445.1598H} show that following a starburst event, the \lir\ can decrease 
more gradually than the true SFR, as dust can still be significantly heated by the recently formed stars.
This can lead to serious overestimations of the true SFR by the \lir, for example, in the case of young post-starbursts. 
This is a plausible scenario for the sources presenting this characteristic in our sample, as the method experimented
and the variability of the explored SFHs allow for such a combination of circumstances. 
Indeed, in the right panels of Fig. \ref{SFR_comp-fig} we see that while the sources with $\sfrbc \approx \sfruvir$
have $t/\tau \leq 1 $ (indicative of very slow change in the SFRs, and hence fulfilling the assumptions of the \cite{K98} calibrations),
the sources that deviate from the one-to-one relation have very large $t/\tau$ ratios 
(larger than the ones deviating in the left panels where the extinction is unconstrained). 
Their SFRs are declining very fast (five-fold decrease in a 50 Myr lapse for example) and hence have many recently
formed stars that contribute in great part to the IR luminosity.

Thanks to the above we conclude that the breaking of the age-extinction degeneracy occurs in a dichotomic way, with most sources finding 
their best energy-conserving fit to be close to Kennicutt's assumptions (77.6\% of the energy conserving sample), and a fraction being best fit with the models 
that allow for the fastest available decline within the choices left by  \lir-constrain, namely $\tau = 30$ Myr, preferred by most of the quenching 
sources. 
In Sect. \ref{s_quen} we attempt to decipher what spectro-photometric features in the quenching sources impose this kind of 
solution in the fits, and whether their discrimination is possible.
 
%
\subsection{Timescales}	\label{s_timesc}		 
%

In this section we will briefly go through our findings in terms of the ages $t$ and the e-folding timescales \taf, and trends/preferences
we observe among them depending on the SFH, and how they help shaping the \mstar -- SFR diagram for the energy conserving sample.

\begin{figure}[htb]
\centering
 \includegraphics[width=0.5\textwidth]{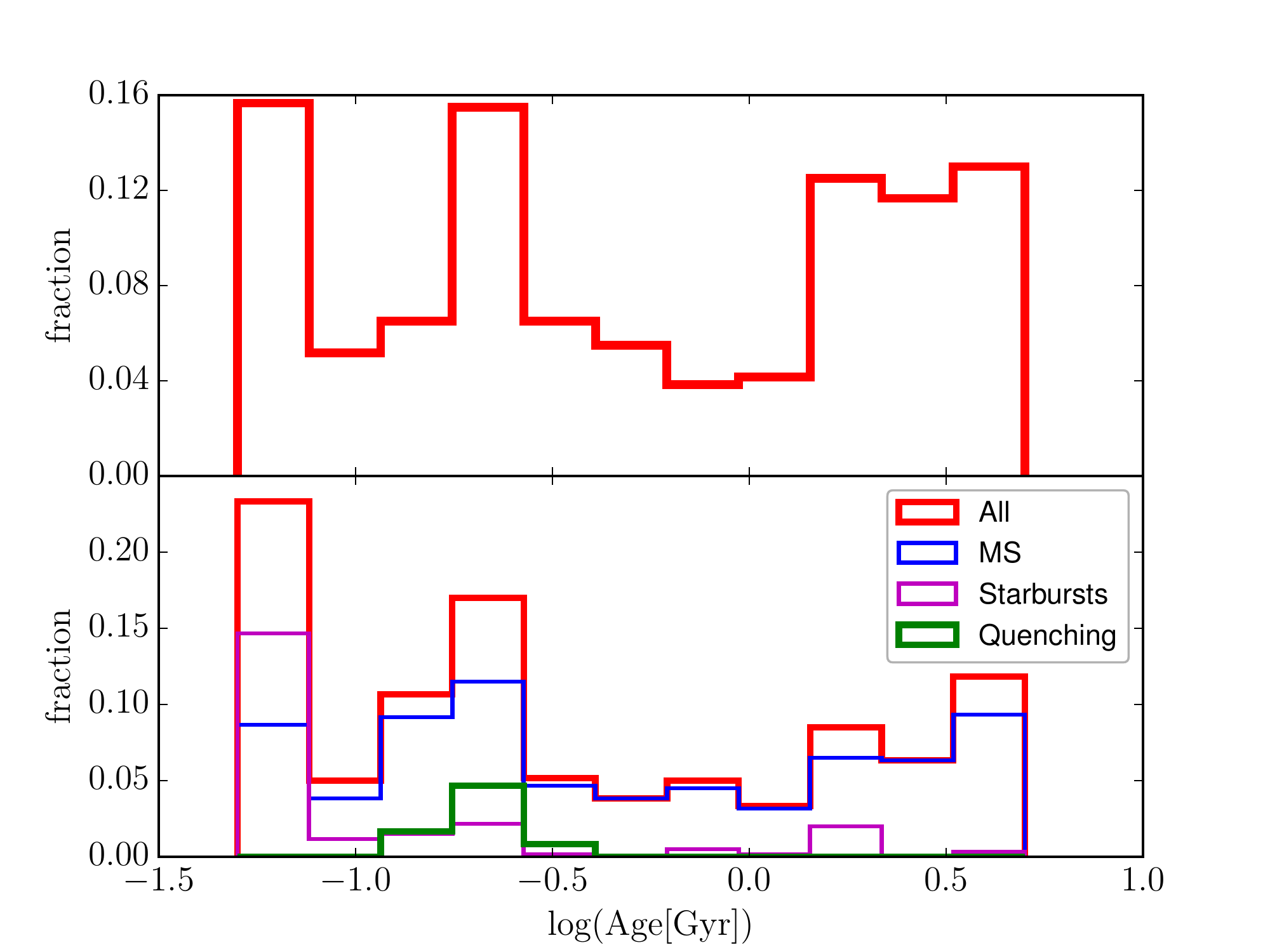} 
  \caption{Age distribution, made in the same way as Figs. \ref{rsb-fig} and \ref{mass_histo-fig}. The notable difference is the higher fraction of sources
  in the lowest age bin, due to more sources being interpreted as starbursts with the help of the \lir\ constrain.
  }
  \label{f_age_hist}
\end{figure}

Figure \ref{f_age_hist} shows the age distribution of the sample, including the distribution for the unconstrained fits, for comparison. Starbursts
dominate the lowest age bin, MS sources occupy the whole parameter space but dominate the large ages, and the quenching sources are 
very localized between $\sim$100 and $\sim$400 Myr.

The largest SFRs (mainly starbursts) tend to have very small \taf\ as well as ages, when considering rising SFHs. The same 
sources when fitted with declining SFHs prefer the largest \taf's but always small ages, indicating that they avoid having a declining SFH as much as possible.
Their ages present them as very young events, and potentially shortlived bursts.

The rest of the sources tend to follow a somewhat bipolar distribution in timescales, similar for all SFHs.
However, as with the starbursts the choice of \taf\ for a given source can vary greatly depending on the SFH. A source fitted with a 
small \taf\ when the SFH is rising will tend to be fitted with a large one when the SFH is declining. 
This said, each SFH taken apart has a majority of sources (55-60\%) fitted with $\taf \geq 0.5$ Gyr. 
When considering the best solutions among all the SFHs a relative majority of 47\% of the sources prefer \taf's greater
than 0.5 Gyr.

\begin{figure}[htb]
\centering
  \includegraphics[width=0.5\textwidth]{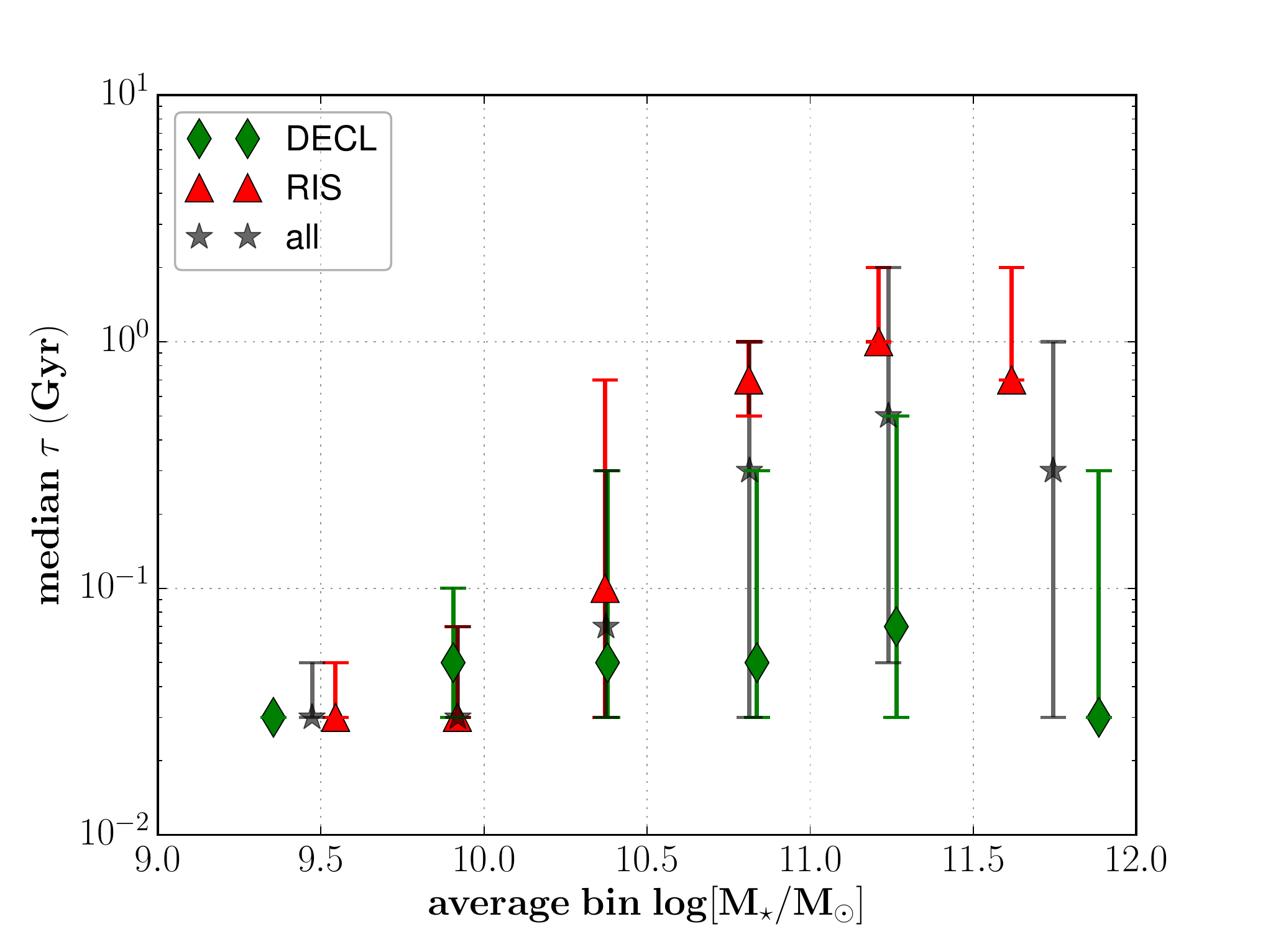}
    \caption{Median \taf\ per mass bin. 
	     Gray stars denote the median \taf\ for all SFHs (excluding constant), and red triangles and green diamonds 
	     are the medians for the rising and declining SFHs respectively, considered separately.
	     Error bars represent first and third quartiles. \taf\ appears to rise for the rising SFHs.} \label{mass_tau}
\end{figure}

We find that \taf\ correlates well with \mstar, in the case of rising SFHs, as is shown in Fig. \ref{mass_tau}, where we plot 
the median \taf\ per mass bin, from the energy conserving fits. The low mass galaxies tend to have very small \taf\ (SFR rising very fast), 
whereas massive galaxies have large \taf, meaning their star-formation tends to become constant. 
On the other hand, the declining
SFH models, when preferred tend to favor a short e-folding timescale of the order of 30-50 Myr, regardless of mass, 
with the exception of high mass (log$[\mstar/\msun] > 10.5 $) main sequence galaxies, for which \taf\ is superior to 100 Myr.
Also, the ratio of preference for rising models versus declining ones shifts with mass, with more sources preferring a rising SFH
than a declining  for log$[\mstar/\msun] \lesssim 10 $, and the opposite for larger masses (this point is discussed further in Sect. \ref{s_sfh_mass},
see Fig. \ref{mass_anatomy-fig}). 
We also remark that this variation in \taf\ for the rising models also correlates with redshift.  Sources at redshifts larger than $z \sim$ 2.5 
tend to very strongly favor the shortest timescales (30 Myr), and the median \taf\ increases with lowering $z$ to reach 1 Gyr at $z \lesssim 1.5$.
This aspect and the eventuality of an evolutionary sequence are discussed further in Sect. \ref{s_evol}.

Given that the fraction of sources preferring the smallest \taf\ allowed in our fits is important (whether we consider the starbursts or the quenching  sources), 
we have briefly explored whether these sources
would opt for even smaller \taf\ if it were available. We thus performed SED fits with $\tau <30 $ Myr to test this, and found that 
some sources were better fitted with $\tau = 10$ Myr, mostly preferring rising SFHs, and a negligible number was best fitted with $\tau = 5$ Myr.
This did not bring any substantial change to the sample's properties.

\subsubsection{The timescale and age parameters on the SFR -- \mstar\  diagram}	\label{s_ms_t}

\begin{figure*}[htb]
\hspace{-0.25 in}
   \includegraphics[width=0.57\textwidth]{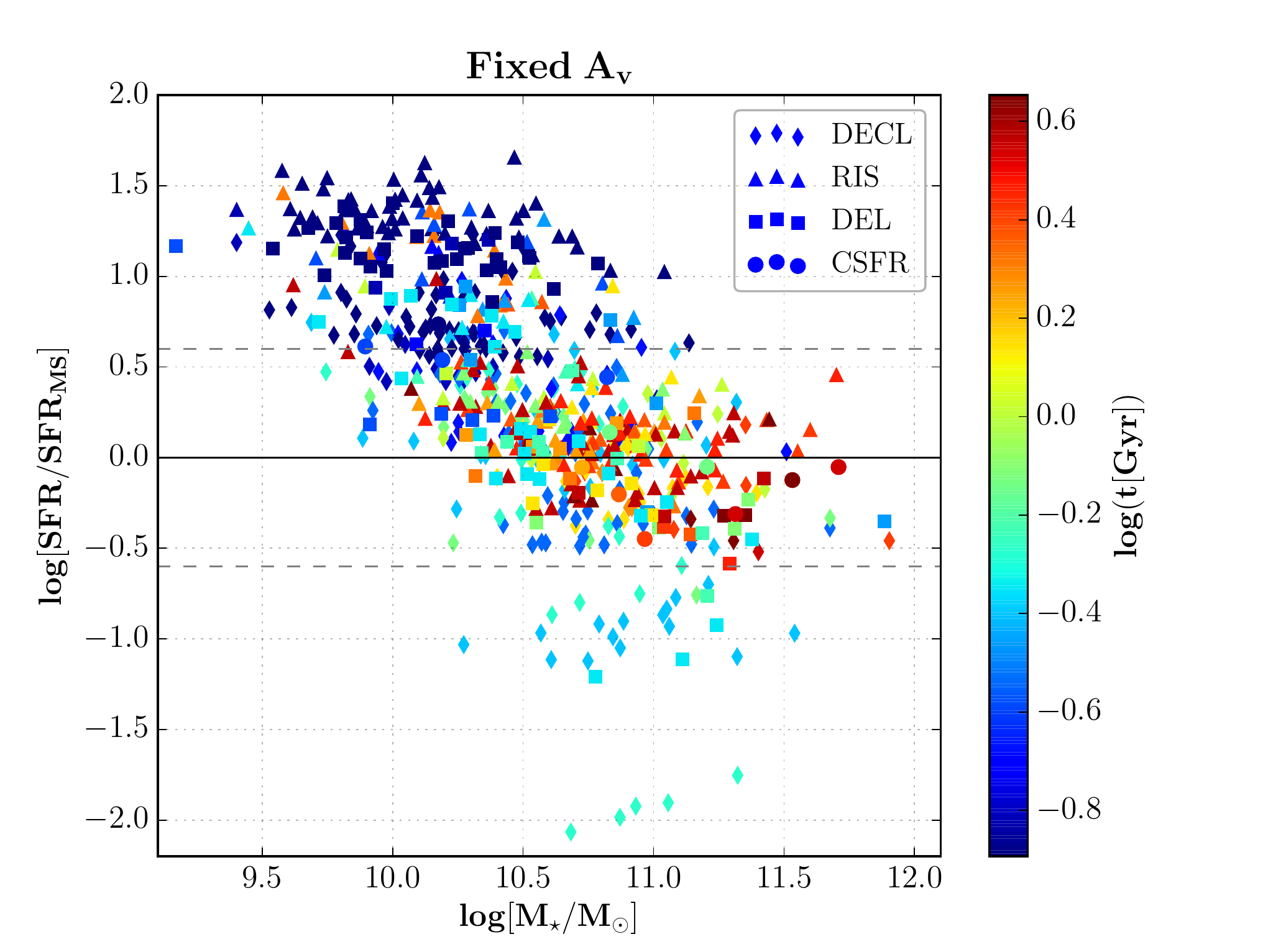} 
\hspace{-0.4 in}    
     \includegraphics[width=0.57\textwidth]{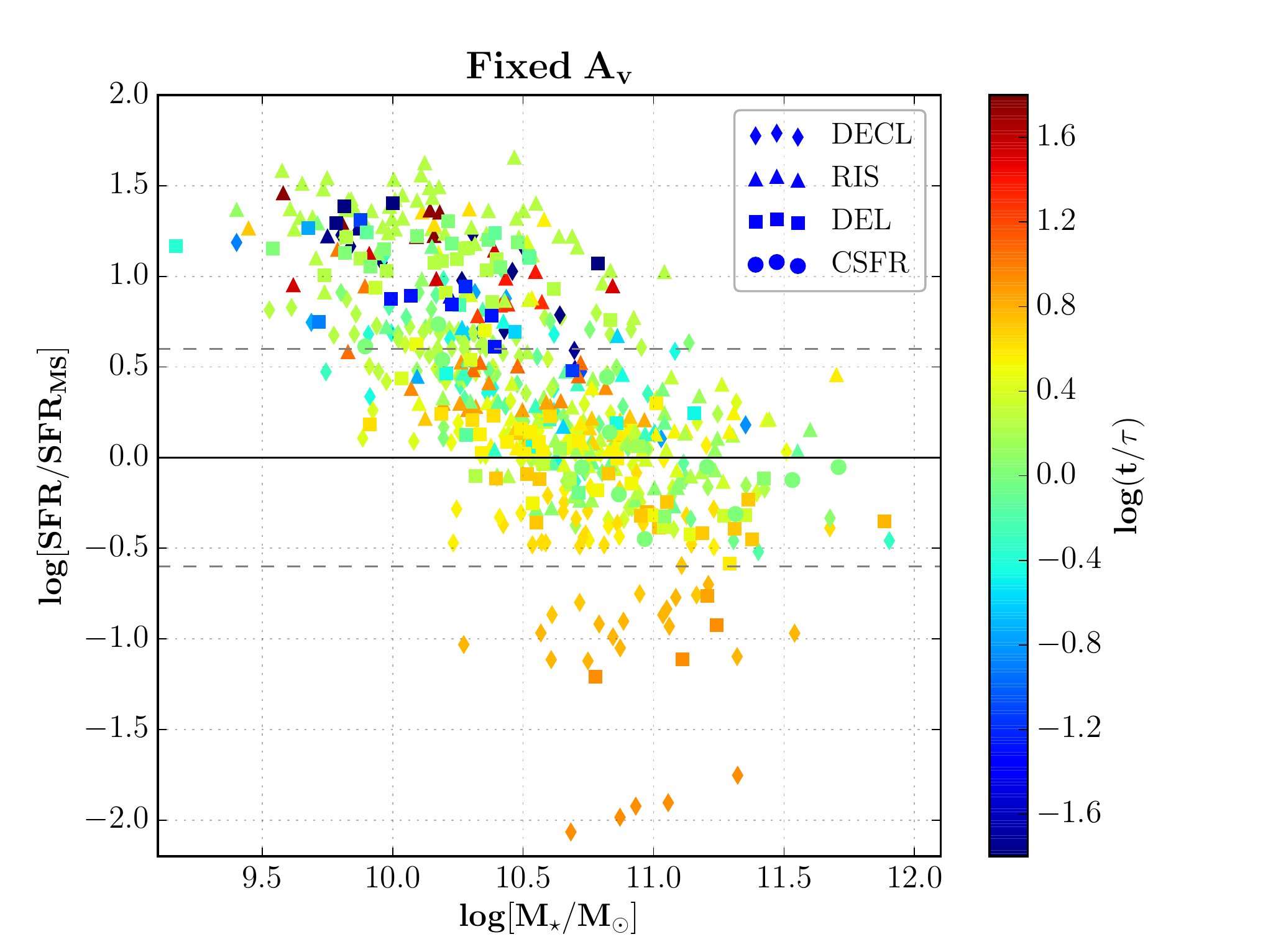}
\caption{Normalized SFR -- mass diagram for
      the energy conserving sample, with in the colorbars, age $t$ (left) and the age over e-folding timescale ratio, $t/\tau$ 
      (right). We see that the majority of galaxies with old ages are on the MS, and are slowly evolving.
      Galaxies below it, have medium ages but have seen their SFRs strongly declining.}	 \label{ms_t}           
\end{figure*}

The combination of population age $(t)$, and the e-folding timescales \taf, plays an important
role in the position of a galaxy on the SFR -- mass plane. In Fig. \ref{ms_t} we revisit the normalized SFR -- mass diagram to highlight 
this fact. The plot on the left colors the data according to age, and the one on the right according to 
the age over e-folding timescale ratio, $t/\tau$. This ratio, in the case of the declining models, is representative of 
the present-over-past SFR ratio. A large $t/\tau$ ( $\gg1$) implies that a lot more of the stellar mass was built in earlier
ages than is being built now. A small $t/\tau$ ($\lesssim 1$) indicates that current star formation is very important 
compared to the present mass. Rising models, have ongoing star formation which by construction is important regardless of the ratio,
but of course very small \taf's can not be sustained for billions of years as they would lead to unrealistically extreme growth
\citep{renzini2009}. 
In this regard the fact that we find very small \taf's in the rising SFHs of the low mass sources only (predominantly on the SB sequence), and 
large \taf's for high masses is encouraging. 

From Fig. \ref{ms_t} (left) the first thing that can be noticed, is that the galaxies that are well on or near the main sequence, are 
majoritarily rather old. Starbursts, with a few exceptions are almost all as young as allowed (50 Myr). The galaxies
found below the main sequence are typically a few hundred Myr old. In Fig. \ref{ms_t} (right) we can see, that both MS and SB galaxies
have $t/\tau$ ratios close to 1. 
Since we find a majority of large ages (0.3 Gyr or more) for the  MS galaxies (and most large ages are found in MS galaxies), 
this means they are interpreted as growing through long episodes of star formation.
Ratios smaller than one are found only in the SB regime. Some large  $t/\tau$ ratios are also encountered
there, only for rising models (triangles) and correspond to ages of $\sim1-2$ Gyr. We cannot address whether such solutions are physically
possible or not, but they might also approximate a long steady growth, followed by an ongoing burst.
The sources below the MS have all large ratios,  $t/\tau \approx 10$, indicating that their SFRs have dramatically declined
over the last tens of Myr. Given this fact, their relative young age, and their medium-to-large masses, they can very well be originating
from the SB, where after a short lived burst their SFR declined steeply, making them potential candidates for quenching.
 We note in passing that no source best fitted with a rising SFH is found below the MS. 
If galaxies located below the MS were to initiate a burst of star formation, they would probably move very fast upwards in SFR by the time they can be detected by \emph{Herschel}.

A very small fraction of the sources best fitted with a rising SFH in the SB regime show large ages ($t \gtrsim 1 $ Gyr) and very large $t/\tau$
ratios, showing that their SFRs have been dramatically increasing for an extended period. Closer examination shows that, given the fact
that the SFR at the onset of star formation is very close to 0, this rapid increase has had significant impact on the mass build-up of these
models only in the last hundreds of Myr, with \mstar\ rising one order of magnitude or more in the same period. Such solutions obtained with exponentially rising SFHs
can be seen as approximations of ongoing recent bursts, as best allowed by the models we have explored. Indeed, if the photometry is dominated
by an ongoing burst, there is little way to constrain how star formation was 1 Gyr before it.

\subsection{On the star-formation histories (SFHs) and observed trends}	\label{s_sfh}

\begin{table}[tb!]
	\resizebox{0.49\textwidth}{!}{\begin{tabular}{l c c c c}
		\hline
		\hline
		  SFHs 	& CSFR(\%)  & DECL(\%) & DEL(\%)  &  RIS(\%) \\\hline
		  Free \av 	& 3.6   & 50.8 & 23.3 &  22.3 \\
		  Fixed \av  	& 1.9   & 43.0 & 17.0   &  38.1 \\
		  Fixed \av, $z>3$ 	&  0  & 22.2  & 18.5    & 59.3   \\	
		\hline   
	\end{tabular} }
	\caption{Percentages of preference in fitting the  various SFHs to the sample, based on the \ki2. 
	In the first line we have the initial sample with $\ki2 < 10 $ for the CSFR models (704 sources).
	The second line	considers only the energy conserving subsample where the \lir's are correctly matched to observation
	(633 sources). In the third line we show the high-$z$ end of the previous (27 sources), where the fraction of solutions
	preferring the rising model has absolute majority.}         
	     \label{tab_sfh}
\end{table}

In this section we briefly discuss the SFH preferences of our sample with the method we explore, and how they are 
affected with respect to the more standard approach.  The robustness of the results shown here, based on 
best fitting SFHs, has been assessed with Monte-Carlo simulations, as described in Appendix \ref{s_errors}. 

Over the entire sample, in the fits with no extinction constraint,  half of the sources prefer the declining SFHs, followed by the delayed and rising
with a little less that 25\% for each, and the rest for the CSFR models (Table \ref{tab_sfh}).
Given that most of these solutions underestimate the observed
\lir, the fixed \av\ solutions increase the extinction on average, thus making the models more star-forming, or even starbursty. 
Doing so, there is an impact on the SFH preferences for the best fits, in favor of the rising SFHs mostly, which allow for maximum
ongoing star formation. 
Of course this is valid in a statistical sense and many sources that successfully reproduce the observed \lir\ can prefer SFHs with strongly decreased
SFR, as discussed earlier. The fraction of solutions favoring the rising models, also increases significantly with $z$ (and exceeds 50\% after
$z$ = 2.5), as is discussed in more detail in Sect. \ref{s_sfh_z}. 

\subsubsection{SFHs versus \mstar}		\label{s_sfh_mass}

\begin{figure}[htb]
    \hspace{-0.2in}
    \includegraphics[width=0.51\textwidth]{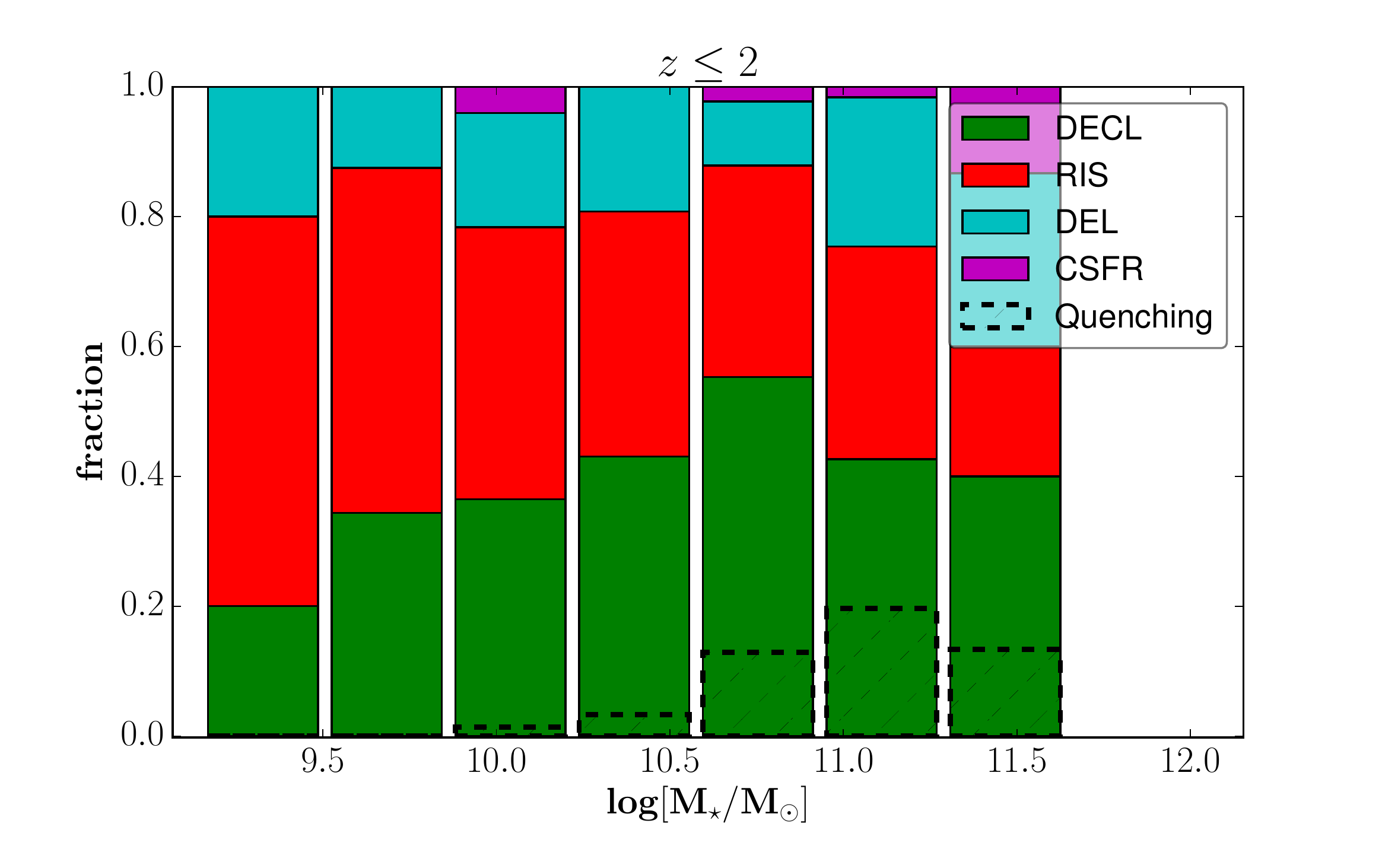}
    \caption{Fractions of each SFH in the mass distribution of the $z\leq 2$ sources where the sample is 
    closer to mass-completeness. The dashed and hatched fractions represent the quenching sources.
    We can see that the main trend  is that the rising models predominate in the low masses and their fraction decreases with increasing mass.
    }
  \label{mass_anatomy-fig}  
\end{figure}

Among the preferred SFHs in the sample we note a trend in SFH preferences according to \mstar\
(shown in Fig. \ref{mass_anatomy-fig}, where we plot the fractions of preferred SFHs in bins of mass for the $z\leq 2$ sources 
where the sample is less biased toward starbursts).
As would be expected and can be guessed also from the SFR -- mass diagrams in Sect. \ref{s_ms} (Fig. \ref{MS-fig}), 
rising SFHs are more preferred in the low-mass end, with 50-60\% of the best fits in the mass range $10^{9.5}-10^{10} \msun$, 
while at the high-mass end this fraction decreases to less than 20\% and the fraction of best fits with  declining models increases.
We note also that the preference of constant SFR models, although marginal, increases also with mass.
Overlaid on the diagram, are the fractions of the quenching sources, found in the largest mass bins mainly ($\mstar\geq 10^{10.5} \msun$),
but remain small overall.
Being that the low mass bins ($\mstar \leq 10^{10} \msun$) contain mainly starbursts and their fraction decreases with mass, the trend in SFHs
is affected by this. When considering MS galaxies only, we see no particular trend.  
On the other hand, extending this diagram to the higher $z$ sources enhances the trend.

The mass dependence of SFH types is not affected by the \lir-constrain. Indeed the diagram of Fig. \ref{mass_anatomy-fig}
shows no qualitative difference in the case of the fits with the unconstrained dust extinction. However, the fraction
of the quenching sources is much larger then, since the SFRs and \lir's tend to be strongly underestimated then 
(as discussed in the previous sections and shown in the SFR -- \mstar\ diagram of Fig. \ref{MS-fig}),
reaching 50\% and more for masses above $10^{11} \msun$.

%
%
\subsubsection{SFHs versus redshift}	\label{s_sfh_z}

An important aspect we wish to address in this work concerns the question whether the evolution of cosmic star-formation history through redshift
is reflected in the evolution of individual galaxies, in a statistical way, of course. This means, qualitatively speaking,
whether star-forming galaxies beyond the peak of star formation at $z \sim 1.5$ tend to be found more often to have their star formation on the rise,
and if in more recent epochs this tendency is inversed in favor of declining star formation.
%

\begin{figure}[htb]
  \hspace{-0.2 in}
  \begin{tabular}{l}
   \includegraphics[width=0.53\textwidth]{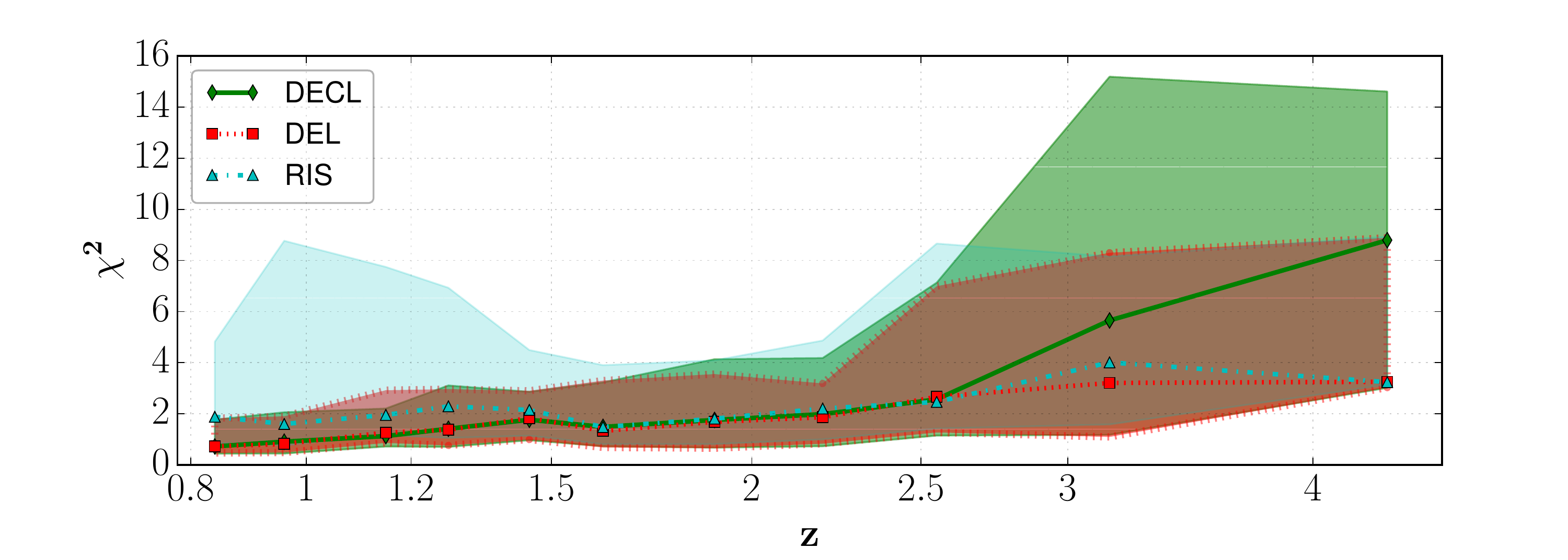}	
  \end{tabular}
  \caption{
 Values of the median \ki2 per redshift bin for the 3 SFHs explored (excluding CSFR),
  for the MS galaxies of the energy conserving sample.  The areas indicate the 68\% interval surrounding the medians.
  (\textit{Colors/patterns:} 
  green diamonds and hatched area : declining; red squares and filled area limited by dotted lines : delayed;
  cyan triangles and open circle-filled area: rising).
  }
  \label{x2-fig}
\end{figure}

To do so, we restrain ourselves to the MS galaxies of the sample, the starbursts being usually very short-lived events, where it makes little
sense to discern this evolution. The threshold for belonging to the MS subsample is set at 0.6 dex above or below the MS \citep[as depicted in][]{2011ApJ...739L..40R},
a rather loose definition, to keep from lowering the already limited statistics. Also, the sample is extended toward lower redshift (down to
$z \sim$ 0.8) to better check the significance of the eventual trend below $z \sim 1.5$. Finally, it is separated in redshift bins, aiming for balanced
distribution in numbers and logarithmic scale. Obviously the low-$z$ bins are more 
populous than the high-$z$ ones, with the two last ones ($ z_{\rm med}\geq 3$) 
containing about 10 sources each, which may hinder somewhat the robustness of the diagnostic.

Fig. \ref{x2-fig} shows the evolution of the median \ki2 obtained for each SFH and redshift bin. 
CSFR was opted out, as it is not strictly comparable to the others because of having one fewer degree of freedom (and 
 does not add anything of value in the plot).
Although no strong conclusions can be drawn,
the trends at the 68\% level give interesting insight.
Below $z \sim 1.5$, 
we see that declining and delayed models prevail while the rising SFHs produce comparatively worse fits. 
The latter have a median \ki2\ about two times larger than the former. 
In the $1.5<z<3$ range there is no 
clear preference, and above $z \sim 3$, the preference shifts toward the rising models. 
Delayed SFHs still work well as they behave similarly to the rising models in their early phase. 
Declining models on the other hand produce very bad solutions on the high-$z$ in our energy conserving fits.

Given the limitations of the sample at high-$z$, we have checked whether the large \ki2's of the declining
SFHs were due to higher $R_{\rm SB}$ within our adopted definition of the MS. The median \ki2's do not change
if we consider  $R_{\rm SB} \geq 2$ in all $z$ bins in order to compare the $z \gtrsim 3$ galaxies with strictly
similar ones at lower $z$, indicating that the change at high $z$ is not due to a bias in $R_{\rm SB}$.

From this exercise, it is tempting to suggest that there is indeed a parallel between the global CSFH, and the statistical behavior of
individual galaxies, with rising SFHs being preferred at high $z$ and declining at lower $z$. 
 Of course, other systematics relating to the detection quality of sources at $z\sim 3-4 $ with \emph{Herschel} that are beyond the scope of this work
might come into play \citep{elbaz2011}.
Larger samples above $z \sim$ 3, as well as gaining access to lower luminosities, might be necessary to conclude 
more robustly.
This said, the findings of this section (or as summarized in Table \ref{tab_sfh}, together with the trends in mass (Sect. \ref{s_sfh_mass}) 
leave room for optimism. Based on what we have seen, lower mass galaxies at high-$z$ that have been out of reach for \emph{Herschel} are expected 
to reinforce the observed trend.	



\section{Comparison with the average cosmic star-formation history} 	\label{s_csfh}

\begin{figure*}[htb]
  \centering
  \begin{tabular}{l r}
     \includegraphics[width=0.48\textwidth]{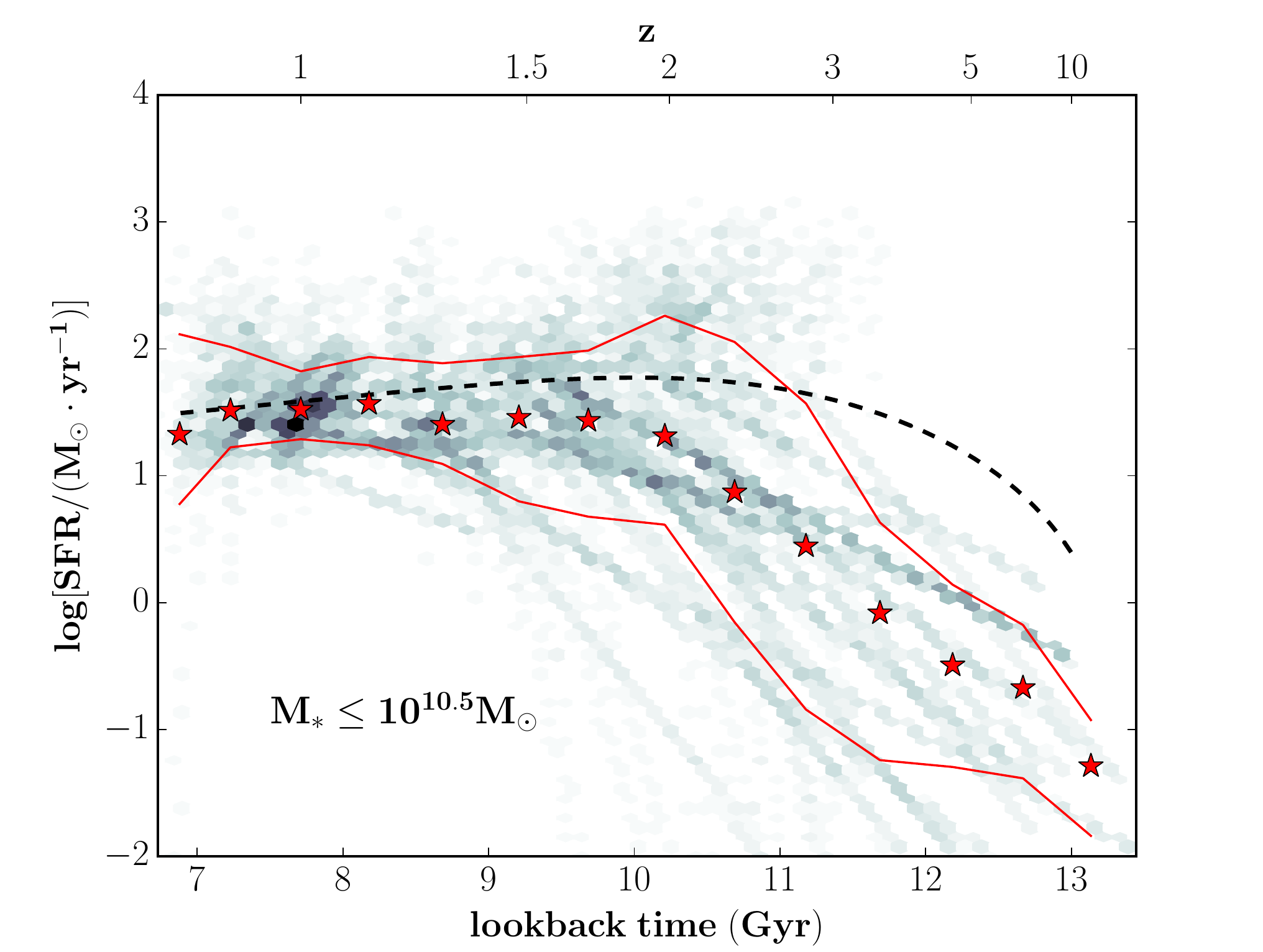} &  \includegraphics[width=0.48\textwidth]{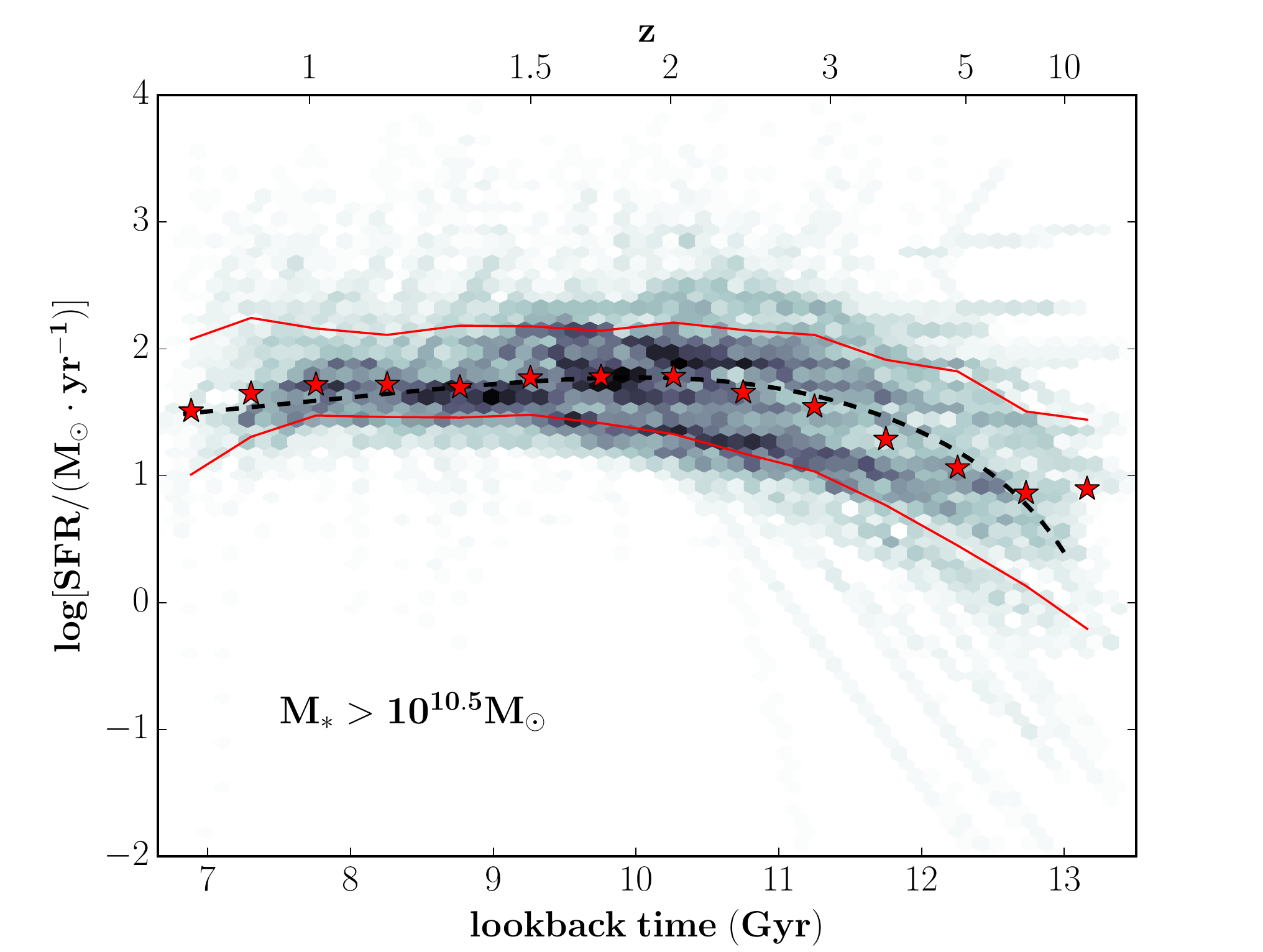}\\
      \includegraphics[width=0.48\textwidth]{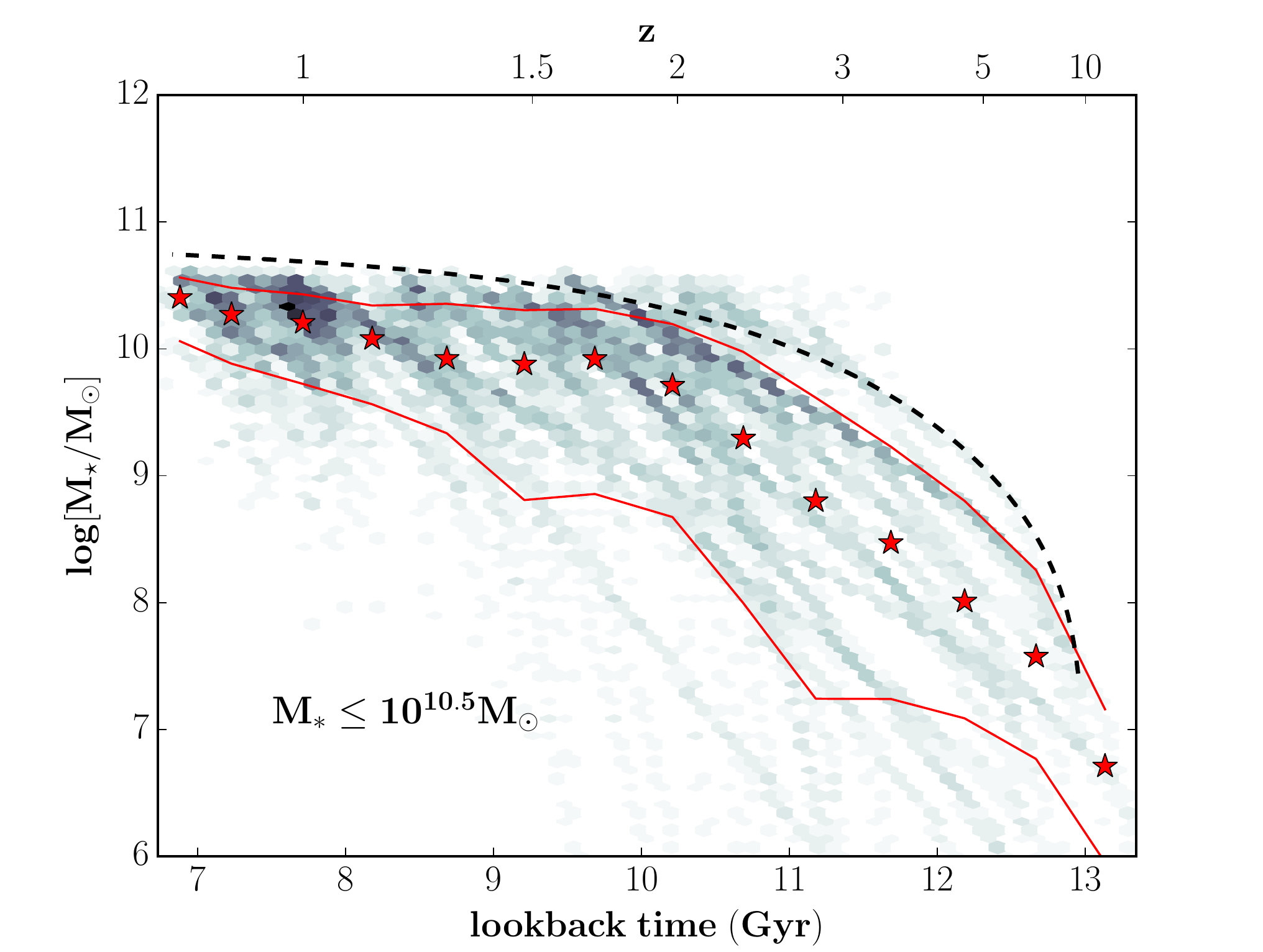} &  \includegraphics[width=0.48\textwidth]{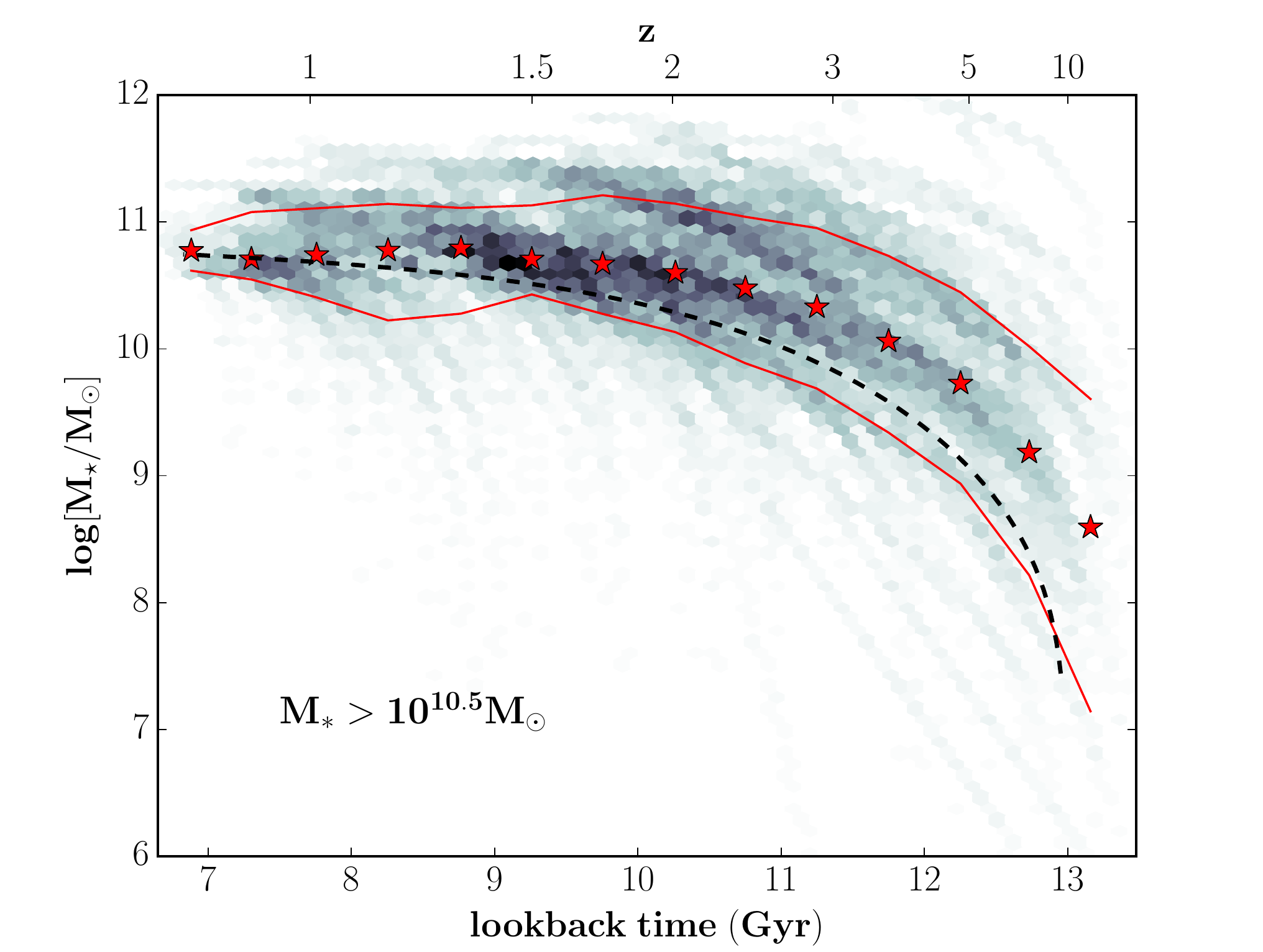}\\   
   \end{tabular}
    \caption{Evolution of the best fitting SFHs of the sample in terms of SFR (top) and stellar mass (bottom), in the 
	     form of density maps. The left panels contain the low mass half of the sample, with $\mstar \leq 10^{10.5}\msun $, and the right panels the
	     high mass half, above $10^{10.5}\msun$. The red stars and lines mark the medians and 68\% intervals in bins of 0.5 Gyr. 
	     On the top panels, the black dashed line corresponds to the CSFH from \cite{madau2014}, arbitrarily scaled to the height of the 
	     heavy mass sample (similarly on the bottom panels the stellar mass density, integrated from the CSFH and rescaled).
	     The general picture obtained is that the more massive galaxies have built their mass earlier and 
	     faster than the lighter ones. 
  }			\label{f_cosm_evo}
\end{figure*}

Presently we explore the past star-formation history of our sample, namely how stellar mass and SFR 
have evolved in our best fitting solutions\footnote{Like in Sect. \ref{s_sfh_z}, we will be using our extended sample down to $z \sim 0.8$, 
to have a better view of the epoch after the peak of cosmic star formation.}.
Namely, following up on the remarks of the SFH preferences versus $z$ in the previous section, a more
direct comparison with the CSFH can be of interest. Also it can be useful to situate our findings in the broader
context of galaxy evolution, such as whether the general picture of star formation we have obtained for this sample is  
 compatible with the  ``downsizing'' scenario \citep{cowie96}.  
In short, this hypothesis proposes that the most massive galaxies form earlier in time and faster \citep{thomas2005}
and is one of the popular galaxy formation scenarios in the community 
\citep[e.g.,][]{2004Natur.428..625H,2005ApJ...619L.135J,2005ApJ...635..260S,perez2005,2007MNRAS.378.1550P,2009ApJ...698L.116P}.

Evidently, our models cannot offer insight as to how star formation would continue past the redshifts of observation, but it is interesting
to trace the mass build up back in time, as it is shaped by the SFH models we have used. 
We recall that our models consist of simple exponential functions (Sect. \ref{sed_mod}). 
As such, the individual SFHs they produce are equally simple, and representative of the broad lines of star formation in a galaxy.
We have seen in Sect. \ref{s_sfh} that the  galaxies of our sample tend to have different preferences among the SFHs 
depending on their properties ($z$, \mstar), but how does the evolution of the sample as a whole look like? 

To build the picture of the sample's star-formation history, we first reconstructed each galaxy's individual SFH from the parameters of their
best fit. From the age of the solution and the redshift of observation, we computed the time of the onset of star formation in each case:
\begin{equation}
  t_{\rm start} = t(z) - {\rm age(SED)}		\label{eq_time}
\end{equation}
Given the variety of ages we have found in the sample, the redshifts corresponding to these times range from $z\sim 1$ up to $z\geq20$. 
It is important to stress that by ``time of onset'' we mean the time when the first stars of the modeled stellar population formed.
This does not prohibit the real galaxy considered from having formed stars earlier than that, in the context of a more complex SFH where
the oldest stars can be outshined by the younger ones, and hence missed by our fits. 
Each one of these times serves as starting point for the track the corresponding galaxy has followed, which is reconstructed using  
\taf, the initial SFR, and the appropriate function of its best fitted SFH. 
In this way we can visualize how the stellar mass has built up with time. 

As the sample is not large enough compared to samples of non IR-selected galaxies, rather than separate it in many mass bins and compare
the evolution of the heaviest versus the lightest in time \citep[as done in][]{2013ApJ...762L..15P} we separated it into two mass bins, 
each bin containing about half of the sample\footnote{The extended sample contains a larger
quantity of lower mass galaxies, bringing the median mass close to $ 10^{10.5}\msun$.}, with the boundary set at $\mstar = 10^{10.5}\msun $.

For the two subsamples, the evolution of their SFRs and mass build-up with cosmic time are plotted as density maps versus lookback time\footnote{We recall
that for the cosmology used in the present work, the age of the Universe is 13.462 Gyr.}(and $z$ in the secondary axis above the plots) in Fig. \ref{f_cosm_evo},
which is organized as follows: the left panels show the galaxies with $\mstar \leq 10^{10.5}\msun $, and the right panels the ones above $10^{10.5}\msun $.
The top panels show the evolution of the SFRs, and the bottom panels the evolution in mass. The red stars show the median values in bins 
of 0.5 Gyr, and the lines the 68\%\ intervals around them. The gray scale indicates the density of the overlapping tracks, and this way allows
to visualize the SFH of the subsamples as wholes, from their individual SFHs. The CSFH of \cite{madau2014} is also depicted, 
arbitrarily rescaled\footnote{Indeed, the comparison here is meant
to be purely qualitative, as the SFRs shown are not in the form of a volume density.} to match the lower $z$ part of the heavy mass diagram.

We can see that the  SFRs are higher at early times in massive galaxies, whereas lower mass galaxies have a steady increase of SFR over
several Gyrs. This follows from the fact that high mass galaxies are majoritarily better fit by declining SFHs, and the low mass galaxies
favor the rising SFHs. So for the former, it is equivalent to pushing their rising phase toward the earliest moments of the Universe, 
something not modeled in the SFHs used (but logically implied in the form of an instantaneous burst 
which gives the declining SFHs their high initial SFR). 

From the median values that trace the areas
of the diagram that are densest, we note that the massive galaxies reach their peak of star formation at redshifts between 1.5 and 2.
while the low mass tend to reach it closer to $z\sim1$, albeit with greater scatter. Interestingly, the global evolution of the massive
galaxies resembles closely the CSFH, with only a relatively small scatter. 

\begin{figure}[htb]
\hspace{-.18 in}
  \includegraphics[width=0.54\textwidth]{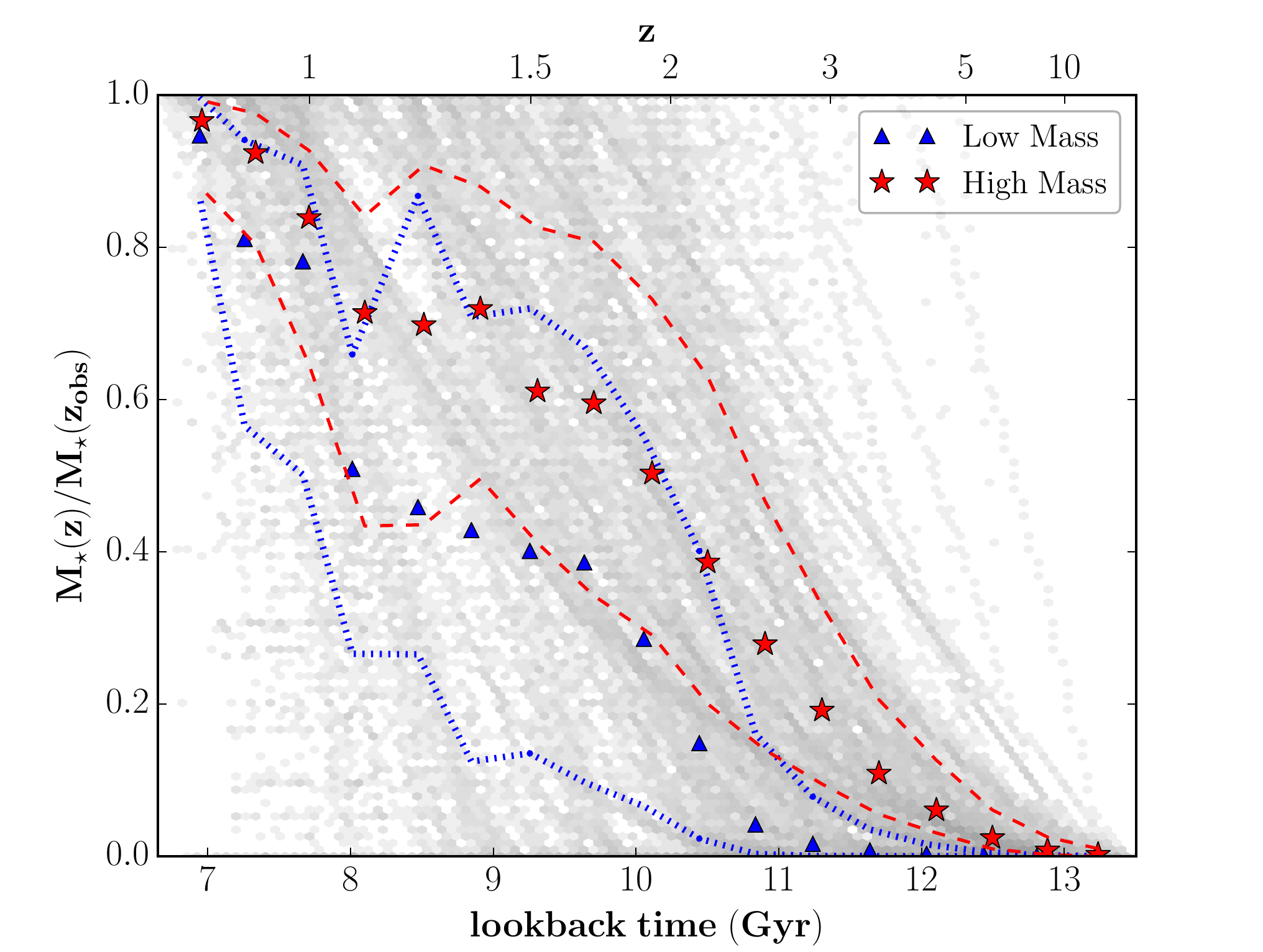} 
    \vspace{-0.1 in}
    \caption{Median evolution of the mass fractions of the high-mass (red stars) and low-mass (blue triangles) galaxies of the sample. The dashed lines
    represent the 68\% intervals, and the individual tracks of each source (as in Fig. \ref{f_cosm_evo}) are shown in the background in gray. 
    We can see that the high-mass sources reach their half-mass by $z\sim2$, while for the low-mass ones this happens close to $z\sim1$. 
  }	\label{f_mfracs}
  \hspace{-.1 in}
  \includegraphics[width=0.52\textwidth]{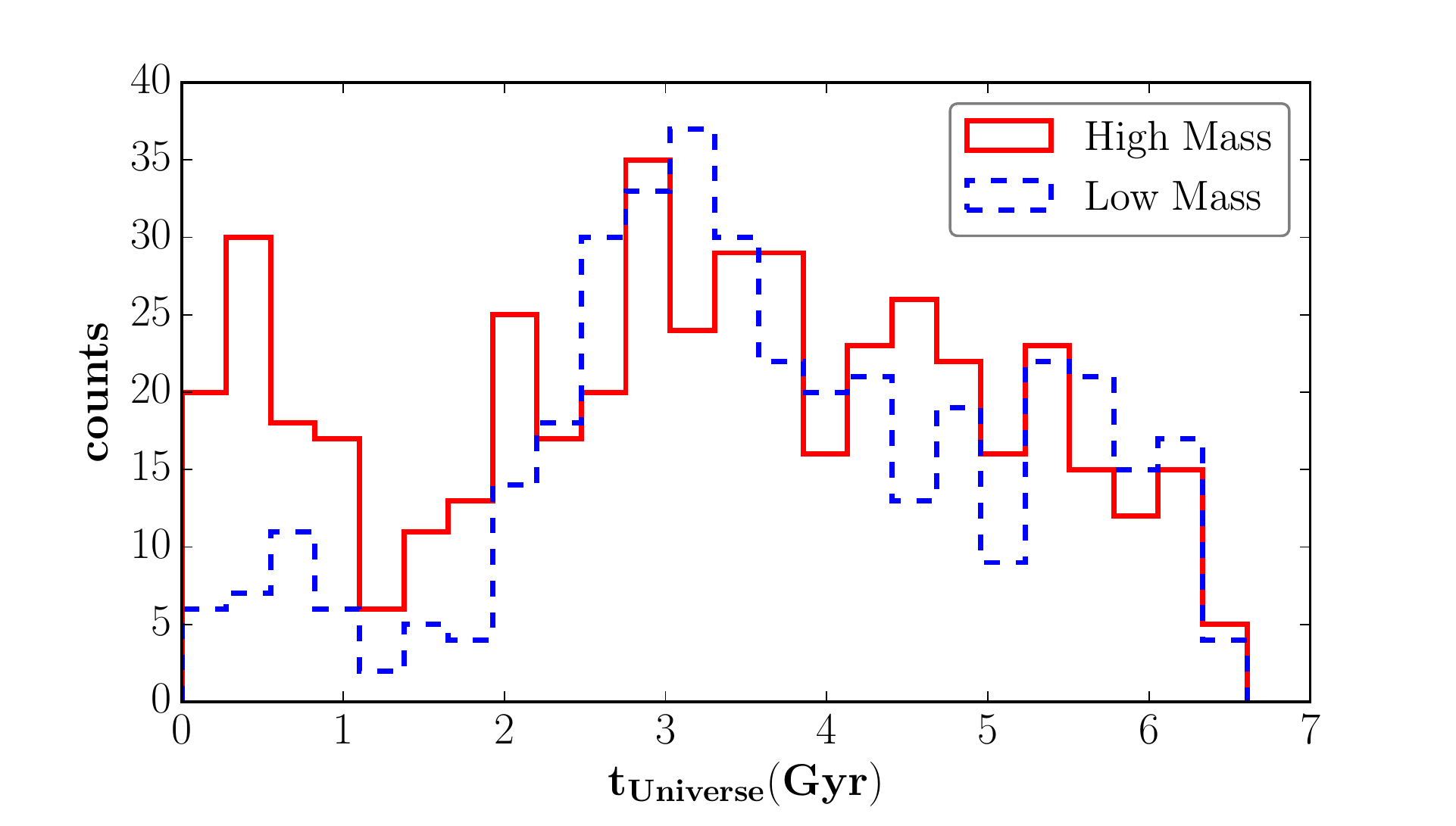} 
    \vspace{-0.1 in}
    \caption{ Distributions of the starting times of the best fitted SFHs for our sample. The red solid histogram corresponds to
    the high mass half of the sample, and the blue dashed one to the low mass. The times are defined as the time of observation minus
    the age of the modeled SED (Eq. \ref{eq_time}). We can see that there are significantly fewer low mass galaxies in the first two
    Gyrs of time.
  }	\label{f_onset}
\end{figure}

When comparing the evolution of the masses, we note that the massive galaxies build up faster, and by $z\sim2 $ have often built half
of their mass. In contrast the lower mass galaxies build up more gradually and 
seem not having reached a collective peak of their star formation
at $z\sim2 $, but rather show mass increase down to $z\sim 1$. 
This contrasted evolution in \mstar\ between the high-mass and the low-mass sources is further illustrated
with Fig. \ref{f_mfracs}, where we show the median evolution of the mass fraction (defined as the mass built at any given redshift divided
by the ``final'' mass at the redshift of observation) with time. The median evolution of the high-mass sources is shown with red stars, while 
that of the low-mass sources is shown with blue triangles (the dashed lines indicate the 68\% intervals, and the individual tracks 
are shown in the background in gray). The contrast is very visible until $z \lesssim 1.5$, at lower $z$ both groups contain increasing numbers
of younger star-formation events that cause some overlap.

It is interesting to compare the times of onset of star formation of the two subsamples, to better assess the presence of downsizing. 
In Fig. \ref{f_onset} we show the distributions of $t_{\rm start}$ for the high mass (red) and low mass (blue-dashed) subsamples.
The striking remark to be made here is the difference of the distributions in the first 2 Gyr of the Universe's age. We see that 
galaxies that started their star formation (as modeled by our SEDs) during that period are about three times more likely to end up in the high mass half
of the distribution by the time they are observed. 

The above remarks imply that the  evolution of the sample in our analysis is compatible with the downsizing scenario
and other recent works.  Relying on simulations, \cite{2013ApJ...770...57B} find that their most massive galaxies ($\mstar\geq 10^{11}\msun$)
peak in SFR at $z\sim3$, while low mass galaxies show an increase in SFR that endures to the present day. Similarly to the findings of
\cite{2013ApJ...762L..15P}, we find that the median sSFR of the low mass galaxies is $\sim 0.5$ dex larger than that of the massive ones for $z\leq 5$ 
(and comparable for higher $z$).
Of course, the limitations of our sample can come into play in this question too, as low mass galaxies are not detected beyond certain $z$, 
and if some galaxies were to form early and reach only a moderate
mass in a short lapse of time, then they would not have been picked up in an IR-selected sample.
That said, we have checked how the bias toward starbursts affects the sample's SFH \citep[we recall that our sample by construction has a larger
contribution from starbursts than the average star-forming galaxies population, as shown in e.g.,][]{2011ApJ...739L..40R, schreib2014}, by using only the 
MS galaxies to trace it. Keeping in mind that the bulk of the sample is observed at $z<2$ and hence the diagrams of Fig. \ref{f_cosm_evo} are dominated
by these galaxies, we find that it changes almost imperceptibly. In particular the high-mass bin whose starbursts account for only $\sim 10\%$ in number
remains about the same. The low-mass bin has a larger number of starbursts and also suffers more from mass-incompleteness, but the general evolution 
we see in Fig. \ref{f_cosm_evo} stays the same.

\section{Implications} 	\label{s_implic}

In the present section we discuss aspects that follow from the analysis and are of particular interest, regarding 
our view of star formation. 
Notably, on the ages, the implications of the existence of a potential evolution in the rising SFHs' timescales, and 
on the quenching sources. 

As discussed in Sect. \ref{s_ms_t}, depending on the position on the SFR -- \mstar\ diagram with respect to the MS, that is, on it, above it,
or below it, the SED models best fitting our sample present some common features. 
In terms of ages, we can distinguish three main groups, that coincide with the three regimes we see on the \mstar -- SFR diagram
(as shown in Fig. \ref{ms_t}):
\begin{itemize}
 \item[1)]   the high-end of the SFRs, that have very young or the youngest ages,
      corresponding mostly to starbursts;
 \item[2)]   modest SFRs (of the order of 100 \msunyr\ or below) that have $\sfrbc \approx \sfruvir$, they have old ages, above
      1 Gyr. In combination with their large \taf\ timescales, can be interpreted as ``normally'' star-forming, and are located 
      on or close to the MS;
 \item[3)]    sources whose \sfrbc\ deviates from \sfruvir, these start from $\sim 120$ Myr and can be up to
      several hundreds of Myr old. This is compatible with being in a post-starburst state, or undergoing quenching.
\end{itemize}
As shown in Table \ref{mass_table}, the three groups also cover different mass ranges, with the SB being lighter, and the quenching heavier than
the MS sources.	We briefly recall how the said groups are defined in this work: MS galaxies are within 4 times of
their corresponding main sequence, starbursts are more than 4 times above it, and the quenching galaxies are labeled so, if their \sfrbc\ is
4 times or more smaller than their corresponding \sfruvir. Because of the definitions used, there is a very small overlap, with 7 sources qualifying
as quenching but are located close to the MS. They are counted with the former, as they are in a rapid transit from above to below the MS. 
A more detailed summary of the masses and other physical parameters can be found in the Appendix \ref{s_errors}.

\begin{table}
 	\centering
	\resizebox{0.49\textwidth}{!}{\begin{tabular}{l c c c c}
	\hline\hline
	 Group & $<\log[\mstar/\msun]>$ & $1\sigma$ & ages (Gyr)& \% \\\hline	
	 Main Sequence & 10.65 & 0.4 & 0.09-2.6  & 69 \\
	 Starbursts & 10.14 & 0.34 & 0.05-0.18 &  24\\		
	 Quenching & 10.89 & 0.27 & 0.13-0.25 &  7 \\
	\hline
	\end{tabular} }
  \caption{ Summary of some physical properties found in our sample after regrouping the sources with respect to their position on the 
	   \mstar -- SFR diagram: mean (sensibly equal to the median) and standard deviation of $\log[\mstar]$, the median 68\% interval of the ages, and the 
	   percentage that each group holds within the sample.}	\label{mass_table}
\end{table}


\subsection{The rising SFHs: a potential evolution in timescales?}		\label{s_evol}

\begin{figure*}[htb]
 \hspace{-0.24 in}
    \includegraphics[width=0.55\textwidth]{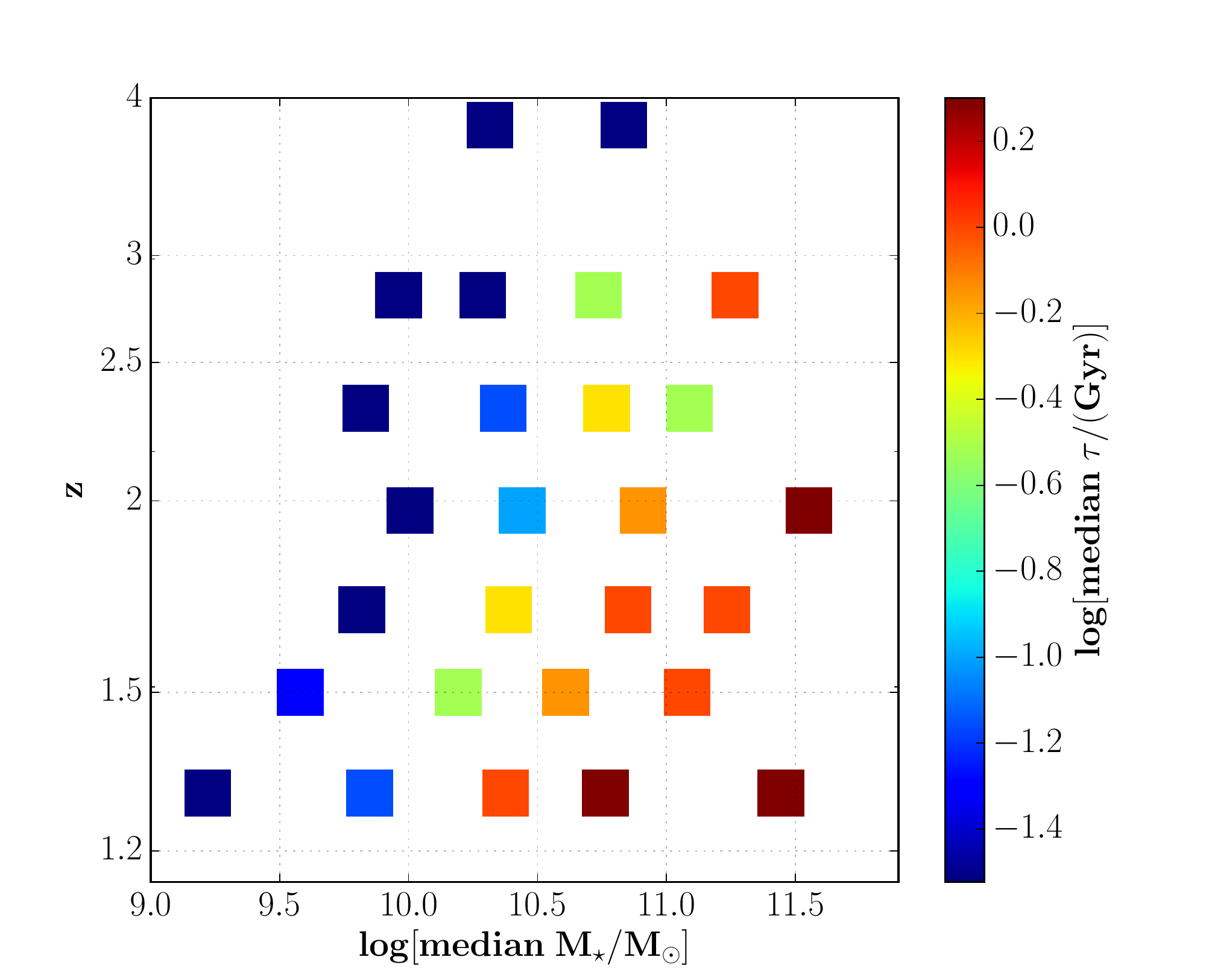}
   \hspace{-0.38 in}  
     \includegraphics[width=0.55\textwidth]{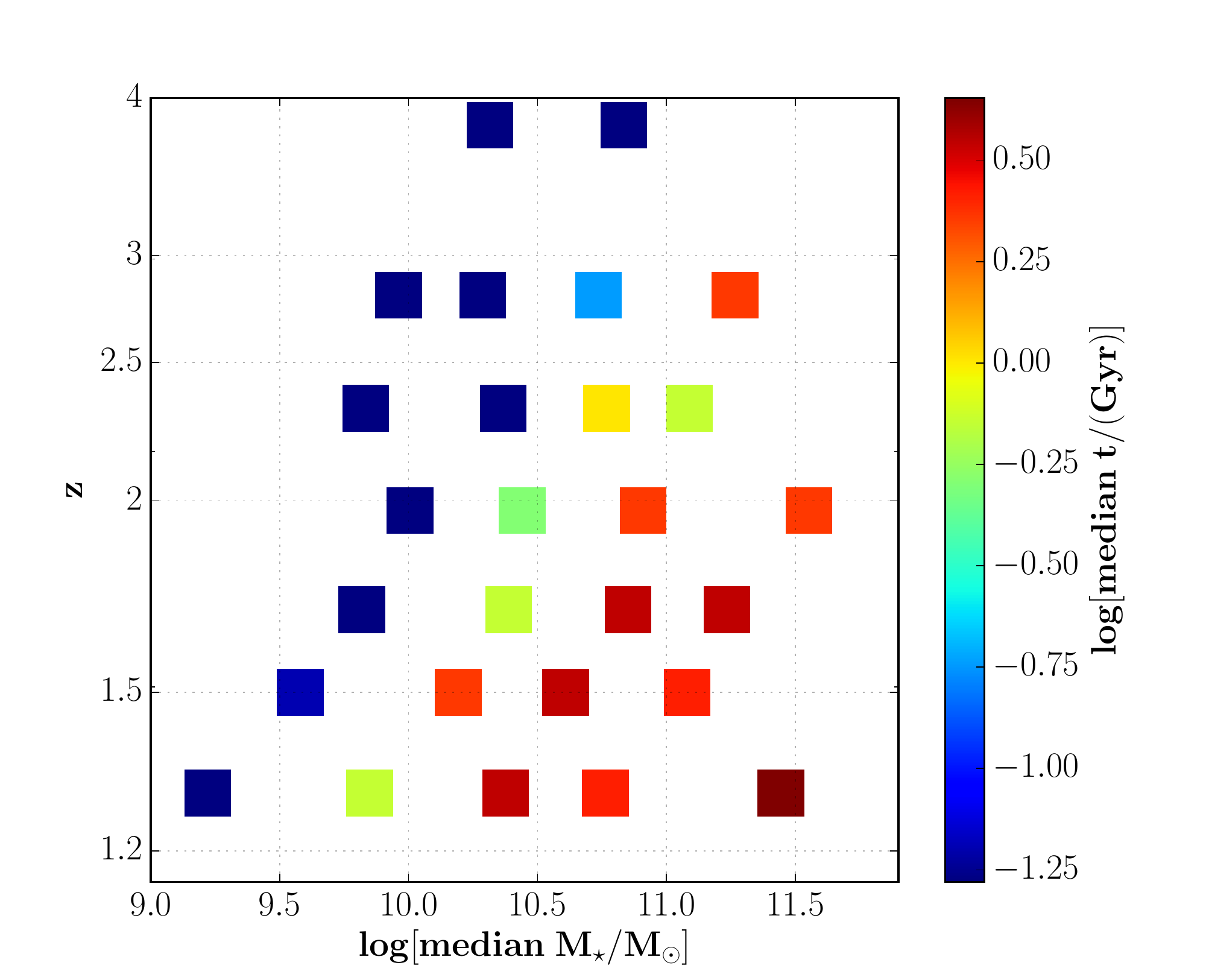}     
    \caption{Binned \mstar\ vs $z$ diagram, for the sources best fitted with the rising SFHs, with the median \taf\ (left) and age $t$ (right) shown in color
    Both \taf\ and $t$ are observed to increase with  \mstar\ and cosmic time (decreasing $z$). 
	     } 		\label{m-z-tau-fig}  
\end{figure*}
 
In Sect. \ref{s_timesc} we have noted that there is a trend in the e-folding parameter \taf\ with both 
\mstar\ and $z$ when considering the fraction of the sample that is best-fitted by models with rising SFHs. 
In the current section we discuss this into more detail, namely we explore the behavior of both \taf\ and $t$ of the rising
SFHs with $z$. 

Our findings are  shown in Fig. \ref{m-z-tau-fig}, where these galaxies are binned both in \mstar\ (in 0.5 dex-wide bins) and $z$ 
(representing time steps of 500 Myr, except for the highest $z$ bin that spans 900 Myr).
In the left panel the bins are colored
according to their median \taf, and in the right one according to their median $t$.
At each redshift bin, 
\taf, the timescale indicating how fast the SFR rises is increasing with mass, meaning that the increase in the SFR  
is slower in the high mass sources than in the low mass. Likewise, when considering a particular range of \mstar, we see 
that the galaxies of that mass have an SFR that is increasing much faster at high-$z$ than at lower $z$.
Closer examination of the distribution on the \mstar\ -- $z$ plane reveals that the lowest masses and \taf\ bins (the blue squares in 
the left of the diagram) contain mainly SB galaxies ($\tau \approx 30 - 50$  Myr), whereas the rest is dominated by MS sources. 
In the right panel of Fig. \ref{m-z-tau-fig} we can see that the ages $t$ have a behavior very resembling to that of \taf, leading to
a consistent possible interpretation that the SFR of the massive galaxies has increased to its current levels earlier on. 
So, ultimately, what we have is that at a given mass range, a higher $z$ galaxy in our sample increases its SFR faster than a lower $z$ one (while the
SFRs themselves are overall higher at high-$z$ compared to low-$z$), while at a given $z$ a higher mass galaxy appears to have been forming stars
for a longer period (and again at higher SFR, exception made for the starbursts which evidently peak at higher SFRs while being very young and have 
low masses). This whole picture is consistently compatible with the downsizing scenario discussed in Sect. \ref{s_csfh}.  
These findings, in particular the evolution of the ages can of course be expected to some extent by the limited age of the Universe, which 
lowers the age cap with increasing $z$. But this cosmological constrain alone does not seem sufficient 
to produce this smooth evolution. 

Indeed, the discussed trends of \taf\ and $t$ with both \mstar\ and $z$ become as pronounced as seen in Fig. \ref{m-z-tau-fig} only in our energy conserving fits.
The same diagram made in case of the unconstrained extinction fits, does not show trends as clear or monotonic as the former.
Absent the IR-constrain, and due to the higher preference for declining SFHs in this case, the bulk of the sample appears to have formed a too important
amount of mass early on (by $z\sim5$), regardless of the final mass, which leaves little room for evolution.
We note in passing that we have explored similarly constructed diagrams for the sources preferring declining SFHs and have observed no significant
trends like the ones discussed for the rising SFHs.

As the choice of the sub-sample used in Fig. \ref{m-z-tau-fig} is based on a relative criterion (sources that are better fitted with rising rather 
than with declining SFHs), we made the same diagram using the absolute criterion of including only sources for which the rising SFH fit yields 
$\ki2 \leq 5$. The result is very sensibly the same as the one shown. The dispersions for the median \taf's are similar to the ones of Fig. \ref{mass_tau}.

 
Further on, the observed behavior of \taf\ leads legitimately to ask the question whether it indicates an evolution in the rising SFHs, 
if the lower-mass, high-$z$ galaxies were to be the ancestors of the higher-mass, lower-$z$ ones. 
This cannot be addressed easily, but one can explore how galaxies would evolve in time and on the SFR -- \mstar\ diagram if their SFH was based on it,
meaning if they would follow a more complex rising SFH, where \taf\ becomes larger with time and stellar mass.
We explored this possibility with the help of a simple toy-model to build-up mass following a prescription 
derived from this trend. 
To do so we started from the analytical expression of the exponentially rising SFH, 
only this time instead of using a constant \taf\ throughout
the evolution, \taf\ will depend on \mstar\ and $z$.
A galaxy is created with an arbitrary initial mass at $z=4$. Different initial SFRs are used, to qualitatively explore the various regimes of star formation
with respect to the MS. A time grid with 2000 steps is defined from $z=4$ to $z=1$, the redshift range roughly covered  by the sample. 
At each step, \taf\ is redefined according to the present mass and redshift, and the increase in SFR and \mstar\ is impacted accordingly. 
Iterating this procedure gives us the time evolution of \mstar\ and SFR and produces tracks that we can compare on the \mstar -- SFR diagram.
 We recall here that for this experiment, we do not consider the declining
SFHs, because for a galaxy to stay on the MS over large periods of time 
its SFR needs to be at least constant, if not increasing
\citep[e.g.,][]{2009ApJ...698L.116P,renzini2009,maraston2010}. 
This holds true despite the fact the MS normalization decreases with time, as established so far up to  $z \sim 4$ \citep[][]{schreib2014}. Indeed,
if a galaxy strictly follows the MS as it is currently parametrized, its SFR will be increasing for most of its growth, and would decrease only once
reaching a high mass (of the order of $10^{11} \msun$) or a low redshift ($z\leq 1$).
A more 
proper model, taking into account the turning off of star formation and the path toward quiescence, would necessitate a more elaborate numerical prescription, and is
beyond the scope of the present work\footnote{For a work focusing specifically on modeling quenching sources, \cite{ciesla2016} is suggested. }.

To create the prescription for $\taf(\mstar,z)$ we performed 2-dimensional planar fits of \taf\ as  function of \mstar\ and $z$.
Higher order polynomial fits were also tested but despite the residuals being slightly reduced, they proved unpractical
as even small extrapolations were unstable.
Two fits were made, one utilizing all the covered area in Fig. \ref{m-z-tau-fig}, including both SB and MS, and one
considering only the MS galaxies. The former assumes that galaxies start at the SB regime as the minimal mass bin is 
made of SB galaxies. Likewise for the latter and initial conditions on the MS. 
The parameterizations obtained are:

\begin{equation}
  \log(\tau/\mathrm{Gyr})  = 0.993  \log(\mstar/\msun) - 0.511  \mathit{z}  -10.070		\label{eq_full_param}
\end{equation}

for the MS+SB fit, and:

\begin{equation}
  \log(\tau/\mathrm{Gyr})  = 0.402  \log(\mstar/\msun) - 0.382  \mathit{z}  -3.770		\label{eq_ms_param}
\end{equation}

for the MS-only fit. 
A lower threshold of \taf = 0.03 Gyr was kept, to avoid having the SFR jumping at non realistic values in a few Myr because of extrapolations.

\begin{figure}[htb]
 \hspace{-.2 in}
	  \includegraphics[width=0.55\textwidth]{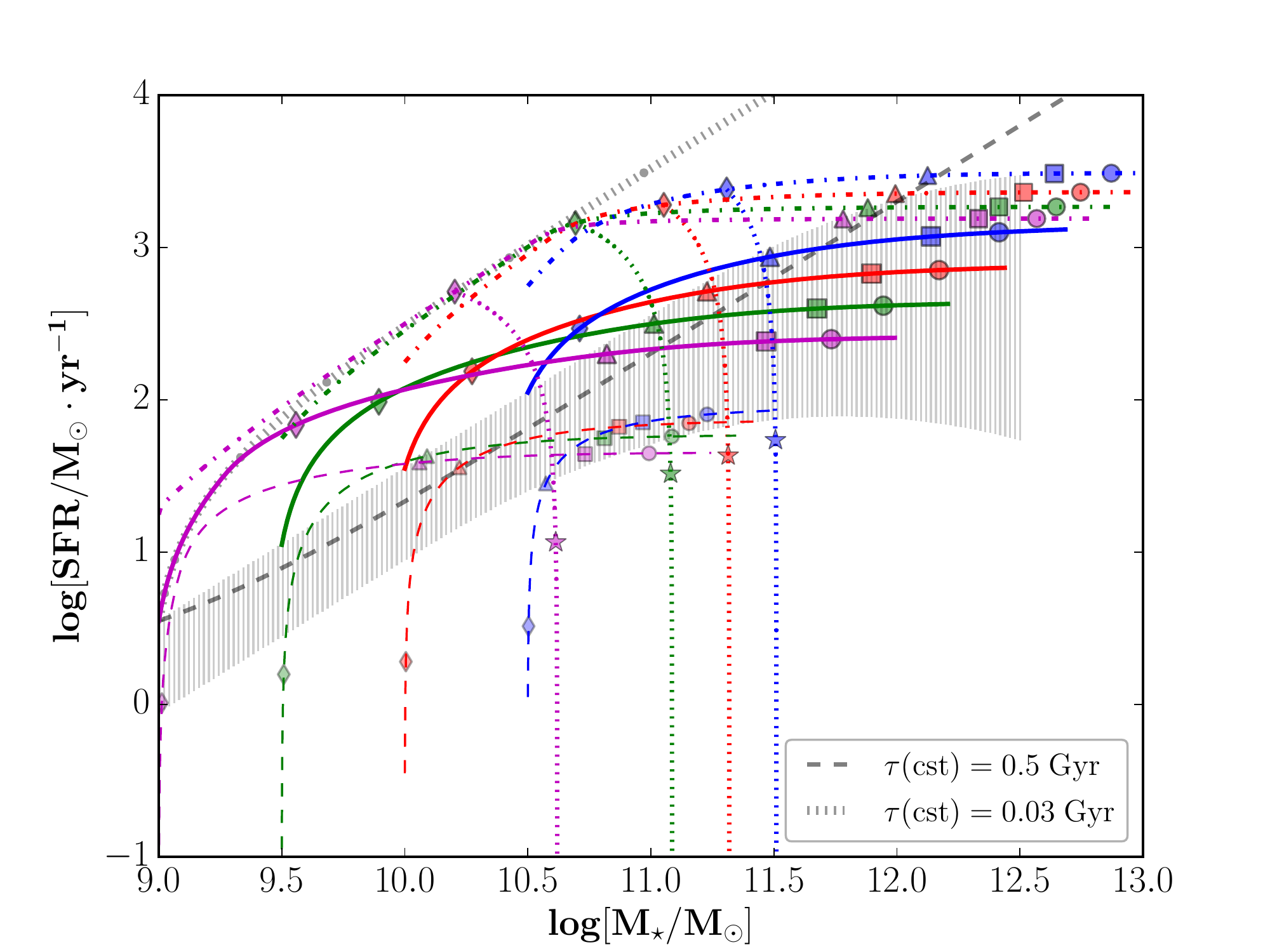}
    \caption{ Mass build-up tracks from a simple toy-model based on the trend of \taf\ with \mstar\ and $z$, plotted
	  on the SFR -- \mstar\ diagram. The colors indicate different initial masses (ranging from $10^9$ to $10^{10.5}$  \msun), 
	  and the line types different initial positions  with respect to the MS, for these masses. The solid and dashed color tracks follow
	  the parametrization of Eq. \ref{eq_ms_param}, and the dash-dotted ones that of Eq. \ref{eq_full_param}. The dotted tracks show how the 
	  starbursts would evolve if at $t = 100$ Myr they would start to rapidly decline in SFR ($\taf = 50 $ Myr).
	  The symbols (diamonds, triangles, squares, and circles)
	  correspond to different times after the beginning of the tracks (respectively, 0.1, 0.5, 1.5 and 2.5 Gyr), while the   
	  stars mark the position on the tracks after 200 Myr of decline.
	  The thick gray lines in the background show how the typical exponentially rising SFHs behave, with a constant \taf\ of 30 and 
	  500 Myr (dotted and dashed, respectively), while the hatched area represents the evolution of the MS from $z=4$ to $z=1$.
	     } 		\label{ms_evol-fig}  
\end{figure}

Using these two parameterizations we produced model tracks that follow a rising SFH with an e-folding parameter that increases with stellar mass
and cosmic time. 
Tracks for four initial masses ($\mstar = [10^9, 10^{9.5}, 10^{10} , 10^{10.5}  \msun]$, covering the lower half of the sample's mass range)
are shown in Fig. \ref{ms_evol-fig}, in different colors. Obviously we had to extrapolate from the parameter space from which the calibrations were produced, as 
for example we do not have sources of $10^9\msun$ beyond $z=2$ (\footnote{The extrapolation was to extend the parameterizations linearly to $z = 4$ and down to
$\mstar = 10^9 \msun$, while keeping the threshold at \taf = 0.03 Gyr. Initial masses below $10^9 \msun$ were not considered.}). 
For the initial SFR's, we show three values per chosen mass: one that corresponds to the MS at given mass at $z=4$, one that is ten times below that
to see how a galaxy below the MS that starts star-forming would fare, and one five times above the MS. 
The two former use the calibration derived
from the MS subsample only (Eq. \ref{eq_ms_param}), and the latter the one from both the SB and MS subsamples (Eq. \ref{eq_full_param}). The hatched area
in the diagram represents the MS from $z = 4$  to $z=1$, according to \cite{schreib2014} (Eq. \ref{eq_ms}).  
%

Along the tracks that emerge from this model, it is relevant to discuss the dimension of time. The symbols (diamonds, triangles, squares, and circles,
with slightly different sizes to distinguish between the three groups of tracks)
mark the different times along each track from the beginning of the evolution (corresponding respectively to 0.1, 0.5, 1.5 and 2.5 Gyr). 
With the SFR rising relatively fast at the beginning, the tracks tend to go above the MS, but within 500 Myr they are back in its
vicinity, with much heavier mass. There, since the increase of their SFR is significantly dampened ($\taf > 1 $ Gyr),
they continue to build up mass for several Gyrs in a steady way. 
The tracks beginning below the MS evolve very similarly to the ones on it, but occupy a lower regime, and ultimately build up less mass
over the considered period. If the initial SFR is set arbitrarily low, the same scheme is reproduced, with always more time spent
for the SFR to reach a value that impacts the mass. But if this is achieved and $z = 1$ is not yet reached, we see the evolution
in mass in such way that their time marks (squares and triangles in particular, marking the 1.5 and 2.5 Gyr positions respectively)
remain aligned with the same ones at higher mass and SFR (this characteristic is inherent to the discussed parametrization of \taf).

The SBs (following Eq. \ref{eq_full_param}) work in the same way, but at much higher regime, 
 meaning that the SFR (and hence the mass) rises faster and to higher values, than with the tracks evolving according to Eq. \ref{eq_ms_param}.
This happens even in the initial phase where the SFR rises relatively fast in all tracks, as can be seen by the greater distance reached
within 100 Myr (diamonds). Initiating the SFRs on the MS rather than 5 times above it brings almost imperceptible changes to the tracks, as long
as the parametrization of Eq. \ref{eq_full_param} is used. 
Evidently, SFRs of the order of 1000 \msunyr\ are not expected to last for billions of years, so at any point of the evolution
a galaxy is expected to see its SFR starting to decline, something not modeled here.

Of course, here the evolution of the tracks past $\sim 10^{12} \msun$ is completely extrapolated with respect to the sample.
It is just the result of allowing the mass to grow continuously for the $\sim 4$ Gyr that separate $z = 4$ from $z = 1$ which has
no reason to occur in reality.	
Given that the fraction of sources preferring 
a declining SFH increases with \mstar\ (Fig. \ref{mass_anatomy-fig}), it can be argued that the heavier a galaxy becomes
the more probable it is for its SFR to start declining. 
Regarding the SB branch in particular, the tracks become extrapolated from the sample already starting at $\mstar = 10^{11} \msun$. 
This means that for the initial masses considered, starbursts should hardly last 100 Myr. 
To illustrate the decline of their SFR, we have arbitrarily added a bifurcation at the 100 Myr marks of their tracks 
(dotted lines) and made it follow an exponentially declining SFH with a minimal $\taf = 50 $ Myr, as found in the quenching subsample.
We observe that from that time, it takes $\sim 200$ Myr for the track to evolve sufficiently below the MS, and to effectively
appear to be quenching (positions marked by stars).  If indeed there is a causal link between the SB galaxies and the quenching ones, 
the cycle of their star-forming episode, is then of the order of 300 Myr, which is compatible with the ages we find in our 
SED fits.

It was not necessary to test a calibration using starbursts only, as they strongly favor the lowest \taf's of the order of 30 -- 50 Myr, regardless of $z$ 
and \mstar. A track showing a standard exponentially rising SFH with a constant \taf\ = 30 Myr is plotted with a gray dotted line, and illustrates how
the very quickly rising models fare on the SFR -- \mstar\ plane. In this case, the SFR reaches $10^4\ \msunyr$ and  \mstar\ jumps 2.5 orders of magnitude
in less than 200 Myr. 
Interestingly, the initial SFR impacts very little on this track, if it is set to begin on the MS or lower, its increase is so rapid that it moves up the diagram
faster than the mass build up, and continues on the same path as illustrated by the dotted line. The only difference is that it reaches $10^4\ \msunyr$
after $\sim 270$ Myr if the initial SFR was on the MS, or $\sim 400$ Myr if it was arbitrarily set 2 orders of magnitude below it. 
If such star-forming events exist they must be extremely rare, but surely should not be lasting longer. 
On the other hand, a more moderately rising SFR (dashed black line, \taf\ = 500 Myr), the same increase in SFR takes place over a few Gyr.

From this experiment -- if indeed a causal relation between \taf\ and \mstar\ and $z$ is real, as seen by our SED fits with rising SFHs -- 
the picture of star formation that emerges is both episodic and steady. It starts off with intense star formation with rapid mass build-up,
and then can evolve rather smoothly close to the  high-mass half of the MS for long-lasting periods,  to the extent that such SFRs can be sustained.

It may be useful to recall that the sample from which 
this toy-model is made, is biased toward high luminosities, and this can in part explain why the tracks  are mostly above the MS.


\subsection{On the quenching sources}	\label{s_quen}

The so-called quenching sources present a particular interest as they distinguish themselves in our analysis.
As a reminder, under the term ``quenching'' we regroup the sources that successfully reproduce the observed
\lir\ in the energy conserving models but have \sfrbc's significantly lower than their \sfruvir, and are rapidly declining.
For most of them, the fits with declining or delayed SFHs yielded significantly lower \ki2 values than the rising fits 
(a factor of 2 or more).

But the best fits obtained (in most cases, with a significant difference in the \ki2) for		
them prefer very rapidly declining SFRs.
They are 43 in total (7\%\ of the energy conserving sample) and are found in the lower-$z$ range of our sample, with half of them
in the interval $z \in [1.2, 1.5]$, and the most distant being at $z\sim 2.4$. They occupy the middle-to-high mass range (Fig. \ref{mass_histo-fig} and \ref{mass_anatomy-fig}).
 We recall that our selection did not consist of purely actively star-forming galaxies, however, according to the color-color criterion
of \cite{williams09} less than one third of the quenching sources qualify as quiescent (13 out of 43). In the \uvj\ diagram they are found close to the area
separating the star-forming galaxies from the quiescent, but more on the side of the former group.
%

\begin{figure}[htb]
 \centering
    \includegraphics[width=0.5\textwidth]{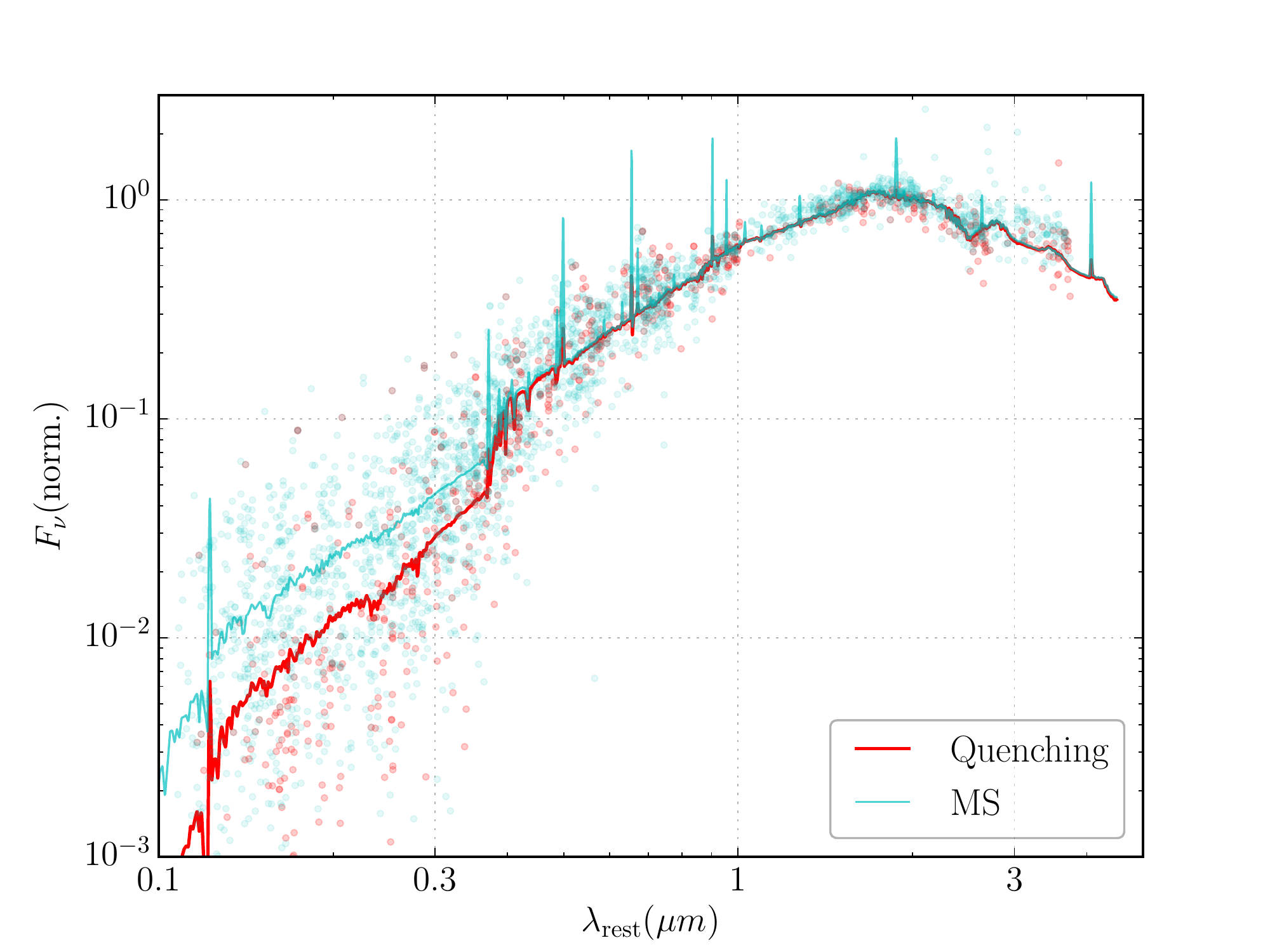} \\	
    \caption{The mean SED of the quenching subsample found in the GOODS sample (red), compared to the mean SED of MS galaxies (cyan) at 
  the same mass range, normalized and de-redshifted. We note the prominent Balmer break ($\lesssim$ 0.4 \micron), indicative of an aging population, and
  the weaker emission lines of the quenching SED. The average extinction is otherwise very similar.
  }
  \label{q_sed_fig}
\end{figure}

Fig. \ref{q_sed_fig} shows the mean SED obtained from the best fits of this subsample (red), compared to the mean restframe SED of 
normal MS galaxies in the same mass range as the quenching (cyan), after having been de-redshifted and normalized to $1.6$ \micron.
They show comparable amounts of dust attenuation,  with a mean $\av = 2$  for the quenching and
$\av = 1.9 $ for the MS galaxies. The main difference between the two is the Balmer break  that is more prominent in the quenching
SED, indicative of an aging population. 
This comes in strong contrast with the actual median ages we obtain for the two subsamples compared here: the quenching subsample has 
a median age of 180 Myr, while the MS comparison subsample has a median age of 1 Gyr. 
This may appear counter-intuitive if one judges by the SEDs of Fig. \ref{q_sed_fig}, but is easily explained by the strong opposition in the nature
of the SFHs that produced these SEDs (very rapidly declining SFRs versus quasi-constant or rising).
It must also be kept in mind that outshining \citep[e.g.,][]{maraston2010} can play an important role here in particular in the case of the MS sources who are actively star-forming.
That means that their median age of 1 Gyr should be interpreted as implying that they have a longer SFH, rather than that their SED is dominated by an old population, 
which is far from being the case as we can see in Fig. \ref{q_sed_fig}. 
Indeed, because of the sharp contrast in the star-forming regime  of the most recent 100-200 Myr in the two groups, the MS SED is being dominated by younger stars
than the quenching SED, while the older stars of the former contribute little to none at all to the emerging emission (which is one major 
source of the uncertainty associated to ages defined as the time elapsed since the onset of star-formation, used in the present work, and of course
some caution in using it is always recommended). On the other hand, the quenching SFH does not produce a sufficient amount of young stars to outshine
the older ones, and so the SED is more dominated by ageing stars.

\begin{figure}[htb]
  \centering
    \includegraphics[width=0.5\textwidth]{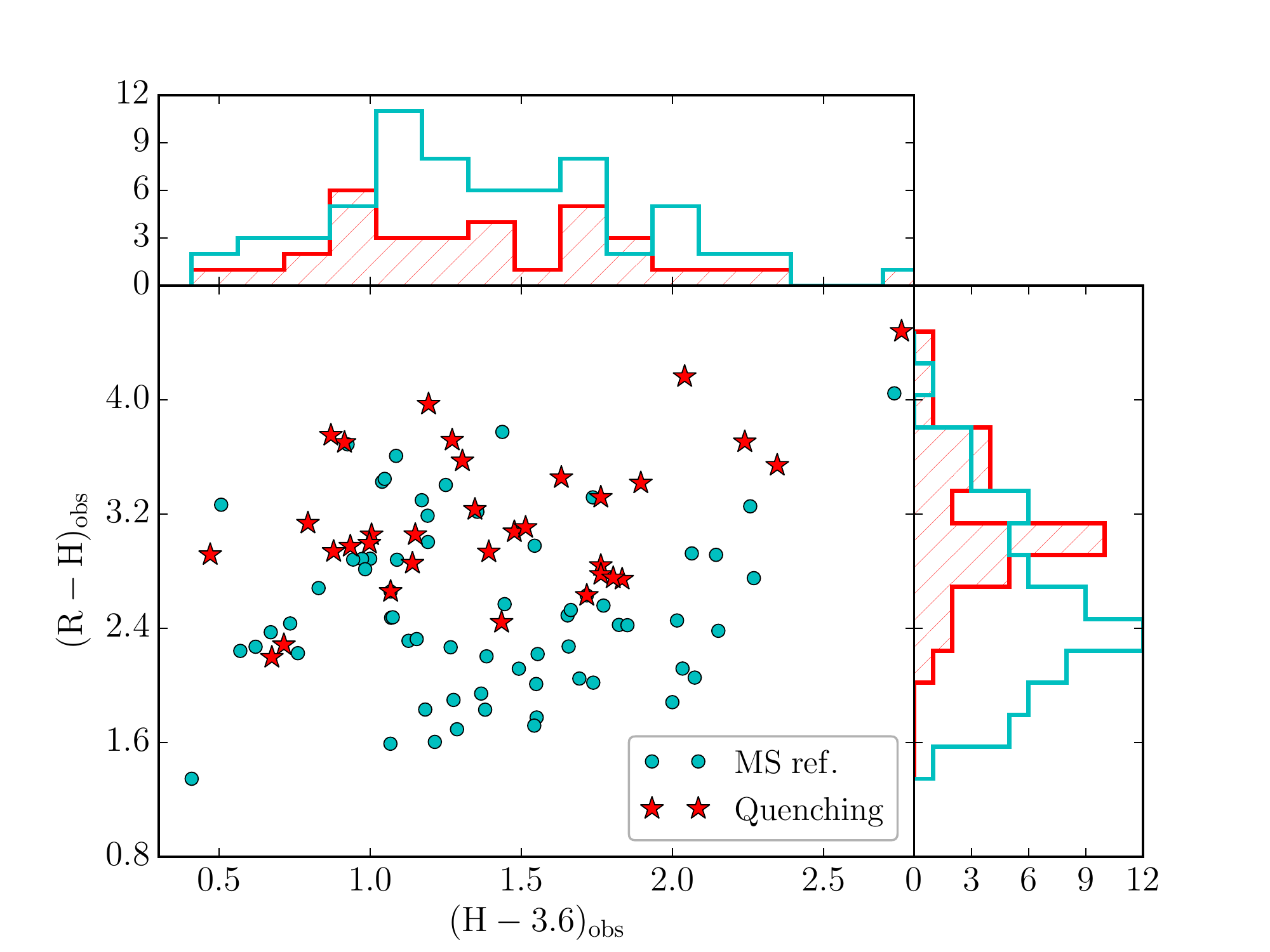}
    \caption{ $R - H$ versus $H - 3.6$ \micron\ color-color diagram and marginal distributions for the quenching sample (red stars, hatched area in the histograms) and the 
    MS reference sample (cyan circles). There is an overall shift upwards of the former compared to the latter, that is due to the more
    important Balmer break of the quenching sources.
  }
  \label{colordiag_fig}
\end{figure}

From this difference observed in the mean SEDs of these sources shown in Fig. \ref{q_sed_fig}, we turned to explore the photometry of the concerned
sources, and the possibility of discrimination with color diagnostics. We opted for a color-color diagram 
using the restframe NUV$-R-J$ bands, \`a la \cite{2013A&A...556A..55I}, following the authors' remark that the NUV$-R$ is more sensitive to
the present-over-past star formation than $U-V$. These bands respectively correspond to 260 nm, 650 nm, and 1.2 \micron.
For the redshift interval from 1.2 to 1.6 (containing 75\% of the quenching subsample) they can be fairly well approximated by the 
choice of the $R$, $H$, and 3.6 \micron\ bands in the 
observed frame. The first two bracket the Balmer break, while  $H - 3.6$ \micron\ is sensitive to the dust attenuation.
Fig. \ref{colordiag_fig} shows the $R - H$ versus $H - 3.6$ \micron\ colors for the quenching subsample (red stars) and the MS control sample (cyan circles), 
together with their marginal distributions with histograms. There is significant overlap between the two groups, with the only important difference
being in the $R - H$ color distribution  that is sensitive to the Balmer break. It is close to impossible to discriminate the 
two based solely on broadband photometry, but we see that for a given extinction, the reddest $R - H$ sources are very likely to be in
the quenching category. 

From the emission line flux predictions obtained in the fits, we observe that the equivalent widths of the quenching
subsample are on average 10 times weaker than in the comparison sample. For the parametrization used in our fits (Calzetti law, no differentiation 
between nebular and stellar attenuation), we do not expect \ha\ equivalent widths of more than 30\AA\ for these galaxies.
This feature is something that can potentially
be confirmed with NIR spectroscopy, to confirm their nature, as was shown for an individual case by \cite{sklias2014}. 

There are studies that explore the dependence of SFR indicators with time \citep[e.g.][for \sfruv]{2012ApJ...754...25R}.
As noted in Sect. \ref{s_timesc}, the commonly used Kennicutt calibration for the IR, assumes continuous SF over $\geq 100$ Myr periods.
Because of that, it may produce SFR values that are more like averages over similar periods, smoothing out episodic events or the evolution
over the last tens of Myr.
In this context, the  work of \cite{2014MNRAS.445.1598H} points out that the \lir-inferred SFR estimates may sometimes significantly 
overpredict the current star formation of galaxies (by e.g., misinterpreting post-starbursts as starbursts). 
We believe that this sub-group of quenching galaxies falls in that category, and its difference
appears thanks to the possibility of rapid decline of the SFR among the SFHs explored.

 
\subsection{On the cases where \lir\ is overpredicted}	\label{s_overpred}

  Now we comment briefly on the small fraction of sources for which our SED models overpredict the observed \lir. 
There are about 25 sources that overpredict it by 0.5 dex or more, regardless of the SFH used, when the extinction is left unconstrained. 
As far as their photometry is concerned, they 
do not seem to differ in any particular way from the bulk of the sample. The majority of them show very red colors, little to 
no evidence of a Balmer break, most are detected down to wavelengths of $\sim 1500 \AA$ restframe, and about half down to $\sim 1200 \AA$, so their
UV emission is well covered. Most are in the LIRG range, with only five above $10^{12} \lsun$.

They are found mostly on the right side of the Takeuchi relation on the IRX-$\beta$ diagrams (Fig. \ref{IRX-beta-fig}), but are not among the ones
we label as quenching (as the latter cause no troubles in the reproduction of the observed \lir's). It has been suggested in the literature 
\cite[e.g.,][]{2009ApJ...698.1273S,2012ApJ...745...86W} that for sources in this situation steeper attenuation laws such as the Small 
Magellanic Cloud (SMC) law \citep{prevot84} are better suited to describe them. 		
It should be noted though, that in the aforementioned works the galaxies studied were primarily situated in the low end of IRX, with UV slopes 
at $\beta \lesssim 0$. In our \emph{Herschel}-selected sample, the sources discussed here are in the IRX range between 1 and 3, and with $\beta > 0 $, for the most part.
This means that regardless of the attenuation law considered, a significant amount of extinction is needed to describe them. 
                                                 
Keeping this in mind, we have performed SED fits of this subsample using the approach described in the present work, and applying the SMC law, and also the 
Milky Way (MW) law of \cite{seaton79} that is moderately steeper than the Calzetti law.

From the fits performed with the SMC law, we observe that the \lir\ overprediction vanishes for most. On the opposite, the declining and delayed SFHs 
now tend to underpredict it. The rising and constant SFR models do a better job in that regard but produce significantly higher \ki2's,
in some cases the fits are really not matching the photometry. When applied, the SMC law tends to change the interpretation of the sources' SEDs,
usually finding older, less intensively star-forming solutions, requiring lesser attenuation \citep[see also][]{2012ApJ...754...25R, sklias2014}.

The fits using the MW law reduce the overprediction very slightly, but succeed rather well in fitting the UV-optical slopes, with the notable
exception of the 2175\AA\ graphite extinction feature which this law contains but for which we see no evidence in the photometry of this subsample. 

This exercise can only tell us that a simple switch of the attenuation law does not suffice to solve the issue, and that probably exploring 
a broader parametrization of the extinction law is needed, eventually with the input of medium-band observations. 
Another factor that might push for an overestimation of the extinction is the single component population in the fits, in case the NIR part of the spectrum
is dominated by very old stars, it can artificially produce a redder slope. Like with the UV-dominated sources discussed in Sect. \ref{s_free} and \ref{s_caveat},
a two-population approach should be more appropriate to produce more flexible SEDs, with a small UV budget for dust to absorb.

Lastly, given the small number of cases we found for this level of overprediction, and since most of them show signs of strong to very strong 
attenuation in their SEDs it is plausible that their FIR fluxes have been under-evaluated.

\section{Summary and conclusions} 	\label{s_conclu}

We have conducted a detailed study of a sample of $z\geq 1.2$ \emph{Herschel}-detected galaxies  
from the GOODS fields, to study their physical properties and star-formation histories, and to examine 
how using the IR luminosity as a constraint in SED fitting impacts the derived physical parameters. 
For that we have used the rich multi-wavelength photometry publicly available from \citep{elbaz2011,2013ApJS..207...24G}, and 
the GOODS-N optical catalog from \cite{pannella2015}, not public at the time of this work.
The initial sample has consisted of 753 star-forming galaxies, individually detected by \emph{Herschel}.

The fitting tool used is a modified version of \emph{Hyperz} \citep{bolzonellaetal2000}, modified to impose energy conservation in the 
SEDs of the stellar emission \citep[cf.][]{sklias2014}. Four SFHs were explored, that produce single component populations with an exponentially declining, rising, delayed, 
or constant SFR (Sect. \ref{sed_mod}).
We recall that inferring the SFRs from stellar population SED fits with varying SFHs leads intrinsically to a large dispersion
in the estimation of this quantity \citep[][]{2011ApJ...738..106W,2012ApJ...754...25R, 2013A&A...549A...4S,2013A&A...554L...3O}, and in the 
case of dusty galaxies the \sfrbc's are often underestimate the \sfruvir\ (Fig. \ref{MS-fig}, left panels).

Our approach has consisted in using the observed \luv\ and \lir\ as estimated in preliminary SED fits to 
constrain \av\ from the IR/UV ratio
 \citep[cf.][]{2005ApJ...619L..51B,2013A&A...549A...4S}, thus reducing by one the free parameters and produce energy-conserving fits.
This works well for 84\%\ of the sample, where we obtain good quality fits (in terms of \ki2), that also reproduce the \lir\ 
within 0.2 dex.  
The fact that we are able to produce  SEDs well fitting the photometry of the majority of the sample shows 
that the IR/UV-constrained extinction is a good proxy for the average extinction over the integrated emission of the
studied sources. 
The age-extinction degeneracy, common in red galaxies and that tends to mostly lead to underestimated
\sfrbc's is successfully broken, and in most cases leads to more dusty and star-forming solutions, coherent with the IR emission. 

The models produced in this manner impact and constrain our view of star formation, with respect to classical approach
(i.e., SED fits ignoring IR emission), in the following
ways:
\begin{itemize}
 \item 
  The weak \sfrbc's obtained with the classical fits are overall increased, leading to solutions with more ongoing star formation, that in terms of \ki2
  fit almost as well as when the extinction is a free parameter.
  In this way, the \sfrbc's become very comparable to the \sfruvir's for most yet not all of the sample (Fig. \ref{SFR_comp-fig}).
 
 \item 
  The estimation of the stellar masses is not much affected by our method.
  The values obtained with our method are on average $\leq 0.1 $ dex lower than with the unconstrained extinction fits. 
  This change, along with the previous point are the result of the IR/UV ratio leading to slightly higher \av\ on average, which
  in turn produces solutions that are slightly younger, less massive, and more star-forming.

 \item
  The instantaneous SFRs inferred by the SEDs (\sfrbc) are comparable to \sfruvir\ for about 80\% of the sample, implying that the 
  corresponding sources are near or at a state of equilibrium as assumed by \cite{K98}. 
  Our approach allows us to identify a fraction ($\sim 20$\%)
  of sources with $\sfrbc \ll \sfruvir$ (labeled quenching throughout this work), whose SFHs indicate a rapid decline
  and for which \lir\ does not properly trace recent SF \citep[cf.][]{2014MNRAS.445.1598H}.
  The quenching galaxies	
  belong to the heavier half of the sample in \mstar\ and 
  present a larger Balmer break than similar mass main sequence galaxies, which explains why they are best fitted with rapidly declining SFRs. 
  Spectroscopic observations should allow to confirm/infirm our interpretation of the data.

\end{itemize}

One main focus of this work has been to explore how the considered star-formation histories behave with this relatively large sample, and 
how the choice among them is affected by the \lir\ constrain.
In this context an important aspect are the involved timescales, specifically
with respect to the distribution of the sample on the SFR -- \mstar\ diagram. It is well summarized by considering the ages of the solutions and 
the  $t/\taf$ ratio, which illustrates how much change occurred within each one. 
We note that the main sequence galaxies have large ages and $t/\taf$  ratios
  close to 1, indicating continuous growth for long periods. The galaxies above the MS (starbursts) have mostly minimal ages, and fast rising SFRs. 
The quenching galaxies (below the MS) have ages superior to 100 Myr, and large $t/\taf$  ratios, meaning that their SFRs have declined rapidly in a short
 period of time, and are consistent with being post-starbursts. 
  
Regarding the trends in the star-formation histories of the studied  galaxies and signs of evolution, we find the following results:
\begin{itemize}
 \item
  There is a notable shift from  
  declining SFHs that are preferred for half of the sample when the extinction is unconstrained, toward rising SFHs in the energy-conserving
  models. This preference increases with redshift, and the fraction of sources best fitted with the rising SFHs exceeds 50\% at $z\geq 2.5$.
 The evolution of the preferred SFHs is suggestive of the cosmic star-formation history, with sources best fitted with declining SFRs 
  at $z\la 1.5$ and rising SFRs at high redshift, although this last remark requires further investigation with larger samples. 
 \item
  There is also a strong trend in the SFH preference with galaxy mass, where the majority of low mass galaxies (50\% or more for $\mstar \la 10^{10.3}\msun$)
   is best fitted with rising SFHs and high mass (50\% or more for $\mstar \ga 10^{10.5}\msun$) ones prefer declining SFHs. 
   
  \item
  By separating the sample in two parts in mass, and tracing the best fitted SFHs back in time, we are able to remark that its star-formation history
  is compatible with the downsizing scenario, despite the simple analytical form of the SFHs, based on exponential functions. 
  Massive galaxies, with typically $\mstar \ga 10^{10.5}\msun$, reach a peak
  of SF between $z = 1.5$ and 2, and follow the CSFH closely with little scatter.
  Lower mass galaxies have a peak of SF closer to $z\sim 1$.
  
 \item
  When considering the solutions with rising SFH among the rising models, we observe a strong trend of the e-folding parameter \taf\ with both $z$ and \mstar.
  The lower masses and higher redshift sources tend to have the smallest \taf, and then it increases with \mstar\ and lowering redshift. 
  From this trend, we have constructed a simple evolutionary toy-model, producing mass build-up tracks that follow this dependency of \taf\ with 
  the mentioned parameters. Applied to initial conditions within our sample it starts first in a bursty manner, rising above the MS for 
  about 500 Myr, then reaching it and evolving steadily for a couple of Gyr along the MS. During the period from $z=4$ to $z=1$ we tend to reach
  masses higher than $10^{12} \msun$, which suggests that at some point star formation must decline, corroborated by the increasing
  number of high mass galaxies that are best fitted by declining SFHs. 

\end{itemize}

Although some of the above findings are limited by statistics, the trends revealed can only gain in robustness when larger IR-selected samples will become available,
going beyond the reach of \emph{Herschel}, both in mass and $z$.

Finally, for the small fraction (16\%) of the sample for which \lir\ was not reproduced well 
by our models we note:
\begin{itemize}
 \item 
The sources where IR-constrained fits yield strongly underpredict \lir\ are very IR-luminous 
and tend to have among the bluest UV slopes of the sample. This indicates that the single component population approach and the condition of energy conservation as
applied here do not hold (that also shows in the fact that their fits have for most very large \ki2's), 
and other methods should be tried.

\item
For the sources that overpredict \lir\ we have tested first whether the choice of the Calzetti
law was inadequate. We have explored other attenuation laws but this alone does not solve the issue. Commonly used alternatives
like the SMC law prove very ill-adapted for these sources, for which their photometry suggests that they are  intrinsically very red. 
We note in passing the absence of the 2175 \AA\ feature in the studied subsample.
An explanation for their IR emission remains to be found.

\end{itemize}

Ways to address the described issues above are being explored, and will be the subject of a future publication. They will include
the use of double-population SEDs, and a broader parametrization of the extinction laws. 
More complex SFHs inspired by the trends discussed in this work can also be built and put to the test in characterizing observations.  

 To further constrain and confirm our findings concerning the quenching subgroup, our next steps will be directed to include spectroscopic
 observations when available, and medium-band survey data such as SHARDS \citep{perez-shards2013}. The latter have proven the capacity of
 bringing robust constraints in $z\geq 1$ galaxies that are one step more evolved than the sources discussed in the present work \citep{2015arXiv150707938D}.
 To strengthen the findings on the $z\sim 3-4$ sources using the constrain of energy conservation, ALMA observations will help improve the numbers,
 detection-quality, and luminosity-limitations for that era, crucial for understanding star formation before the cosmic peak.

\begin{acknowledgements}

We thank Prof. Veronique Buat for her insightful comments on the UV-related measurements and aspects.
This work was supported by the Swiss National Science Foundation and the University of Geneva.
P.S. acknowledges funding from the Swiss National Science Foundation under grant P2GEP2\_165426.

\end{acknowledgements}

\bibliographystyle{aa}
\bibliography{refs,references_ds}

\appendix
\section{On the robustness of the best fitting SFHs and estimation of uncertainties}	\label{s_errors}

Given the fact that most of the present analysis relies on the discrimination between different SFHs 
based on \ki2 minimization, we now will provide some elements regarding the robustness of 
the best fitting solutions and the uncertainties regarding the derived parameters.
To this end, we have conducted SED fits on Monte-Carlo (MC) variations of the stellar photometry 
(1000 variations per source, obtained by perturbing each photometric measurement within its corresponding error bar) for each SFH.
From this we obtain median values and 
68\% confidence intervals of the relevant physical parameters to illustrate the uncertainties.
To best assess the stability of the presented star-forming picture, and to provide 
meaningful insight on the physical parameters of the study's sample, 
the MC catalog was separated into the three discussed groups (MS, SB, and quenching),
based on the solutions found in the best fits of the original photometry.  
 
We first comment on the robustness in SFH preferences over the full sample and the subgroups, 
and subsequently discuss the physical parameters more quantitatively.

\paragraph{Remarks on the best fits.}

\begin{figure*}[htb]
 	\centering
 	\hspace{-.3 in}
	     \includegraphics[width=0.35\textwidth]{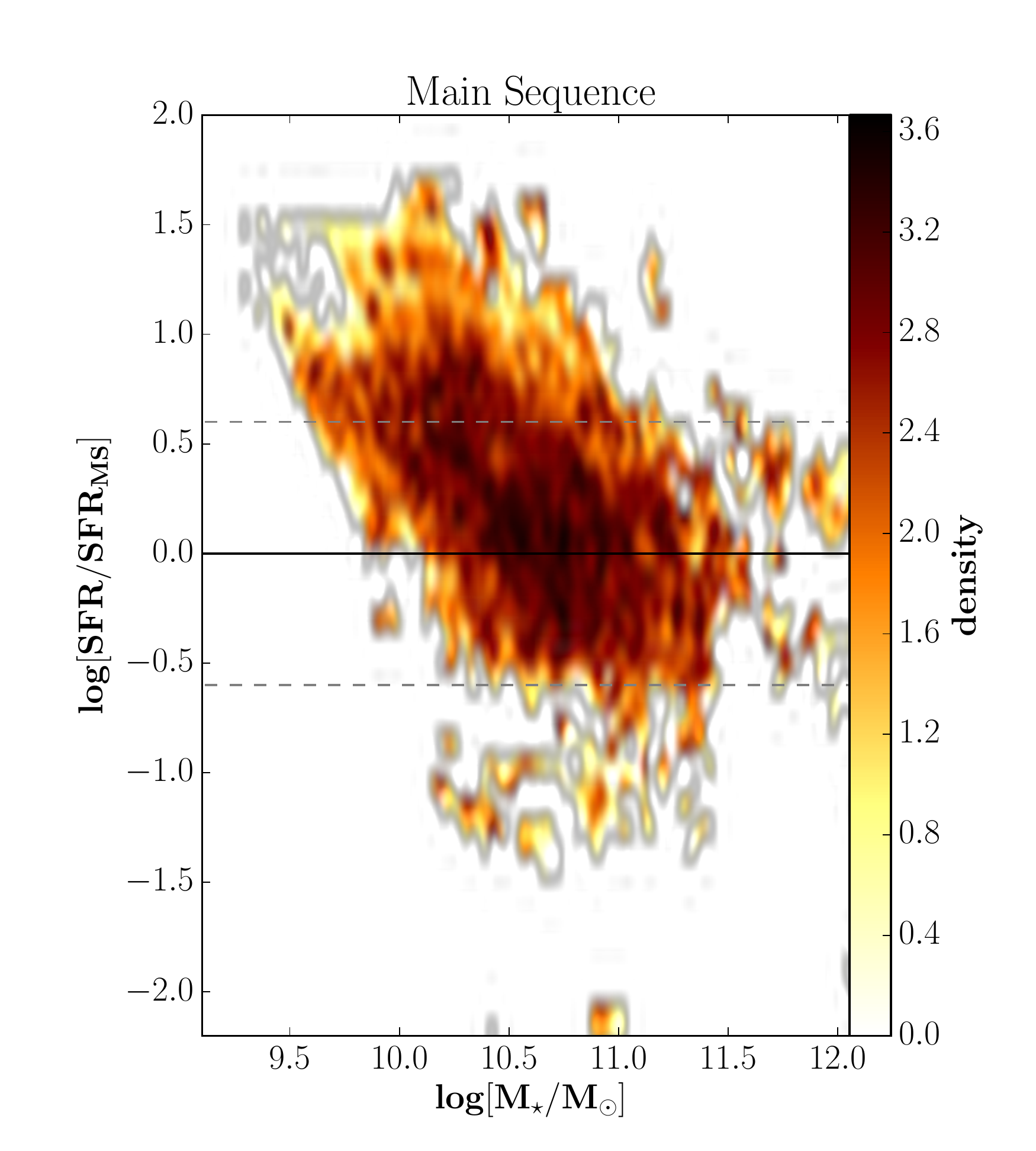}	\hspace{-.2 in} \includegraphics[width=0.35\textwidth]{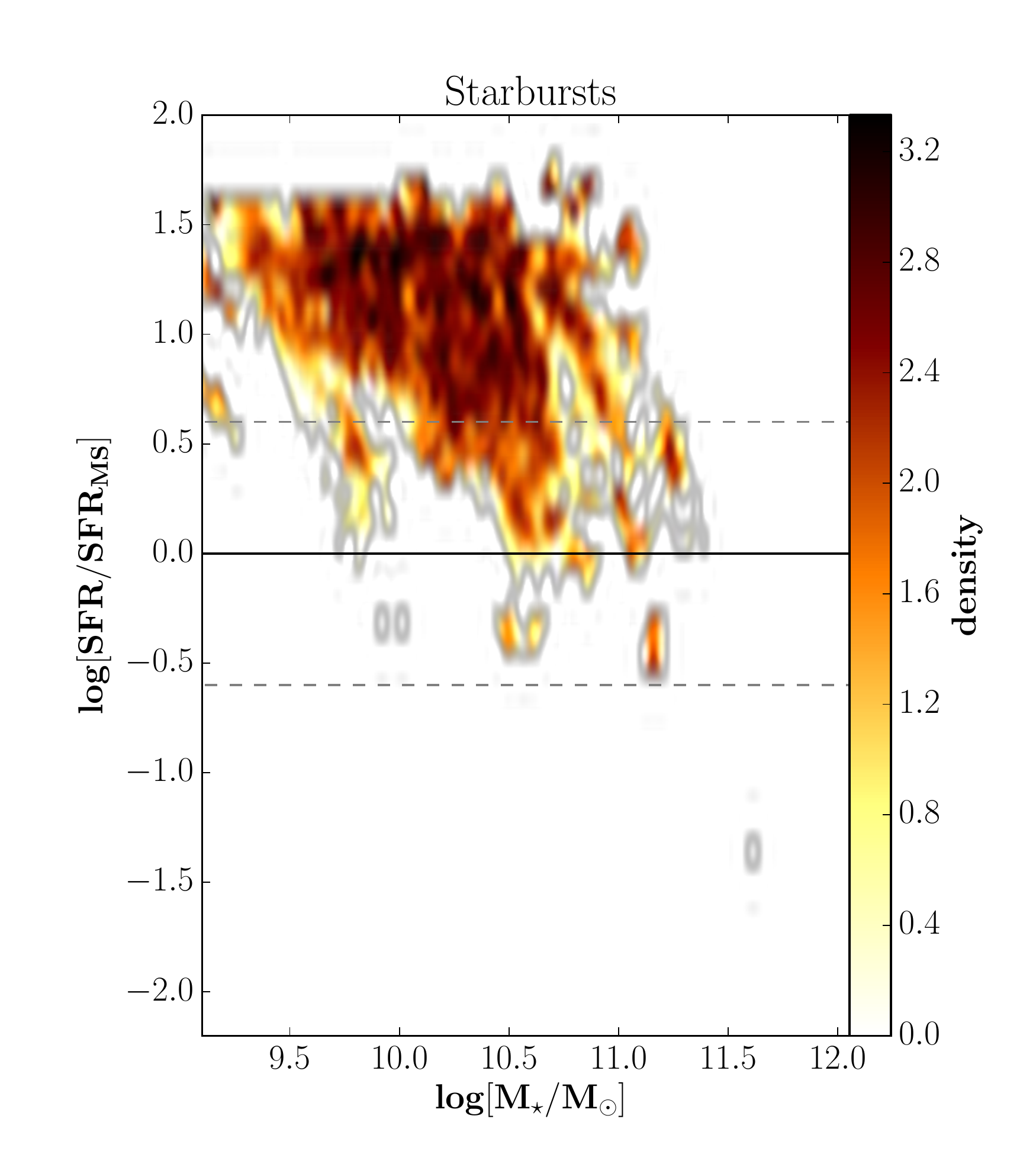} \hspace{-.2 in} \includegraphics[width=0.35\textwidth]{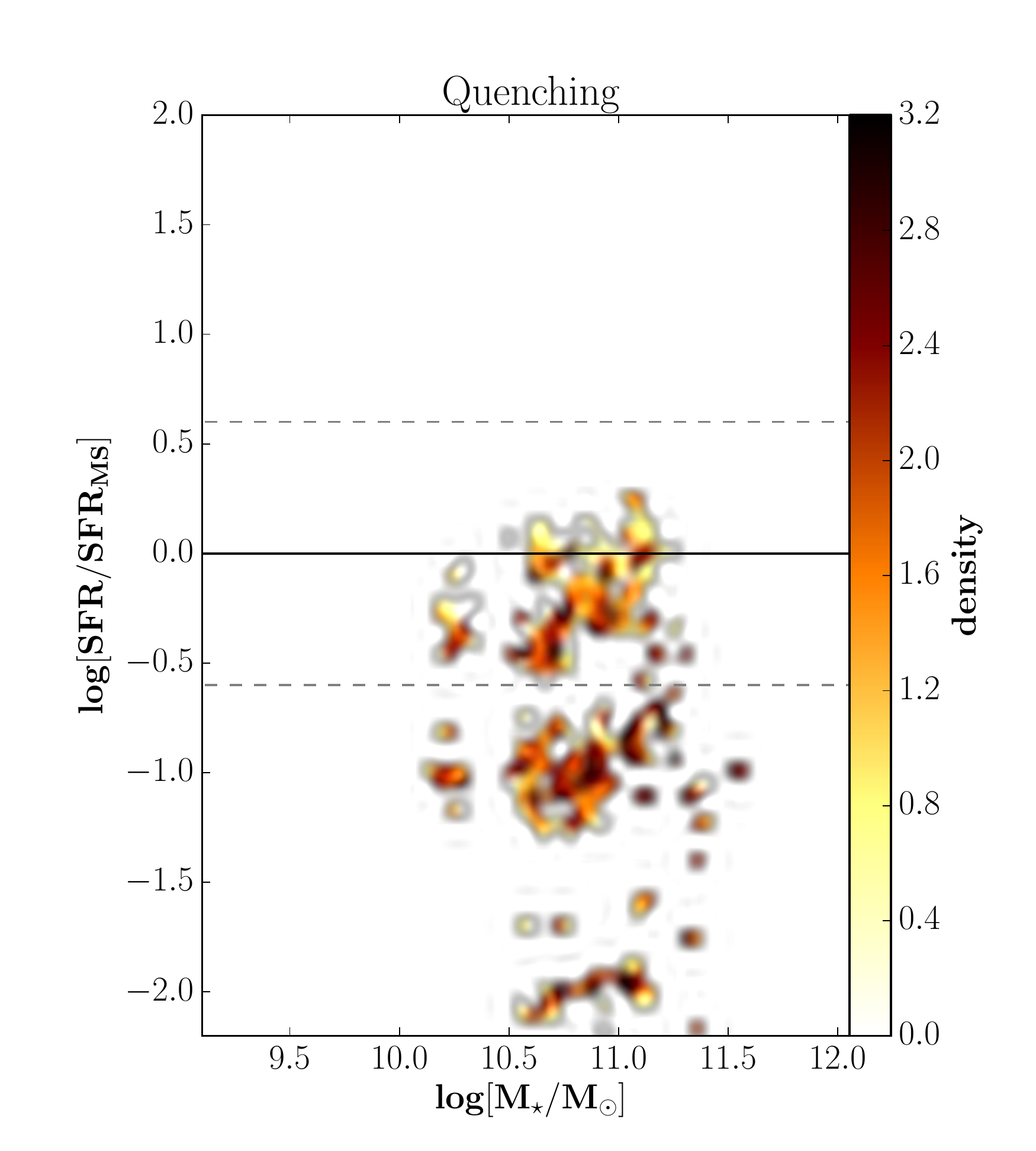}  
	\caption{Density plots in the SFR -- \mstar\ plane for the MC variations of the sample, separated as MS, SB, and quenching
	(left, center, and right, respectively) according to the best fits of the original photometry. Densities are in log. 
	In all groups, a large majority of the MC  best fits is found occupying the area where the original sources were lying. 
	}	\label{f_ms_density}
\end{figure*}

For each galaxy, the most common SFH type among the 1000 variations matches the best-fitting SFH of 
the original photometry in 78.3\%\ of the sample (83.3\% in GOODS-South which has the deepest and best constrained photometry), with
preference percentages explicited below.

When comparing the \ki2 of every fit obtained with the declining set of SFHs 
to the \ki2 of the same fitted variations from the rising set, we find that for 
every source, a large majority of the MC variations prefers one SFH over the other, usually with 
a rate close to 100\% (the most ``disputed'' case found has 67.6\% 
of its variations favoring the declining SFHs vs 32.4\% for the rising, and fewer than
10 sources present a split more balanced than 70-30). Evidently, when considering a comparison 
including the delayed SFH, the "scores" become more evenly split, as these models 
can behave both like the declining or the rising depending on every particular case.

Always based on the \ki2's, the best fitting SFH for every MC variation was kept to check
their prevalence among the distinguished subgroups, and to assess how well confined the area 
they occupy on the SFR -- \mstar\ diagram is. 
The SFH preferences are summarized in Table \ref{t_sfhs_mc}, where we can see that the percentages
of each SFH can vary substantially among the three groups. The last row, showing the preferences for the
whole energy-conserving sample, are very comparable with those from Table \ref{tab_sfh} (fixed $A_V$), indicating that 
the MC catalog behaves in the broad lines similarly to the original photometry sample.

\begin{table}[htb]
 	\centering
	\begin{tabular}{l c c c c}
	\hline\hline
	 Group & CSFR(\%)  & DECL(\%) & DEL(\%)  &  RIS(\%) \\\hline	
	 Main Seq. &3.53  &47.21  & 17.05  & 32.21   \\
	 Starbursts &  1.03  & 12.70 &  26.06 &  60.21   	\\	
	 Quenching  &  $<$ 0.01  & 75.4  & 24.6 &  0.0    \\
	 All  &2.68  &40.65  &19.76  &36.91  \\
	\hline
	\end{tabular}
  \caption{SFH preferences for the three subgroups of the sample (and all summed together in the last row) 
  as obtained by the SED fits of the  MC variations of the photometry.}	\label{t_sfhs_mc}
\end{table}

Fig. \ref{f_ms_density} shows how the fitted MC variations are distributed on the SFR -- \mstar\ plane, in the form of a density plot
to assess the robustness of the labeling (MS, SB, and quenching) given to the studied sources in our work. 
Each panel shows the results of the MC fits belonging to sources whose original photometry fits casted them to belong in one of the
aforementioned categories (i.e., the left panel shows the data from the MC fits of sources initially labeled as MS, the middle panel those initially
labeled as SB, and the right one for the quenching).  
This allows us to visualize how stable the obtained SFHs are in a statistical manner. 

From this we remark the following: 81.5\% of the MC fits from MS sources is found within the MS, with the rest mainly being interpreted as more starbursty;
92\% of the MC fits from SB sources is found above the MS; finally, 66.1\% of the MC fits of quenching sources are found below the MS, but with 95\% of them 
favoring $\taf \leq 100$ Myr (the 33.9\% found within the MS is to be contrasted with the 7 sources out of the 43 labeled as quenching that were found
within the MS in the original photometry analysis).
The marginal overlap between the MS and SB is explained by the fact that some of the MC variations can point toward solutions involving slightly lower masses
and higher SFRs (and hence more starbursty, like when a rising SFH is preferred over a declining SFH), and vice versa. But as discussed, this happens for 
a small minority of cases.

\paragraph{Median values and 68\% confidence intervals of physical parameters.}

In Table \ref{t_uncertainties} we present the median values and 68\% confidence intervals of the main 
physical parameters describing our SED models. We recall that the extinction is obtained by the IRX ratio and served as input.
With respect to the physical parameters of the original photometry fits, we find no particular deviations in the 
quantities obtained in the MC fits.  

\begin{table*}[htb]
 	\centering
	\begin{tabular}{l c c c c c}
	\hline\hline
 		& $t$ (Gyr) & \taf\ [Gyr] & \mstar\ [$10^{10} \msun$]  & SFR [\msun$\cdot$ yr$^{-1}$] & \av\ (mag) \\ \hline
Main Sequence	&  0.36 (0.09--2.6)    & 0.3 (0.03--2.0)   & 4.47 (1.72--11.2) 	 	& 79.84 (33.51--211.2)	       &  1.7 (1.2--2.4) \\
Starbursts	& 0.052 (0.052--0.255) & 0.03 (0.03--0.3)  & 1.57 (0.62--3.51)  	& 340.1 (143.2--894.5)         &  2.24 (1.5--3.2) \\
Quenching	& 0.255 (0.128--0.255) & 0.03 (0.03--0.05) & 7.96 (4.42--14.7)  	& 11.28 (1.57--35.24)	       &  2.0 (1.5--2.5) \\	
 	\hline
	\end{tabular}
  \caption{Median values of relevant physical parameters and 68\% confidence intervals (in parentheses), as obtained by the SED fits 
  of the MC variations.}	\label{t_uncertainties}
\end{table*}


\section{Examples of SED fits from our analysis}		\label{s_sed_figs}

In the present section we show SED plots for a number of selected sources, with the best fitting solutions when
the extinction is unconstrained and when assuming energy conservation, plotted together for visual comparison. 
To economize space on the plots, the expression $\log(L_{\rm IR,predicted} /L_{\rm IR,observed}) $ is shortened
to $\Delta \rm IR$. All figures are shown in the $\nu F_\nu$ formalism, that reduces the range in the y-axis and 
offers more detail.

\begin{figure*}[htb]
  \centering
  \begin{tabular}{l r}
     \includegraphics[width=0.5\textwidth]{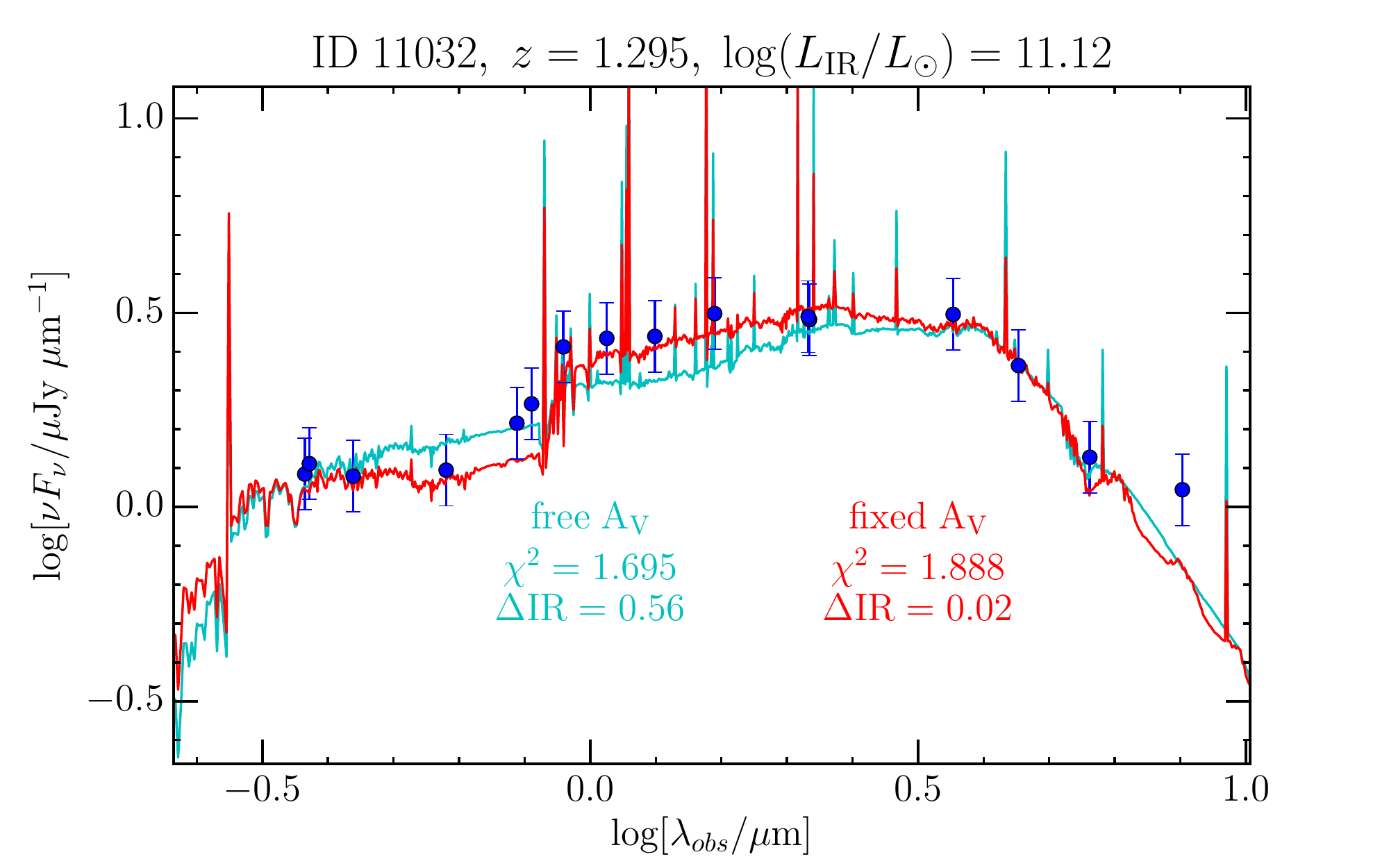} &  \includegraphics[width=0.5\textwidth]{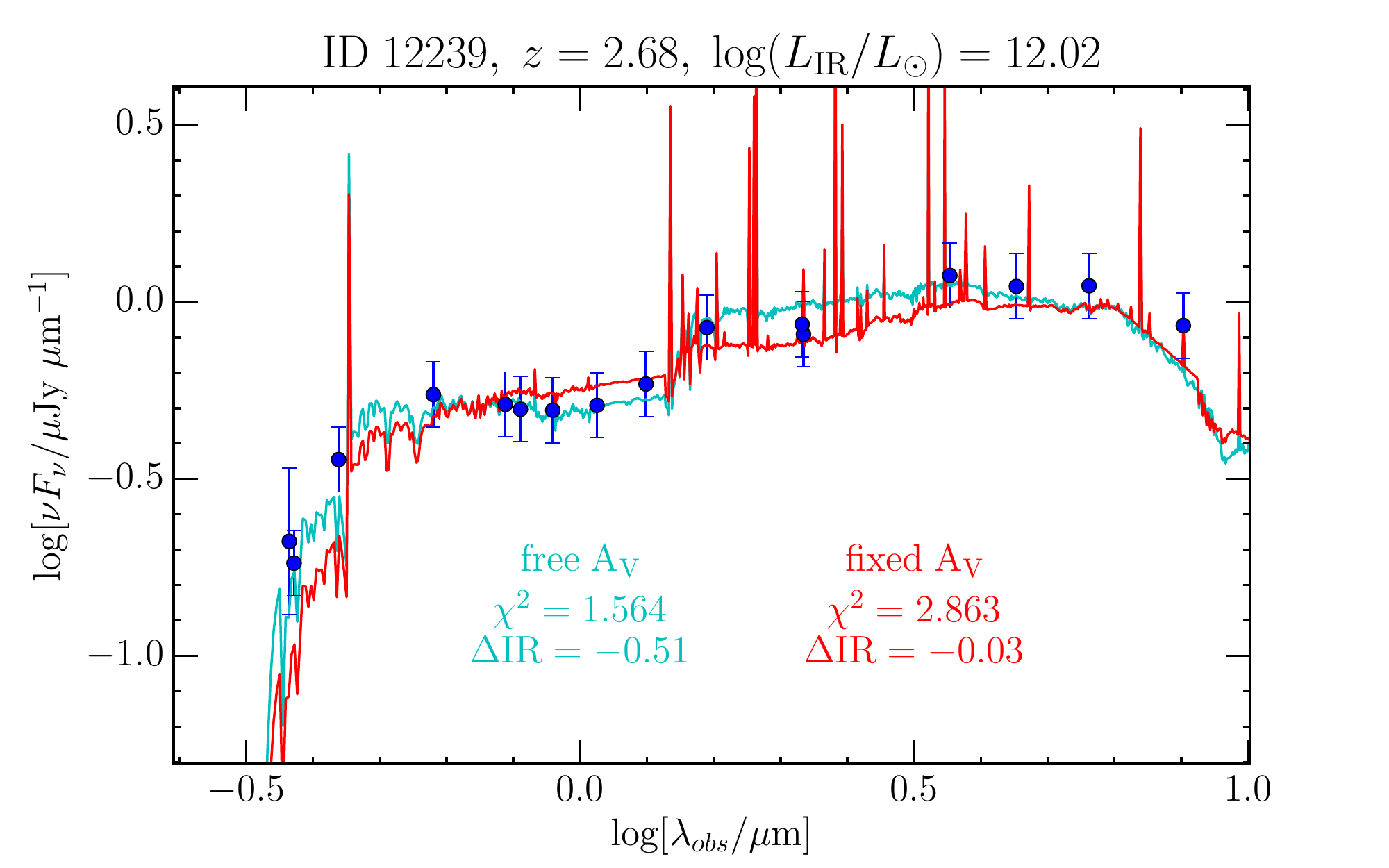}\\
      \includegraphics[width=0.5\textwidth]{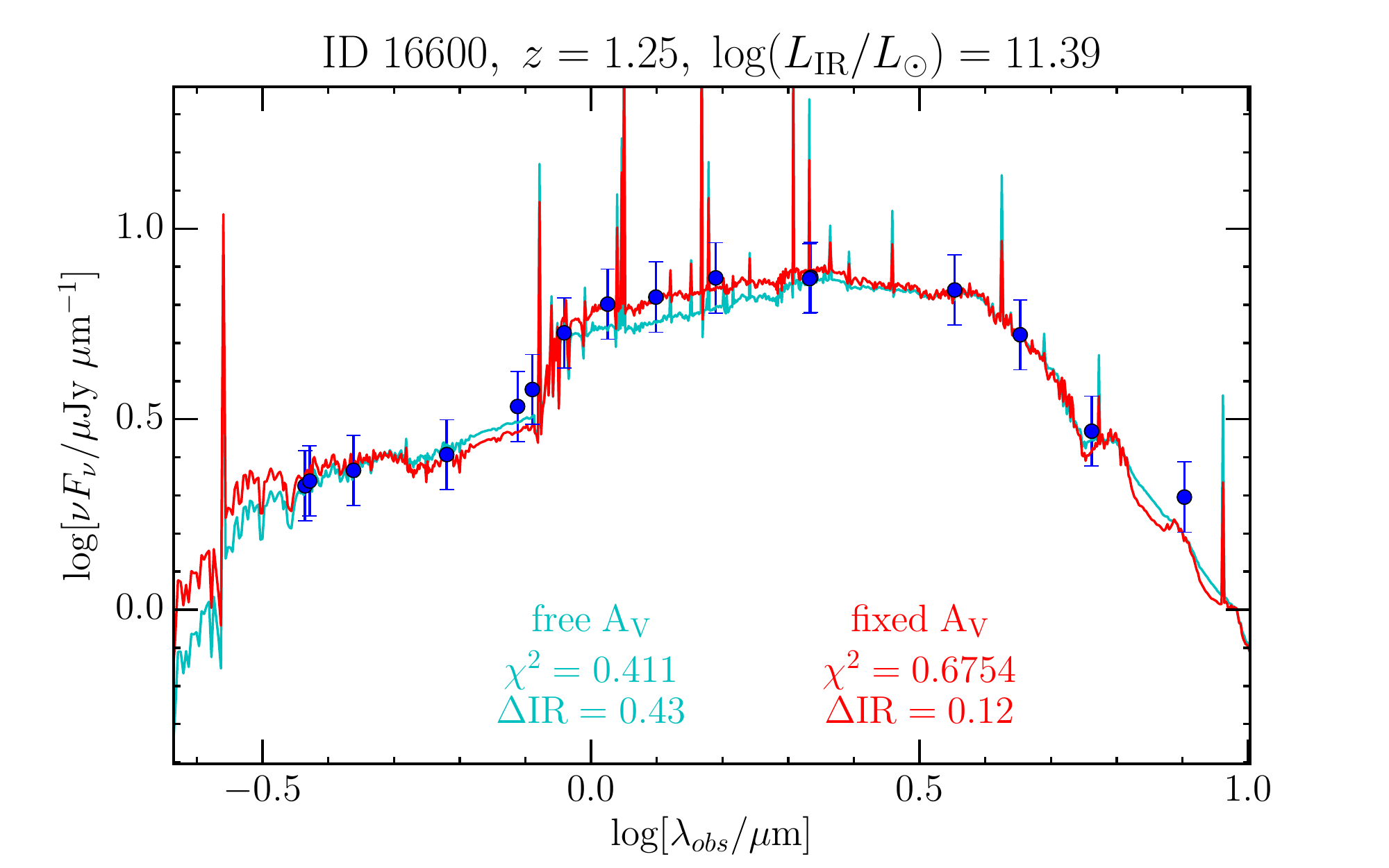} &  \includegraphics[width=0.5\textwidth]{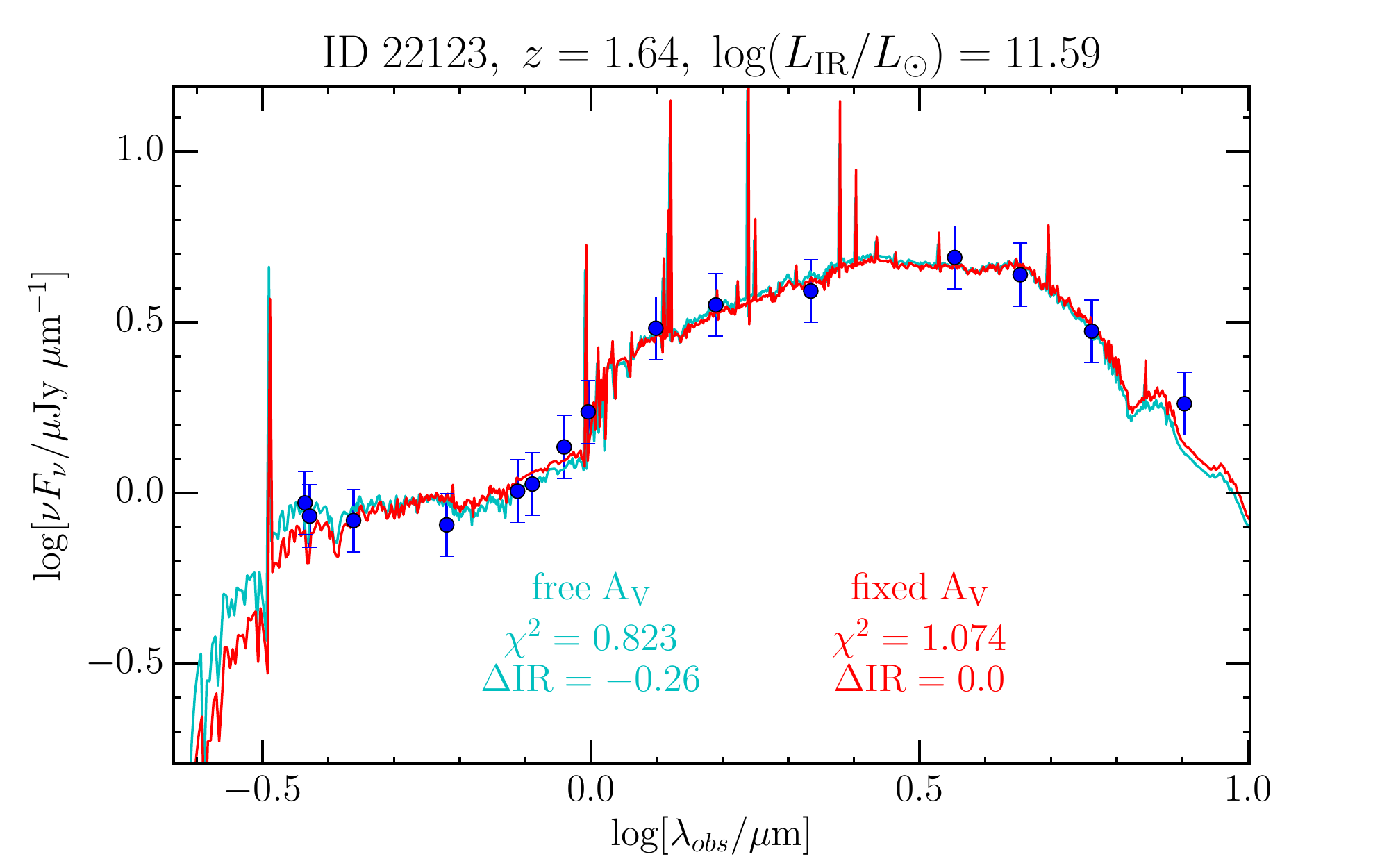}\\
      \includegraphics[width=0.5\textwidth]{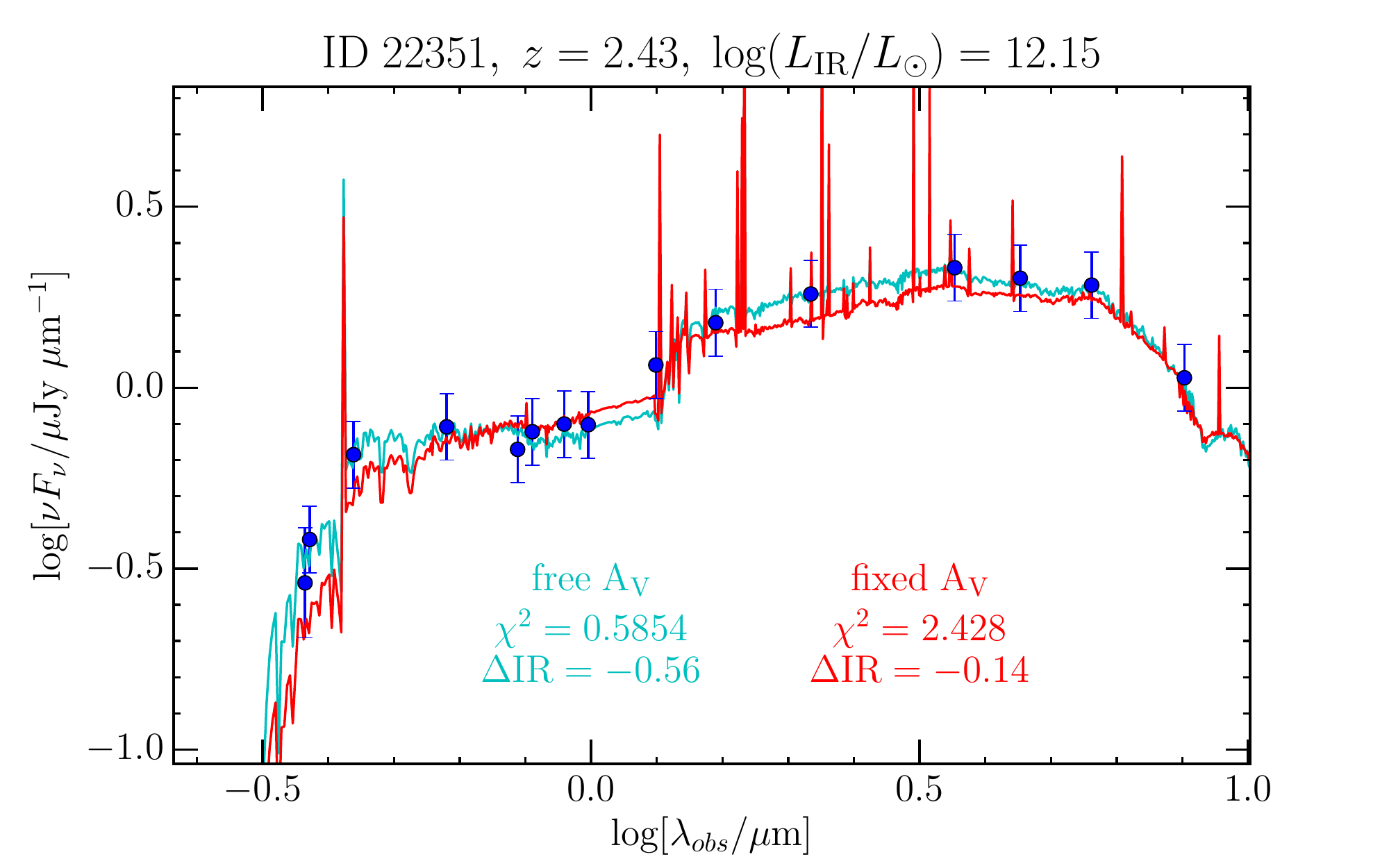} &  \includegraphics[width=0.5\textwidth]{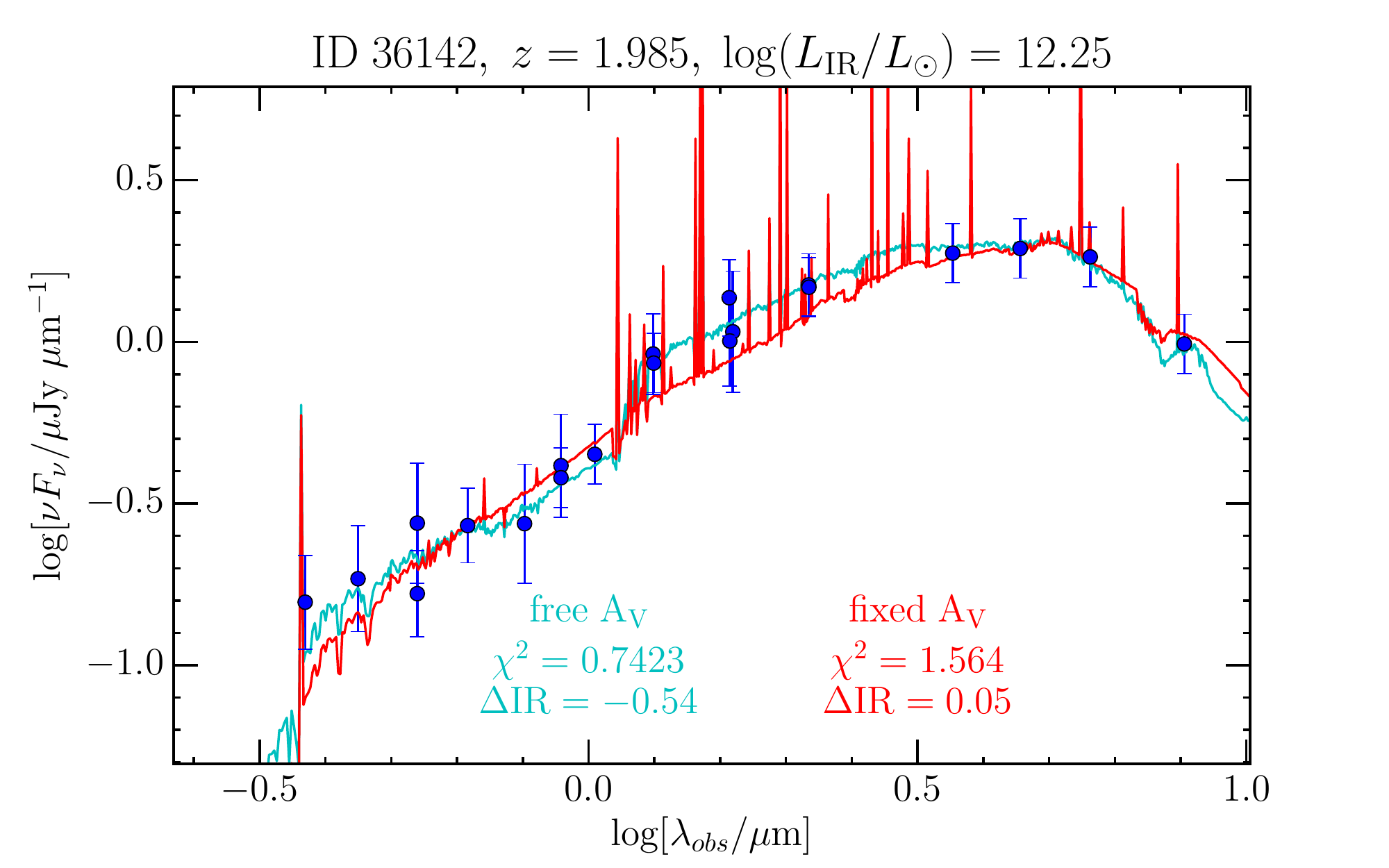}\\
    \end{tabular}
    \caption{Examples of  galaxies fitted with the standard approach (``free \av'', cyan), and with the energy conserving approach (``fixed
    \av'', red). The photometry is shown with blue circles, and the error bars are 1$\sigma$.
    For each fit, we show the reduced \ki2\ and the ratio of predicted over observed \lir\ ($\Delta {\rm IR} = \log(L_{\rm IR,predicted} /L_{\rm IR,observed})$).
    Although in some cases the ``before and after'' SEDs look similar in first order, they can represent different sets of parameters, illustrating
    how the \lir\ constrain helps in finding solutions that are more consistent with the observations.
  }			\label{f_seds_good}
\end{figure*}

\begin{figure*}[htb]
  \centering
  \begin{tabular}{l r}
       \includegraphics[width=0.5\textwidth]{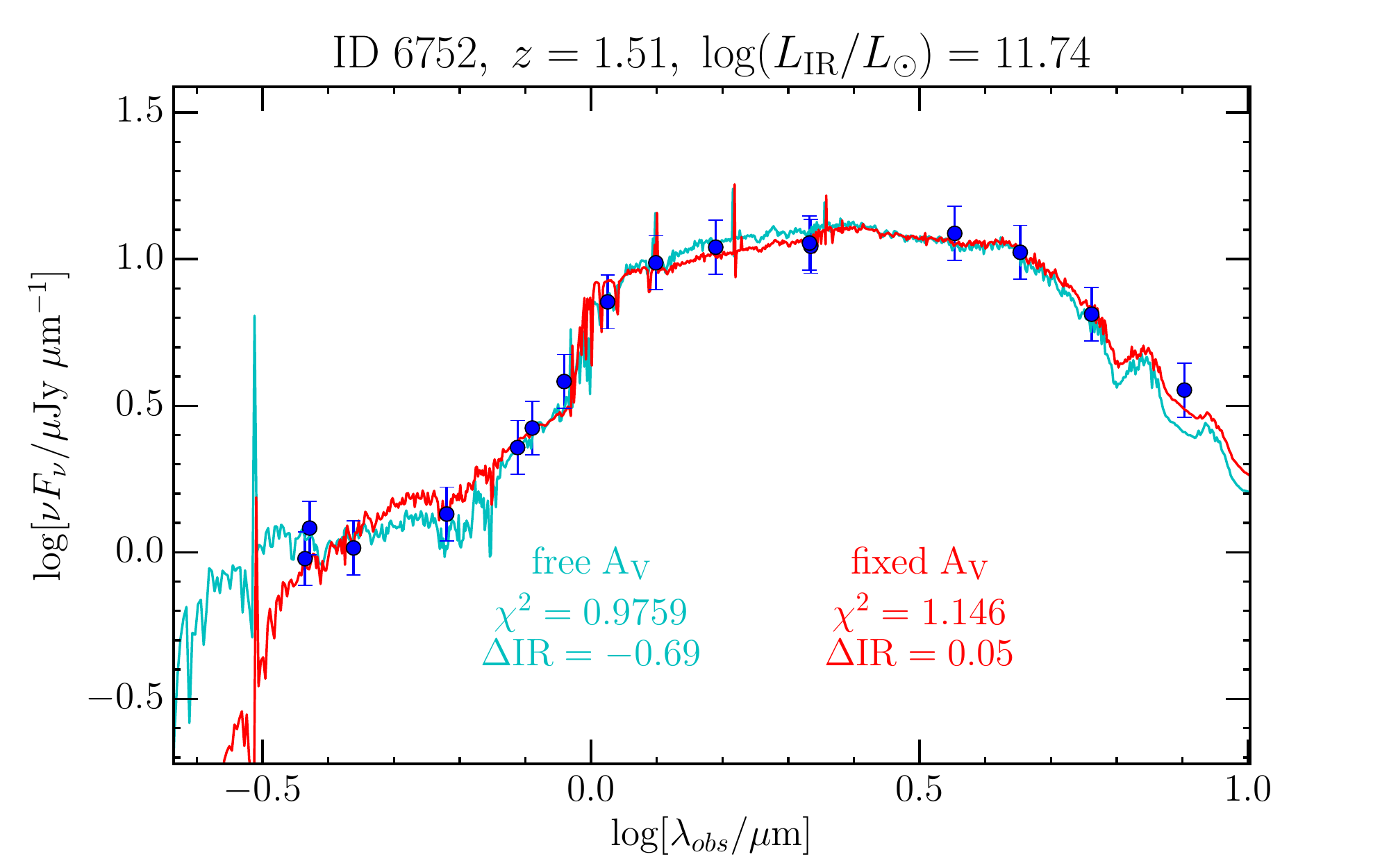} &  \includegraphics[width=0.5\textwidth]{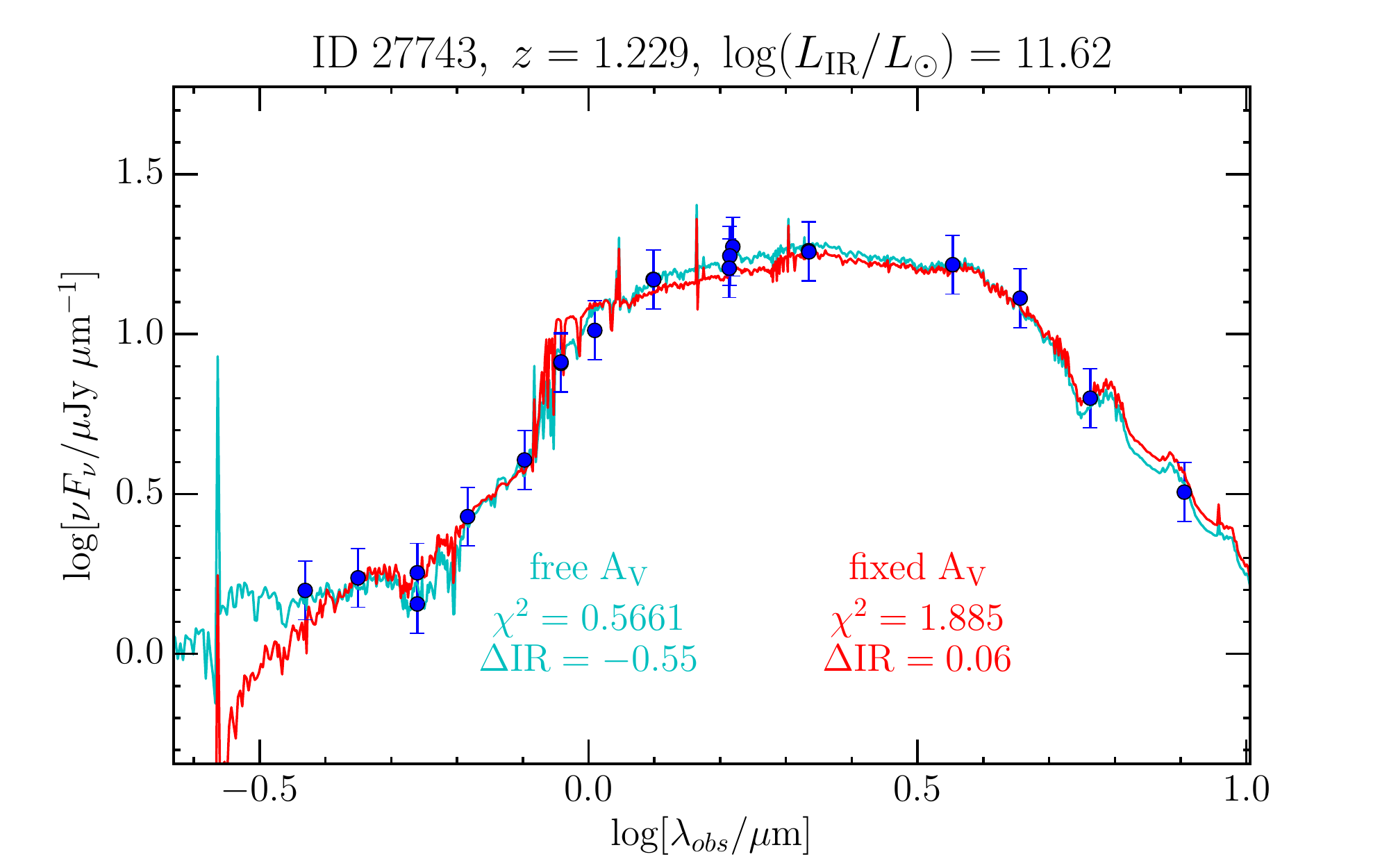}\\
      \includegraphics[width=0.5\textwidth]{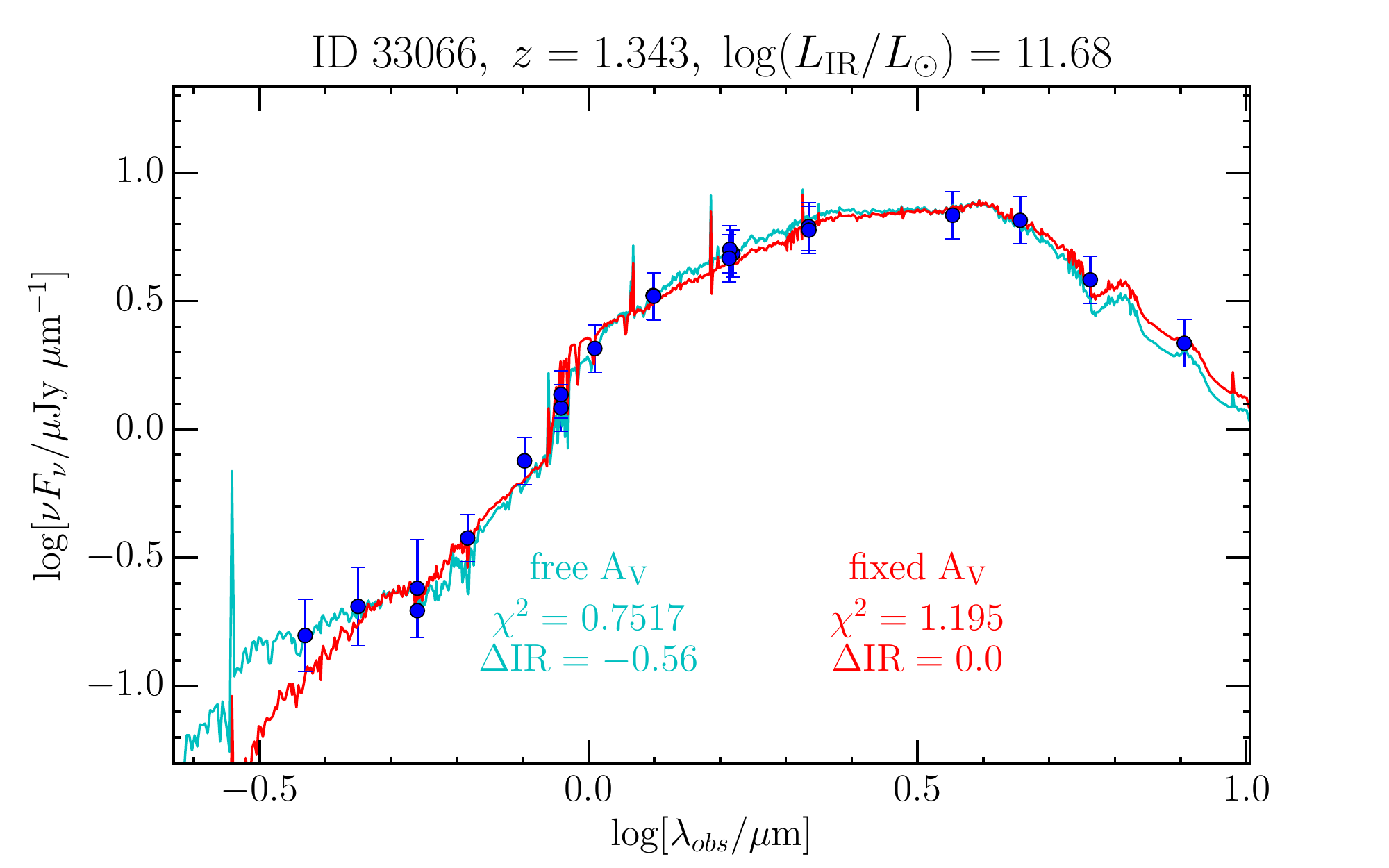} &  \includegraphics[width=0.5\textwidth]{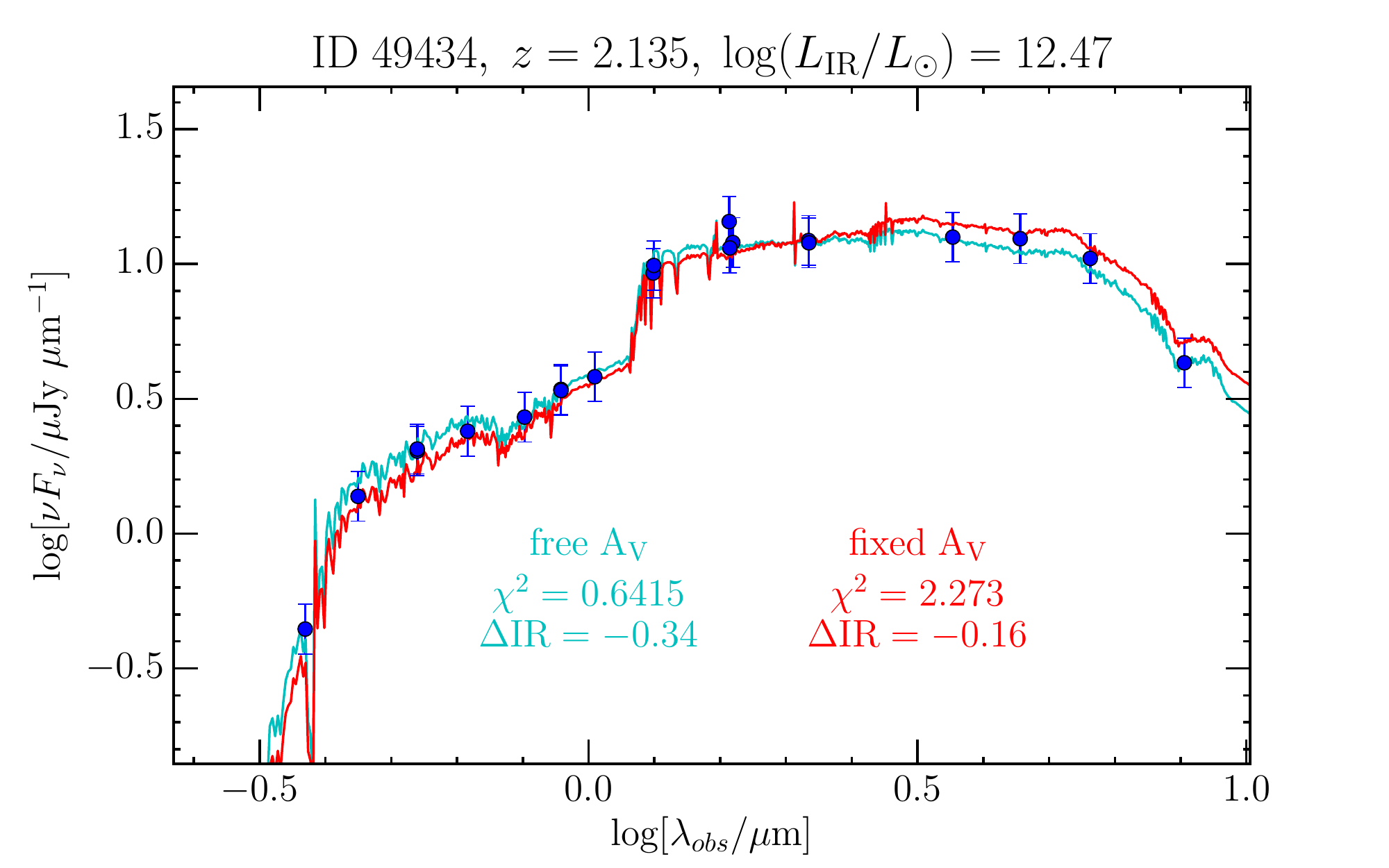}\\
   \end{tabular}
  \caption{
  Same as Fig. \ref{f_seds_good}, only the present SED fits are taken from the quenching subsample only.
   }			\label{f_seds_q}
\end{figure*}

\begin{figure*}[htb]
  \centering
  \begin{tabular}{l r}
     \includegraphics[width=0.5\textwidth]{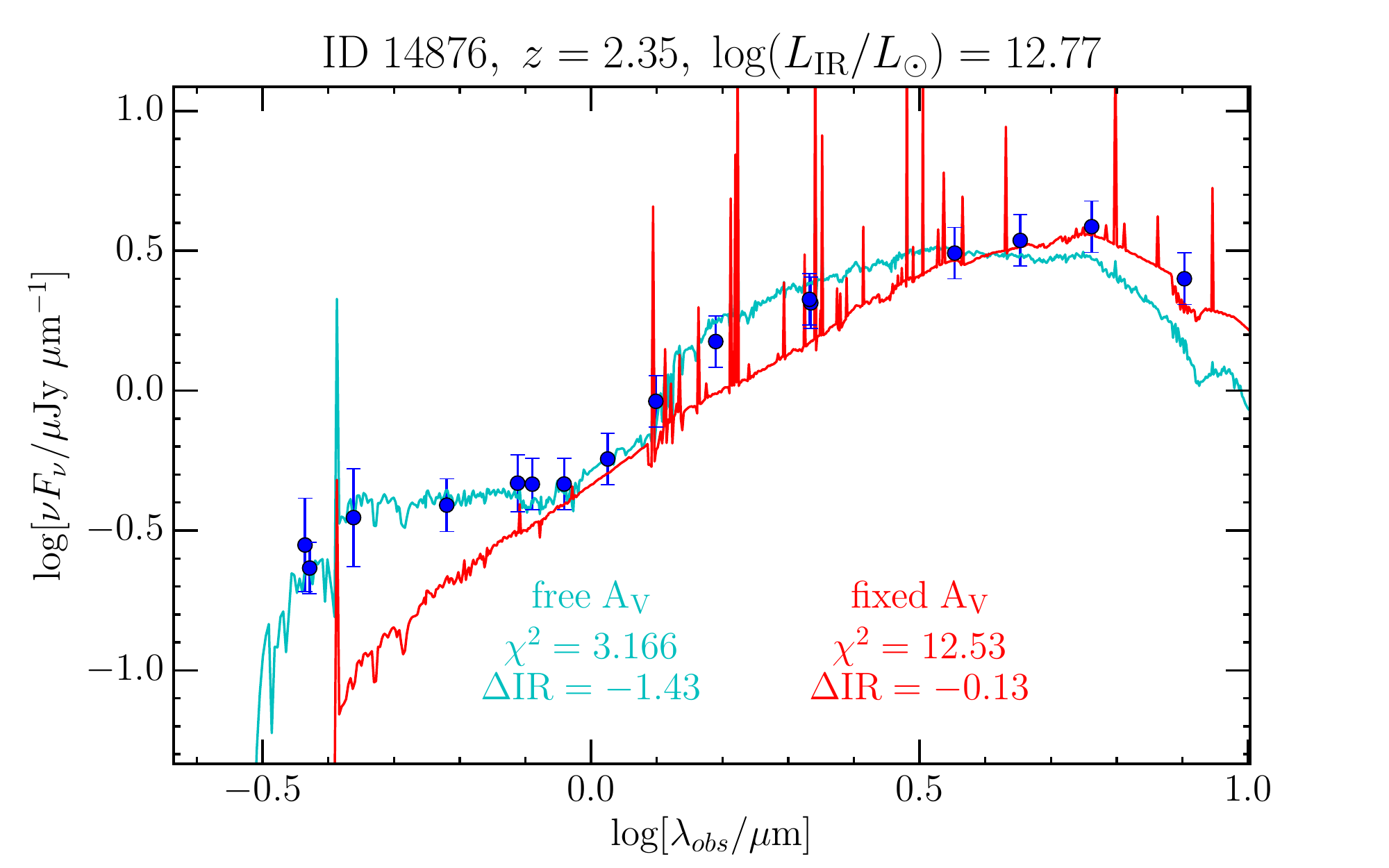} &  \includegraphics[width=0.5\textwidth]{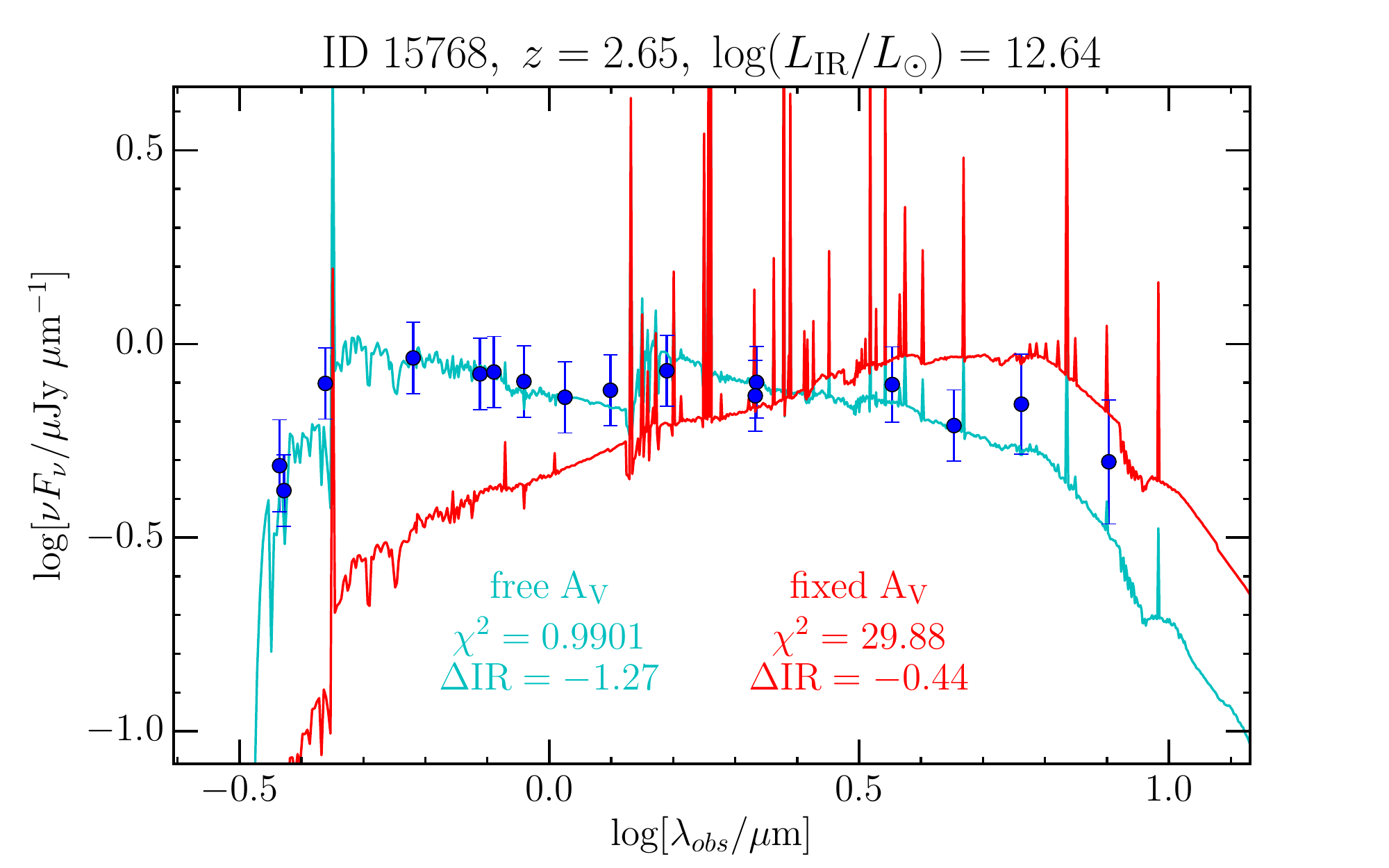}\\
      \includegraphics[width=0.5\textwidth]{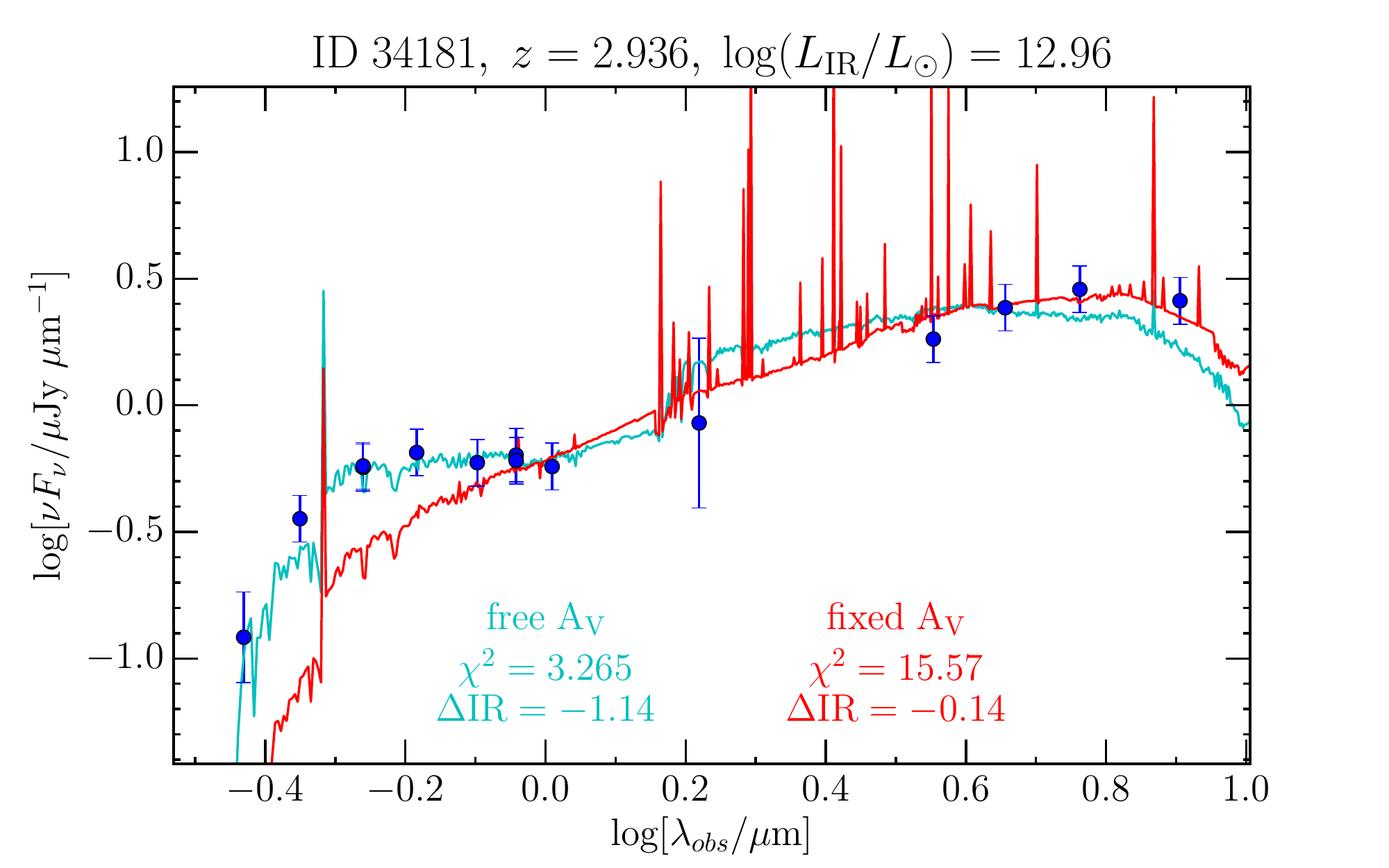} &  \includegraphics[width=0.5\textwidth]{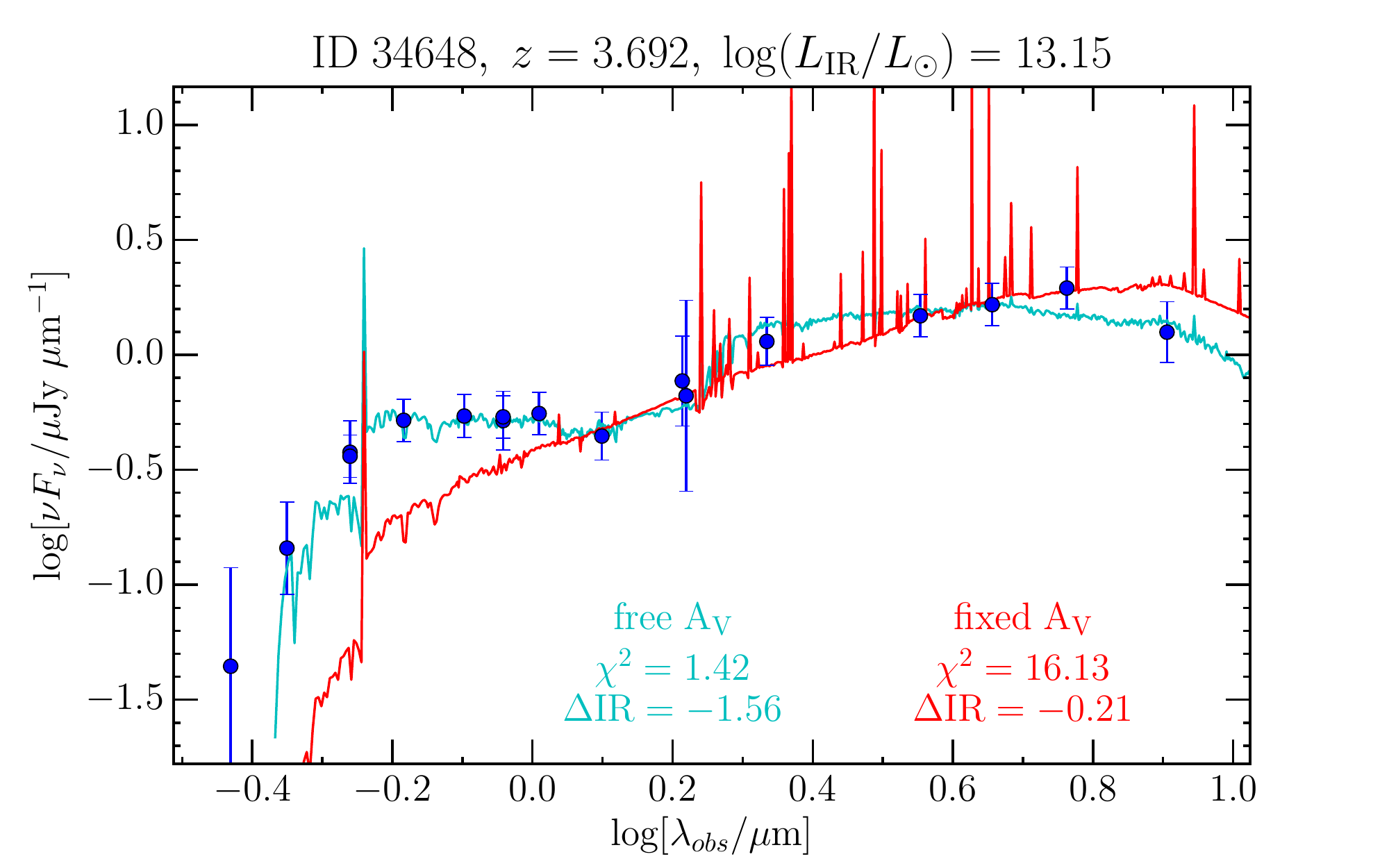}\\
   \end{tabular}
  \caption{A selection among the sources for which the energy conserving approach does not produce good quality fits. We recall that
  those sources lie on the high \lir\ end, and have bluer slopes than what is inferred by their IRX ratios. As a consequence, the
  difference between the free and fixed \av\ solutions are significant. While the unconstrained fits (cyan) range from moderately good to very good,
  they underpredict the \lir\ by more than 1 dex. The fixed \av\ fits (red) present SEDs that are incompatible with the photometry, and in cases
  still do not account well for the \lir.
  Colors and symbols as in \ref{f_seds_good}.
   }			\label{f_seds_blu}
\end{figure*}

Fig. \ref{f_seds_good} shows examples representative of the majority of the sample, where our  energy conservation produces
 good to very good fits, and finds solutions consistent with the observed IR luminosities.  Fig. \ref{f_seds_q} shows the 
sources tagged as quenching in particular, where most often what are perceived as old-passive populations in the unconstrained fits are 
re-modeled as younger post-starbursts with more residual star formation and enough dust to account for the IR observations. 
Fig. \ref{f_seds_blu} shows some of the most striking cases of the sources discussed in Sect. \ref{s_caveat}, where galaxies are simultaneously
too blue and IR bright, and can't be modeled properly using our simple approach.
 

\end{document}